\documentclass[aps,prd,reprint,10pt,longbibliography,nofootinbib]{revtex4-2}
\usepackage{textcomp}
\usepackage{amsmath}
\usepackage{amsfonts}
\usepackage{calc}
\usepackage{mathrsfs}
\usepackage{amssymb}
\usepackage{mathtools}
\usepackage{array}
\usepackage{color}
\usepackage{bm}
\usepackage{graphicx}
\usepackage{hyperref}
\usepackage{tensor}
\usepackage[italicdiff]{physics}
\usepackage{bbding}
\renewcommand{\star}{\text{\tiny \FiveStarOpen}}
\usepackage{xcolor}

\renewcommand{\S}{{\rm S}}
\newcommand{\R}{{\rm R}}
\newcommand{\e}{\epsilon}

\newcommand{\beq}{\begin{equation}}
\newcommand{\eeq}{\end{equation}}

\DeclareMathOperator*{\STF}{STF}
\DeclareMathOperator*{\sym}{Sym}

\newcommand{\E}{\mathcal{E}}
\newcommand{\B}{\mathcal{B}}

\newcommand{\Edot}{\mathcal{\dot{E}}}
\newcommand{\Bdot}{\mathcal{\dot{B}}}
\newcommand{\nhat}{\hat{n}}
\newcommand{\gzero}{\mathring{g}}
\newcommand{\lie}{\mathcal{L}}
\newcommand{\hronehr}{\,{}^0 h^{\mathrm{R1}}}
\newcommand{\hronehrpl}[1]{\,{}^0 h^{\mathrm{R1}(#1)}}
\newcommand{\inthrhr}[1]{\int\!\hronehr#1\dd{t}}
\newcommand{\hrstar}{h^{\mathrm{R1*}}}
\newcommand{\hrtext}[1]{h^{\mathrm{#1}}}

\newcommand{\egamma}{\big|_{\gamma}}
\newcommand{\rone}{\mathrm{R1}}

\newcommand{\lieZ}{\pounds}
\newcommand{\hdm}{h^{\delta m*}}
\newcommand{\rtilde}{\tilde{r}}
\newcommand{\gobj}{g^{\text{obj}}}
\newcommand{\gobjp}[1]{g^{\text{obj},#1}}
\newcommand{\fullg}{\mathsf{g}}

\newcommand{\That}[1]{{\hat{T}}^{(#1)}}
\newcommand{\Xhat}[1]{{\hat{X}}^{(#1)}}
\newcommand{\Yhat}[1]{{\hat{Y}}^{(#1)}}
\newcommand{\Zhat}[1]{{\hat{Z}}^{(#1)}}
\newcommand{\Ahat}[1]{{\hat{A}}^{(#1)}}
\newcommand{\Bhat}[1]{{\hat{B}}^{(#1)}}
\newcommand{\Chat}[1]{{\hat{C}}^{(#1)}}
\newcommand{\Dhat}[1]{{\hat{D}}^{(#1)}}
\newcommand{\Ehat}[1]{{\hat{E}}^{(#1)}}
\newcommand{\Fhat}[1]{{\hat{F}}^{(#1)}}
\newcommand{\Ghat}[1]{{\hat{G}}^{(#1)}}
\newcommand{\Hhat}[1]{{\hat{H}}^{(#1)}}
\newcommand{\Ihat}[1]{{\hat{I}}^{(#1)}}
\newcommand{\Khat}[1]{{\hat{K}}^{(#1)}}

\newcommand{\calP}{\mathcal{P}}
\newcommand{\calR}{\mathcal{R}}
\newcommand{\deltaG}{\delta G}
\renewcommand{\O}{{\cal O}}

\begin{document}

\title{Second-order gravitational self-force in a highly regular gauge}
\author{Samuel D.\ Upton}
\author{Adam Pound}
\affiliation{School of Mathematical Sciences and STAG Research Centre, University of Southampton, Southampton, United Kingdom, SO17 1BJ}
\date{\today}

\begin{abstract}
Extreme-mass-ratio inspirals (EMRIs) will be key sources for LISA. However, accurately extracting system parameters from a detected EMRI waveform will require self-force calculations at second order in perturbation theory, which are still in a nascent stage.
One major obstacle in these calculations is the strong divergences that are encountered on the worldline of the small object.
Previously, it was shown by one of us [\href{https://doi.org/10.1103/PhysRevD.95.104056}{Phys. Rev. D 95, 104056 (2017)}] that a class of ``highly regular'' gauges exist in which the singularities have a qualitatively milder form, promising to enable more efficient numerical calculations.
Here we derive expressions for the metric perturbation in this class of gauges, in a local expansion in powers of distance $r$ from the worldline, to sufficient order in $r$ for numerical implementation in a puncture scheme.
Additionally, we use the highly regular class to rigorously derive a distributional source for the second-order field and a pointlike second-order stress-energy tensor (the {\em Detweiler stress-energy}) for the small object.
This makes it possible to calculate the second-order self-force using mode-sum regularisation rather than the more cumbersome puncture schemes that have been necessary previously.
Although motivated by EMRIs, our calculations are valid in an arbitrary vacuum background, and they may help clarify the interpretation of point masses and skeleton sources in general relativity more broadly.
\end{abstract}

\maketitle

\section{Introduction}\label{sec:intro}

Given the approaching launch of the Laser Interferometer Space Antenna (LISA)~\cite{LISA2017,LISAWeb}, it is essential that we are able to efficiently and accurately model the gravitational waves emitted by potential sources in the LISA frequency band.
A class of sources of particular interest are extreme-mass-ratio inspirals (EMRIs)~\cite{Babak2017}.

In an EMRI, a compact object of mass \(m \sim 1\text{--}10^2 M_{\odot}\) spirals into a supermassive black hole of mass \(M \sim 10^5\text{--}10^7M_{\odot}\), performing roughly \( \e^{-1} \sim 10^5\) intricate orbits while in the LISA band~\cite{AmaroSeoane2015,BHroadmap}, where \(\e \coloneqq m/M\) is the binary's mass ratio.
These $\sim 10^5$ cycles are typically spent within 10 Schwarzschild radii of the supermassive black hole, and they  provide a precise map of its strong-field geometry.
This will allow numerous tests of general relativity, usually with one or more orders of magnitude greater precision than other planned experiments~\cite{Gair2013,BHroadmap}.

However, the long life of the inspiral also imposes stringent accuracy requirements on our models. To make use of a complete signal, and extract all the information it encodes, we require theoretical waveforms that are accurate to much less than one radian of error over the duration of the signal.
Because errors accumulate secularly over the signal's $\e^{-1}$ cycles, they are effectively multiplied by $\sim 10^5$. This means our relative errors in the system's slowly evolving orbital frequencies  must be much smaller than $\e\sim 10^{-5}$.

\subsection{Second-order gravitational self-force}\label{sec:gsf}

The most viable method of modelling an EMRI system to the required accuracy is with gravitational self-force theory~\cite{Poisson2011,Pound2015Small,Wardell2015,Barack2018,Pound2021}. In this approach, the small object is treated as the source of a small perturbation, \(h_{\mu\nu}\), on a background metric, \(g_{\mu\nu}\), enabling us to write the full spacetime metric as
\begin{equation}
    \mathsf{g}_{\mu\nu} = g_{\mu\nu} + h_{\mu\nu}, \label{eq:fullmetric}
\end{equation}
where
\begin{equation}
    h_{\mu\nu} = \sum_{n\geq 1}^{\infty} \e^{n}h^{n}_{\mu\nu}[\gamma]. \label{eq:pert_expansion}
\end{equation}
Here the coefficients depend on $\gamma$, a worldline representing the small object's motion in the background spacetime $g_{\mu\nu}$. In the context of an EMRI, $g_{\mu\nu}$ is taken to be the metric of the central Kerr black hole.

At zeroth order, $\gamma$ is a geodesic of the background spacetime. At subleading orders, the metric perturbation alters the small object's motion, exerting a {\em self-force} that accelerates the object away from geodesic motion:
\beq\label{eq:EOM generic}
\frac{D^2z^\alpha}{d\tau^2} = \e f_1^\alpha + \e^2 f_2^\alpha +\O(\e^3),
\eeq
where $z^\alpha$ are coordinates on $\gamma$, $\tau$ is proper time in $g_{\alpha\beta}$, $\frac{ D}{d\tau}:=u^\alpha\nabla_{\alpha}$ is the covariant derivative (compatible with $g_{\alpha\beta}$) along the worldline, and $u^\alpha:=\frac{dz^\alpha}{d\tau}$ is the four-velocity.

The self-forces $f_n^\alpha$ have both conservative and dissipative effects. Roughly speaking, in an EMRI the conservative pieces of the self-force determine the instantaneous orbital frequencies. In an expansion $\Omega=\Omega^{(0)} + \e\,\Omega^{(1)}+\O(\e^2)$ of the azimuthal frequency, for example, the zeroth-order term is that of a geodesic in $g_{\mu\nu}$, and the corrections $\Omega^{(n)}$ are determined by $f^\alpha_{n,\rm cons}$. The dissipative pieces of the self-force govern the slow evolution of the frequencies, with $f^\alpha_{n,\rm diss}$ determining $\frac{d\Omega^{(n-1)}}{d\tau}$ ~\cite{Hinderer2008,Miller2020,Pound2021}. Hence, to correctly track the orbital frequencies to within a relative error much smaller than $\e$, we must compute all of $f^\alpha_{1}$ and the dissipative piece of $f^\alpha_2$. In other words, we must carry self-force theory to second order in $\e$.

Currently, at first order it is possible to simulate full inspirals, driven by the first-order self-force, from a spinning small object on a generic orbit in a Schwarzschild background~\cite{Warburton2012,Osburn2016,Warburton2017,vandeMeentWarburton2018} or the adiabatic inspiral of an object in the equatorial plane of a Kerr black hole~\cite{Fujita:2020zxe}.
While no analogous inspirals have yet been computed for generic, inclined orbits in Kerr, it is also now possible to calculate the first-order self-force on any fixed bound orbit in Kerr~\cite{vandeMeent2018}.

At second order, it is only recently that numerical calculations of physical quantities have been performed, and they have been specialized to quasicircular orbits around a Schwarzschild black hole~\cite{Pound2020,Warburton:ICERM}. Much work remains to bring second-order calculations to the same state as current first-order calculations before the expected launch of LISA in 2034.

It should be emphasised that second-order calculations do not merely provide an improvement in accuracy over first-order calculations. The dissipative piece of $f^\alpha_2$ is {\em equally important} as the conservative piece of $f^\alpha_1$. This means that calculations of $f^\alpha_{1,\rm cons}$, which represent the bulk of the self-force community's efforts over the past two decades, do not improve the accuracy of a waveform unless they are complemented with calculations of $f^\alpha_{2,\rm diss}$. Both ingredients are equally crucial for performing useful science with LISA data.

Second-order self-force calculations also have applications to other binary models.
Information from gravitational self-force has been used to determine high-order terms in post-Newtonian theory, provide guidance for both post-Newtonian and post-Minkowskian theory, and refine effective one-body (EOB) models~\cite{Damour2010,Barack2010,Antonelli2020}. Second-order self-force results could be used to fully determine two-body dynamics through fifth post-Newtonian order (one order beyond the state of the art~\cite{Blanchet2013}) and through sixth post-Minkowskian order (two beyond the state of the art~\cite{Bern2021})~\cite{Damour2020}.

There is also an increasing body of evidence that the self-force formalism may be directly applicable to binaries well outside the EMRI regime~\cite{LeTiec2011,LeTiec2013,Rifat2020}.
In fact, self-force models may be reasonably accurate even for comparable mass ratios, \(\e \approx 1\), at least in certain areas of the parameter space~\cite{vandeMeent2020,Warburton:ICERM}.

This is particularly relevant after the recent detection of binaries with mass ratios \(\sim 1:4\)~\cite{LIGO2020} and \(\sim 1:10\)~\cite{LIGO2020June}, which indicate that gravitational self-force models could be used for current LIGO-Virgo sources.

\subsection{Self-force theory, singularity structure, and the problem of infinite mode coupling}\label{sec:puncture}

One of the main challenges at second order is coping with the strong divergence of $h^2_{\mu\nu}$ on the small object's worldline. At a practical level, it creates a major numerical burden, which we describe below. At a more foundational level, the strength of the singularity is intimately related to the fact that in a generic gauge, the field equations for $h^2_{\mu\nu}$ are not globally well defined: it contains source terms that are distributionally ill defined on any domain intersecting $\gamma$.\footnote{For introductory references on distribution theory see, e.g., Refs.~\cite{Richards1990,Hormann2009}.}
Our goal in this paper is to reduce the practical challenge by overcoming the foundational one.

To begin, in this section we review how the problem arises. We consider a generic spacetime containing a small object, which may be a material body, a black hole, or something more exotic. We assume the body is compact, with a diameter comparable to its mass $m$, and that $m$ is much smaller than an external length scale ${\cal R}$; in a binary, ${\cal R}$ could be the large mass $M$ or a characteristic orbital separation, but our analysis is not restricted to binary systems.  $\e$ will now be a formal expansion parameter we use to count powers of $m/{\cal R}$, and it can be set to \(1\) at the end of the calculation. Outside the body, we assume there is a vacuum region at least of size ${\cal R}$. All of these assumptions can be relaxed as long as the body's mass and diameter are much smaller than ${\cal R}$. 

Self-force theory provides a framework for solving the Einstein equations in this generic scenario. Its core result is a skeletonization of the small body~\cite{Pound2015Small}, in which (i) the body is reduced to a singularity equipped with the body's multipole moments, and (ii) the singularity moves as a test body immersed in a certain effective spacetime.

The derivation of this skeletonization is based on the method of matched asymptotic expansions~\cite{Pound2015Small}. Sufficiently near the small object, the object's own gravity dominates over the external background, and the expansion~\eqref{eq:fullmetric}--\eqref{eq:pert_expansion} fails. Hence, we assume that the metric in this region is instead approximated by a second expansion that zooms in on the body, and that this second expansion appropriately matches onto the expansion~\eqref{eq:fullmetric}--\eqref{eq:pert_expansion}. When combined with the vacuum Einstein equations, this matching condition determines the form of the metric perturbation $h_{\mu\nu}$ in a local neighbourhood of the object (but {\em outside} the object itself). That form splits conveniently into two fields~\cite{Pound2012SmallBody}:\footnote{There is a standard division into $\hrtext{S}_{\mu\nu}$ and $\hrtext{R}_{\mu\nu}$ at first order~\cite{Detweiler2003}, but in general the division is not unique. For the purpose of our discussion we can adopt any split in which $h^\R_{\mu\nu}$ is smooth on $\gamma$, the effective metric $\tilde g_{\mu\nu}=g_{\mu\nu}+h^\R_{\mu\nu}$ is a vacuum metric on and in a neighborhood of $\gamma$, and the equation of motion is that of a geodesic in $\tilde g_{\mu\nu}$ (through second order in $\e$ for a nonspinning, spherical body). These conditions do not select a unique S--R split~\cite{Pound2014}, but they are satisfied by the split(s) in Refs.~\cite{Pound2012,Pound2012SmallBody,Pound2017}. The alternative S--R divisions in Refs.~\cite{Gralla2012,Harte2011} each violate two of the conditions.}
\begin{equation}
    h_{\mu\nu} = \hrtext{S}_{\mu\nu} + \hrtext{R}_{\mu\nu}. \label{eq:regsing_split}
\end{equation}
The {\em self-field} (or {\em singular field}) \(\hrtext{S}_{\mu\nu}=\sum_{n>0} \e^n h^{{\rm S} n}_{\mu\nu}\) carries the local information about the object's multipole structure. In the local neighbourhood, at small proper distance $r$  from the worldline $\gamma$, the self-field has the schematic form
\begin{align}
    h^{\S 1}_{\mu\nu} &\sim \frac{m}{r},\label{eq:hS1form} \\ 
    h^{\S 2}_{\mu\nu} &\sim \frac{m^2 + M^a + S^a}{r^2}, \\
    h^{\S 3}_{\mu\nu} &\sim \frac{m^3 + m (M^a+S^a)+ M^{ab}+S^{ab}}{r^3},\label{eq:hS3form}
\end{align}
and so on to higher orders, where $m$ is the object's mass, $M^a$ its mass dipole moment (describing the position of its center of mass relative to $\gamma$), $S^a$ its spin, and $M^{ab}$ and $S^{ab}$ its mass and current quadrupole moments. If we choose $\gamma$ such that $M^a\equiv0$, then $\gamma$ represents the object's center of mass at first order~\cite{Gralla2008,Pound2010}. Conditions on corrections to $M^a$ enforce this mass-centredness at higher orders~\cite{Pound2017}. The proper spatial distance $r$, as well as the curve $\gamma$ itself, is defined in the external background spacetime with metric $g_{\mu\nu}$; if the object is a black hole or has a nontrivial internal topology, then $\gamma$ is not a curve in its true interior, but instead a curve in the smoothly extended external background manifold.

The other piece of the perturbation is an {\em effectively external field} or {\em regular field}, \(\hrtext{R}_{\mu\nu}=\sum_{n>0} \e^n h^{{\rm R} n}_{\mu\nu}\). It has the local form of a Taylor series around $\gamma$, 
\beq
h^{{\rm R} n}_{\mu\nu} = \sum_{\ell\geq0} A^n_{\mu\nu a_1\cdots a_\ell}(t)x^{a_1}\cdots x^{a_\ell}, \label{eq:hRn form}
\eeq
where $x^a$ are local spatial coordinates centered on $\gamma$, and the time $t$ outside the body is synchronized with a time parameter along $\gamma$.  $\hrtext{R}_{\mu\nu}$ is a vacuum solution that carries no local information about the object's moments, but instead contributes to the external tidal moments that the object feels.
It can be combined with the external background to define an effective metric,
\begin{equation}
    \tilde{g}_{\mu\nu} = g_{\mu\nu} + \hrtext{R}_{\mu\nu}, \label{eq:eff_metric}
\end{equation}
which is a vacuum metric, and which governs the motion of the body; we return to these points momentarily.

The fields $h^\S_{\mu\nu}$ and $h^\R_{\mu\nu}$ are initially defined in the vacuum region outside the object. However, if we analytically extend them into the object's effective interior, down to all points $r>0$, then several things happen. First, the fields $h^{\S n}_{\mu\nu}$ diverge on $\gamma$; this is the singularity mentioned above. Second, the coefficients $A^n_{\mu\nu a_1\cdots a_\ell}$ in $h^{\R n}_{\mu\nu}$ become trivially identified with the value of $h^{\R n}_{\mu\nu}$ and its derivatives on $\gamma$, such that Eq.~\eqref{eq:hRn form} becomes
\beq
h^{{\rm R} n}_{\mu\nu} = \left.h^{{\rm R} n}_{\mu\nu}\right|_\gamma+x^a\partial_a \!\!\left.h^{{\rm R} n}_{\mu\nu}\right|_\gamma+\frac{1}{2}x^a x^b \left.\partial_a\partial_b h^{{\rm R}n}_{\mu\nu}\right|_\gamma+\O(r^3).\label{eq:hRn Taylor}
\eeq
Moreover, the equation of motion~\eqref{eq:EOM generic} for $\gamma$, which is otherwise written in terms of the fields $A^n_{\mu\nu a_1\cdot a_\ell}$ defined {\em outside} the body, becomes identical to the equation of motion of a test body in $\tilde g_{\mu\nu}$; this result has been established at linear order in $\e$ for an arbitrary compact object~\cite{Gralla2008,Pound2010} and up to second order in $\e$ for a nonspinning, spherical body~\cite{Pound2012,Pound2017}. The latter result reads
\beq
    \frac{D^{2}z^{\alpha}}{\dd{\tau^2}} = - \frac{1}{2}P^{\alpha\mu}(g_{\mu}{}^{\rho} - h^{\R}_{\mu}{}^{\rho})(2h^{\R}_{\rho\beta;\gamma} - h^{\R}_{\beta\gamma;\rho})u^{\beta}u^{\gamma}+\order{\e^3}, \label{eq:2nd_eom}
\eeq
where $P^{\alpha\mu}:=g^{\alpha\mu}+u^\alpha u^\mu$, a semicolon denotes the covariant derivative compatible with $g_{\mu\nu}$, and all fields are evaluated on $\gamma$.
This is an expanded form of the geodesic equation in $\tilde g_{\mu\nu}$~\cite{Pound2015}.

All of the above follows from the matching condition and the vacuum field equations outside the body. The extended fields $h^n_{\mu\nu}$ satisfy those equations for all points $x$ away from $\gamma$,
\begin{multline}
    \delta G^{\mu\nu}[h] +\delta^2 G^{\mu\nu}[h,h]=\order{\e^3}\quad \text{for }x\notin\gamma.\label{eq:total EFE off gamma}
\end{multline}
Here $\delta G^{\mu\nu}$ and $\delta^2 G^{\mu\nu}$ are the linear and quadratic terms in the expansion of the Einstein tensor $G^{\mu\nu}[g+h]$ in powers of $h_{\mu\nu}$, given explicitly in Appendix~\ref{sec:Einstein tensors}.

In this way, we have effectively eliminated the body's small scale from the problem and replaced it with a singularity, without altering the curve $\gamma$ or the metric in the region $r\gg m$. The mathematical problem of solving the Einstein equations with a small extended source has been replaced with a reduced problem of solving the vacuum field equations~\eqref{eq:total EFE off gamma} subject to the conditions 
\begin{itemize}
    \item[(i)] in the limit of small $r$, the solution agrees with the form~\eqref{eq:regsing_split} derived from matched expansions
    \item[(ii)] $\gamma$ obeys Eq.~\eqref{eq:2nd_eom} (in the case of a nonspinning, spherical object).
\end{itemize}
For all $r\gg m$, the solution to this problem will be identical to the solution to the original problem.

The literature on second-order self-force, going back to Ref.~\cite{Rosenthal2006}, has focused on solving the reduced problem via a {\em puncture scheme}~\cite{Detweiler2012,Pound2012,Gralla2012,Pound2012SmallBody,Pound2017,Pound2020,Miller2020} (see also Refs.~\cite{Barack2007,Vega2008,Dolan2012,Wardell2015,WardellWarburton2015,Thornburg2017}). The puncture is obtained by truncating the local expansion of the singular field at some finite power of $r$, so that \(h^{\mathcal{P}}_{\mu\nu} \approx h^{\mathrm{S}}_{\mu\nu}\), and then transitioning it to zero at some arbitrary, finite distance from $\gamma$.
Rather than solving the for the physical field $h_{\mu\nu}$, one solves for the residual field
\begin{equation}
	h^{\mathcal{R}}_{\mu\nu} \coloneqq h_{\mu\nu} - h^{\mathcal{P}}_{\mu\nu} \label{eq:res_field}
\end{equation}
which satisfies \(h^{\mathcal{R}}_{\mu\nu} \approx h^{\mathrm{R}}_{\mu\nu}\) near $\gamma$ but becomes identical to $h_{\mu\nu}$ outside the support of $h^\calP_{\mu\nu}$. Writing the puncture as $\sum \e^n h^{\calP n}_{\mu\nu}[\gamma]$, we move it to the right-hand side of Eq.~\eqref{eq:total EFE off gamma} and split the field equations into a hierarchy of equations for the residual fields:\footnote{The form and behavior of these equations are slightly different than in the puncture scheme detailed in previous papers by one of us (e.g.,~\cite{Pound2012SmallBody,Pound2017,Pound2020}). We adopt this form here to streamline the discussion. We discuss the differences in the conclusion.}
\begin{align}
    \delta G^{\mu\nu}[h^{\calR 1}] &= - \delta G^{\mu\nu}[{h}^{\calP 1}] &\text{for } x\notin\gamma, \label{eq:punctureP1} \\
    \delta G^{\mu\nu}[h^{\calR 2}] &= - \delta^2 G^{\mu\nu}[h^1,h^1] - \delta G^{\mu\nu}[h^{\calP 2}] &\text{for } x\notin\gamma. \label{eq:punctureP2}
\end{align}
As written, these equations do not uniquely determine $h^{\calR n}_{\mu\nu}$, even if sufficient boundary conditions are prescribed in the external spacetime, because they do not tell us whether or not there are delta function sources supported on $\gamma$. However, in order for the total field $h^\calR_{\mu\nu}+h^{\calP}_{\mu\nu}$ to agree with the form dictated by the matched asymptotic expansions, $h^\calR_{\mu\nu}$ must be a $C^{k}$ function at $\gamma$ if $h^\calP_{\mu\nu}$ is truncated at order $r^k$. This implies that the correct field equations on the full domain, including $\gamma$, are
\begin{align}
    \delta G^{\mu\nu}[h^{\calR 1}] &= - (\delta G^{\mu\nu}[{h}^{\calP 1}])^\star, \label{eq:puncture_1st} \\
    \delta G^{\mu\nu}[h^{\calR 2}] &= - (\delta^2 G^{\mu\nu}[h^1,h^1] + \delta G^{\mu\nu}[h^{\calP 2}])^\star, \label{eq:puncture_2nd} 
\end{align}
where we use a $\star$ to indicate that a quantity defined on $r>0$ is promoted to the domain $r\geq0$ as a locally integrable function. This means derivatives in $\delta^n G_{\mu\nu}$ are evaluated in the ordinary strong sense for $r>0$, and then the starred quantities are simply left undefined on the measure-zero set $r=0$ or defined at $r=0$ by taking the limit $r\to 0$ (if it exists). For example, the Euclidean Laplacian acting on $1/r=1/\sqrt{x^2+y^2+z^2}$ evaluates to $\delta^{ab}\partial_a\partial_b\left(\frac{1}{r}\right)=0$ for $r>0$, meaning its $\star$-promotion is $\left[\delta^{ab}\partial_a\partial_b\left(\frac{1}{r}\right)\right]^\star=0$ for all $r\geq0$; this contrasts with the result if we treat the Laplacian in the sense of distributional derivatives, in which case we have the distributional identity $\delta^{ab}\partial_a\partial_b\left(\frac{1}{r}\right)= -4\pi\delta^3(x^i)$.

The form of the puncture guarantees that the sources in Eqs.~\eqref{eq:puncture_1st} and \eqref{eq:puncture_2nd} are locally integrable at $\gamma$. Additional conditions on the puncture arise if we wish to replace $h^\R_{\mu\nu}$ with $h^\calR_{\mu\nu}$ in the equation of motion~\eqref{eq:2nd_eom}. Such a substitution requires that on $\gamma$, \(h^{\mathcal{R}}_{\mu\nu}\) and its first derivatives are identical to \(h^{\mathrm{R}}_{\mu\nu}\) and its first derivatives. Ensuring this generically requires
\begin{gather}
	\lim_{x \to z} \pqty{h^{\mathcal{P}}_{\mu\nu} - h^{\mathrm{S}}_{\mu\nu}} = 0, \label{eq:punc_sing} \\
	\lim_{x \to z} \pqty{h^{\mathcal{P}}_{\mu\nu,\rho} - h^{\mathrm{S}}_{\mu\nu,\rho}} = 0, \label{eq:pun_sing_deriv}
\end{gather}
for all points $z^\mu$ on $\gamma$. With these conditions, Eqs.~\eqref{eq:puncture_1st}, \eqref{eq:puncture_2nd}, and \eqref{eq:2nd_eom} form a coupled set of equations that can be conveniently solved, in the case of a binary inspiral, in a two-timescale expansion~\cite{Hinderer2008,Pound2020,Pound2021}. 

However, at first order there is an alternative, more commonly used method: rather than replacing the object with a puncture in the spacetime, one can (equivalently) replace it with a point mass. If we return to Eq.~\eqref{eq:total EFE off gamma} and now {\em do} treat derivatives in $\delta G^{\mu\nu}$ as distributional derivatives, then the form of the singular field, $h^{\mathrm{S}1}_{\mu\nu}$, determines~\cite{DEath1975,Gralla2008,Pound2010,Pound2012SmallBody}
\beq
\e\delta G^{\mu\nu}[h^1] = 8\pi \e T^{\mu\nu}_1 +\order{\e^2},\label{eq:EFE1 point mass}
\eeq
where
\begin{equation}
    T_1^{\mu\nu}(x) = m \int_\gamma u^\mu u^\nu \delta^4(x,z)\dd{\tau} \label{eq:stress1}
\end{equation}
is the stress-energy of a point mass in the background metric \(g_{\mu\nu}\). The quantity
\begin{equation}
    \delta^{4}(x,z) \coloneqq \frac{\delta^4(x-z)}{\sqrt{-g}} \label{eq:covDirac}
\end{equation}
is the covariant delta function. With this formulation, instead of solving the field equations directly for the regular field, one can solve Eq.~\eqref{eq:EFE1 point mass} for the full field $h^1_{\mu\nu}$ and then extract $h^{\R1}_{\mu\nu}$ using mode-sum regularisation; see Refs.~\cite{Barack2009,Wardell2015} for reviews of this method. Such calculations are often significantly easier to implement and more efficient than puncture schemes, and they have been the basis for most calculations of the first-order self-force.

Equation~\eqref{eq:EFE1 point mass} is a more traditional form of skeletonization than the puncture scheme~\cite{Mathisson1937}. If the field equations were linear, then we could extend it to all multipole orders. The local form~\eqref{eq:hS1form}--\eqref{eq:hS3form} would directly correspond to a skeleton stress-energy~\cite{Pound2012SmallBody}
\begin{multline}
T^{\mu\nu} = \int_\gamma\left[m u^\mu u^\nu \delta^4(x,z) +u^\mu u^\nu M^\rho\nabla_{\rho}\delta^4(x,z)\right.\\ +\left.u^{(\mu}S^{\nu)\rho}\nabla_\rho \delta^4(x,z)+\cdots\right]d\tau, 
\end{multline}
where $S^{\mu\nu}:=-\epsilon^{\rho\mu\nu\sigma}u_\rho S_\sigma$. But the nonlinearity of the field equations spoil this simple correspondence. At second order, the difficulty arises because the second-order Einstein tensor, \(\delta^{2}G^{\mu\nu}\), shown in Eq.~\eqref{eq:delta2G}, has the schematic form \(\delta^2 G^{\mu\nu}[h,h] \sim h\partial^{2}h + \partial h \partial h\). Given the first-order field's behavior \(h^{1}_{\mu\nu} \sim 1/r\), the second-order Einstein tensor diverges like \(\delta^2 G^{\mu\nu}[h^1,h^1] \sim 1/r^4\) at the worldline. This is not locally integrable at $\gamma$, and because it is constructed from a quadratic rather than linear operation on integrable functions, it does not have a unique definition as a distribution on any region intersecting $\gamma$. As a consequence, we cannot obviously write an analogue of Eq.~\eqref{eq:EFE1 point mass} for $h^2_{\mu\nu}$ or define a unique second-order stress-energy.

This strongly divergent behaviour of \(\delta^{2}G^{\mu\nu}\) also introduces the dominant computational burden in a numerical implementation at second order: the problem of \emph{infinite mode coupling}, first described in Ref.~\cite{Miller2016}. In concrete applications to binaries, we typically decompose the fields $h^n_{\mu\nu}$ into a basis of angular harmonics, say $h^n_{\mu\nu} = \sum_{ilm}h^{n}_{ il m}(t_{\rm BL},r_{\rm BL})Y^{ilm}_{\mu\nu}(\theta,\phi)$, where $(t_{\rm BL},r_{\rm BL},\theta,\phi)$ are Boyer-Lindquist coordinates centered on the large black hole; here for concreteness we have written the expansion in terms of Barack-Lousto-type tensor spherical harmonics~\cite{Barack2005}, which would not be used in Kerr calculations in practice, but the problem we describe is insensitive to the details of this decomposition. Given this mode expansion, a single mode of $\delta^{2}G^{\mu\nu}[h^{1},h^{1}]$ becomes an infinite sum of products of modes of $h^1_{\mu\nu}$:
\beq
\delta^{2}G_{ilm}[h^{1},h^{1}] = \sum_{\substack{i_1 l_1 m_1\\i_2 l_2 m_2}}D^{ilm}_{i_1 l_1 m_1 i_2 l_2 m_2}[h^{1}_{ i_1 l_1 m_1},h^{1}_{ i_2 l_2 m_2}],\label{ddGilm}
\eeq
where $D^{ilm}_{i_1 l_1 m_1 i_2 l_2 m_2}$ is a bilinear differential operator involving $t_{\rm BL}$ and $r_{\rm BL}$ derivatives. Since  $\delta^{2}G^{\mu\nu}\sim 1/r^4$ and decomposing into modes involves integrating over two dimensions, modes of $\delta^{2}G^{\mu\nu}$ behave like 
\beq
\delta^{2}G_{ilm}\sim \frac{1}{|r_{\rm BL}-r_o(t_{\rm BL})|^2},\label{eq:d2G_ilm generic}
\eeq
where $r_o(t_{\rm BL})$ is the orbital radius at time $t_{\rm BL}$. On the other hand, the first-order modes $h^1_{ilm}$ are finite at  $r_{\rm BL}=r_o(_{\rm BL})$. This means that the mode sum~\eqref{ddGilm} must recover a strongly divergent function by summing up products of finite modes. In practice, to achieve a fixed accuracy arbitrarily close to the worldline, this requires an arbitrarily large number of first-order modes.

In a numerical calculation, each mode $h^1_{ilm}$ must be found by solving the first-order field equations, meaning there is a practical limitation on the number of modes we can add to the sum. This makes it impossible to calculate even a single mode $\delta^2 G_{ilm}$ in a region around $\gamma$. \textcite{Miller2016} provide a way to circumvent this problem using knowledge of the local four-dimensional $h^{\calP 1}_{\mu\nu}$ near the worldline. That method, which is used in the only extant second-order implementation~\cite{Pound2020,Miller2020}, involves performing two-dimensional numerical integrations of the four-dimensional $h^{\calP1}_{\mu\nu}$ on a grid of $r_{\rm BL}$ values around $r_o$. Such a procedure will be the overwhelming computational expense in any second-order calculations using current methods.\footnote{This point is starkly illustrated with an example. For a quasicircular orbit at a single orbital radius in Schwarzschild, computing the necessary inputs for $\delta^2 G_{ilm}$ up to moderate values of $l$  takes 2 to 3 days on a 40-core machine.
All other aspects of the calculation of $h^{\calR2}_{\mu\nu}$ represent a marginal additional runtime. As a point of comparison, a decade ago an analogous first-order calculation at a single orbital radius with comparable precision could be performed in approximately 10 minutes on an ordinary desktop~\cite{Akcay2010}.} 

\subsection{This paper: highly regular gauges and the Detweiler stress-energy}\label{sec:outline}

In Ref.~\cite{Pound2017} (hereafter Paper I), one of us showed that there exists a class of \emph{highly regular gauges} that are qualitatively more regular than the generic behavior~\eqref{eq:hS1form}--\eqref{eq:hS3form}.
In this class, the most singular piece of $h^2_{\mu\nu}$, $\sim m^2/r^2$, identically vanishes (and likewise, the most singular, $\sim m^n/r^n$ piece of $h^n_{\mu\nu}$ vanishes for all $n>2$). Accordingly, in these gauges the most singular piece of the second-order source, \(\delta^{2}G^{\mu\nu}[\hrtext{S1},\hrtext{S1}]\), is significantly mollified, diverging as $\sim 1/r^2$ rather than $1/r^4$. This implies that the individual source modes in these gauges will behave,  at worst, like 
\beq
\delta^{2}G_{ilm}\sim \log|r_{\rm BL}-r_o(t_{\rm BL})|.\label{eq:ddG_ilm HR}
\eeq
A mildly divergent function of this form should be dramatically cheaper to compute than the much more strongly divergent generic behavior~\eqref{eq:d2G_ilm generic}.

However, Paper I only provided the leading-order term of $h^{\S2}_{\mu\nu}$ in the class of highly regular gauges. Our first goal in this paper is to extend the derivation through linear order in \(r\), the order required to ensure the conditions~\eqref{eq:punc_sing}--\eqref{eq:pun_sing_deriv} are satisfied. The derivation, which closely follows Paper I, is contained in Secs.~\ref{sec:perturbations} and \ref{sec:HRgaugetransfo}.

Paper I also pointed out that because $\delta^2 G^{\mu\nu}[h^1,h^1]$ is well defined as a distribution in these gauges, it is possible to write down a field equation for $h^2_{\mu\nu}$ that is valid on the entire domain $r\geq0$ and to identify a unique second-order stress-energy tensor. In Sec.~\ref{sec:o2stress}, we derive that stress-energy, showing that in highly regular gauges Eq.~\eqref{eq:EFE1 point mass} extends to second order,
\beq
\delta G^{\mu\nu}[\e h^1+\e^2 h^2] + \e^2\delta^2 G^{\mu\nu}[h^1,h^1] = 8\pi \tilde T^{\mu\nu} +\order{\e^3},\label{eq:EFE2 point mass}
\eeq
with
\begin{equation}
    \tilde{T}^{\mu\nu} = \epsilon m\int_{\gamma}\tilde{u}^{\mu}\tilde{u}^{\nu} \frac{\delta^4(x-z)}{\sqrt{-\tilde{g}}} \dd{\tilde{\tau}}. \label{eq:Detweiler stress-energy}
\end{equation}
This is nothing more than the stress-energy tensor of a point mass in the effective metric \(\tilde{g}_{\mu\nu}\). We discuss some of its properties in Sec.~\ref{sec:o2stress}. 

To emphasise the significance of Eq.~\eqref{eq:EFE2 point mass}, we stress that in self-force theory we {\em cannot} freely prescribe a stress-energy tensor.
The assumptions of matched asymptotic expansions uniquely determine the local form of the metric in terms of a set of multipole moments. In cases where all terms in the Einstein field equations are well defined as distributions on the domain $r\geq0$, this local structure encodes the same information as (and uniquely determines) the skeleton stress-energy tensor.
Yet Eq.~\eqref{eq:EFE2 point mass} does not appear here for the first time: in Ref.~\cite{Detweiler2012}, Detweiler posited that this equation, with the stress-energy~\eqref{eq:Detweiler stress-energy}, holds in {\em all} gauges, and we therefore call $\tilde{T}^{\mu\nu}$ the {\em Detweiler stress-energy}. Our derivation shows rigorously that it is valid in the class of highly regular gauges. In Sec.~\ref{sec:T_lorenz}, we extend our analysis to the Lorenz gauge. In that case, the stress-energy tensor is not uniquely determined because not all quantities in the field equations have unique distributional definitions.\footnote{We note that this lack of uniqueness has no bearing on whether a unique global solution can be found. As described in Sec.~\ref{sec:puncture}, the global solution is uniquely determined by the local form of the metric together with the field equations and global boundary conditions.} But we show that there exists a canonical distributional definition of $\delta^2 G^{\mu\nu}[h^1,h^1]$  under which Eq.~\eqref{eq:EFE2 point mass} holds true. We conjecture that this extends to all gauges compatible with the assumptions of matched asymptotic expansions. However, outside the highly regular gauges, these distributional definitions explicitly involve $h^{\S2}_{\mu\nu}$; to use such definitions, one must explicitly solve the second-order field equations locally before one can solve them globally.

After these foundational calculations, in Sec.~\ref{sec:application} we sketch how our results could be used to implement a puncture scheme in a highly regular gauge. Building on Eq.~\eqref{eq:EFE2 point mass}, we also describe how one could solve for $h^2_{\mu\nu}$ and then extract $h^{\R2}_{\mu\nu}$ using mode-sum regularisation.

Except in portions of Sec.~\ref{sec:perturbations}, we specialize to the case of a spherical, nonspinning small object. Throughout, we leave the external background arbitrary.

\subsection{Conventions and definitions}\label{sec:definitions}

We work in geometric units with \(c=G=1\).
Greek indices run from \(0\) to \(3\) and are raised and lowered with the background metric, \(g_{\mu\nu}\), which has signature \((-,+,+,+)\).
Lowercase Latin indices run from \(1\) to \(3\) and are raised and lowered with the flat-space Euclidean metric, \(\delta_{ab}\).
Uppercase Latin indices denote multi-indices, as in \(L \coloneqq i_{1}\ldots i_{l}\).

Terms written in a serif font are exact quantities, e.g., \(\mathsf{g}_{\mu\nu}\) is the full, exact metric describing the physical spacetime.
A prime symbol on the perturbation, \(h^{n'}_{\mu\nu}\), denotes quantities in the lightcone rest gauge, and a star, \(h^{n*}_{\mu\nu}\), denotes quantities in the Lorenz gauge.
No prime, \(h^{n}_{\mu\nu}\), indicates terms in the highly regular gauge.
A prescript, \(\prescript{n}{}{\!A}^{\mu_{1}\ldots}_{\nu_{1}\ldots}\), on a tensor counts the power of $\epsilon$ coming from substituting the acceleration $a^\mu = \sum_{n\geq1}\epsilon^n f^\mu_{n}$ into $A^{\mu_{1}\ldots}_{\nu_{1}\ldots}$.
An overset ring, \(\mathring{A}^{\mu_{1}\ldots}_{\nu_{1}\ldots}\), indicates terms that have been re-expanded for small acceleration and then re-collected at each order in \(\e\), i.e., \(\mathring{h}^{n}_{\mu\nu} = \sum_{i=0}^{n}\prescript{i}{}{h}^{n-i}_{\mu\nu}\) (where, for this purpose, \(h^{0}_{\mu\nu} \coloneqq g_{\mu\nu}\)).
Tildes placed over a tensor, \(\tilde{A}^{\mu_{1}\ldots}_{\nu_{1}\ldots}\), denote quantities defined with respect to the effective metric.

Parentheses and square brackets around indices denote symmetrisation and antisymmetrisation, respectively. Angled brackets, such as \(\langle L \rangle\), denote the symmetric trace-free (STF) combination of the enclosed indices with respect to \(\delta_{ab}\).
In some cases, we additionally use the notation \(\sym_{L}\) and \(\STF_{L}\) to denote symmetrisation and the STF combination over the indices \(L\), respectively.
The covariant derivative (given by \(\nabla\) or a semi-colon) is compatible with \(g_{\mu\nu}\) unless otherwise stated and the partial derivative is denoted by a comma.

A number of calculations in this paper were done using \textsc{Wolfram Mathematica}~\cite{Mathematica} and the tensor algebra package \textsc{xAct}~\cite{xAct,xPerm,xCoba,xPert}.

\section{Perturbations from matched asymptotic expansions in the lightcone gauge}\label{sec:perturbations}

\subsection{Matched expansions and the existence of a highly regular gauge}

We begin with a more detailed review of matched asymptotic expansions and how it leads naturally to the existence of highly regular gauges. The discussion here reiterates material in numerous references, and we specifically follow Paper I. We refer to Refs.~\cite{Kevorkian1996,Eckhaus:79} for a broader introduction to the method of matched expansions.

Our discussion of the local form of the metric will involve some subtleties because we use the self-consistent framework of gravitational self-force theory~\cite{Pound2010,Pound2012,Pound2012SmallBody} (see Ref.~\cite{Pound2015Small} for an overview).
In this approach we expand the perturbation \(h_{\mu\nu}\) while holding the accelerated worldline fixed. This means that $\gamma$ is $\e$ dependent, and the coefficients \(h^n_{\mu\nu}\) inherit that \(\epsilon\) dependence; Eq.~\eqref{eq:pert_expansion} is {\em not} a Taylor series in $\e$.\footnote{More precisely, we treat the metric as a function ${\sf g}_{\mu\nu}(x^\mu,z^\mu,u^\mu,\e)$ and expand for small $\e$ while holding the other arguments fixed. The function on the enlarged manifold that includes the phase space coordinates $(z^\mu,u^\mu)$ becomes equal to the physical metric on the spacetime manifold when $z^\mu$ and $u^\mu$ obey the ($\e$-dependent) equation of motion.} In an ordinary Taylor series, $\gamma$ itself is expanded in powers of $\e$. The puncture or skeleton stress-energy then diverge on the zeroth-order worldline, which is a geodesic of  the external background metric. The corrections to the motion are then encoded in mass dipole moments in the perturbations $h^{n>1}_{\mu\nu}$. This treatment is prone to large cumulative errors because in physical scenarios such as a binary, the body secularly deviates from the background geodesic, causing the dipole moments to grow large with time. The self-consistent treatment circumvents that problem. 

As discussed in Refs.~\cite{Miller2020,Pound2021}, to avoid other, similar errors, in general we should also allow the coefficients $h^n_{\mu\nu}$ to depend on $\e$- and time-dependent external physical parameters. Examples of such parameters are perturbations to the large black hole's mass and spin in a binary. In the analysis below, we can freely allow $h^{\R}_{\mu\nu}$ to depend on such parameters without altering the discussion. We hence leave the dependence on these parameters implicit.

To understand the form of the perturbations near the worldline, we adopt Fermi--Walker coordinates \((t,x^a)\) that are tethered to \(\gamma\)~\cite{Poisson2011}.
The spatial coordinates are defined such that \(x^i = rn^i\), where \(r\) is the proper distance from the worldline along a spatial geodesic orthogonal to $\gamma$, and \(n^i\) is a unit vector giving the direction the geodesic is sent out from \(\gamma\).
The time coordinate, \(t\), gives the proper time on \(\gamma\). Here proper lengths, and orthogonality, are defined with respect to the background metric.

Because these coordinates are tied to an $\e$ dependent worldline, they introduce additional $\e$ dependence into the metric. In particular, the background metric in these coordinates takes the form~\cite{Pound2017}
\begin{subequations}
    \label{eq:metric_background}
    \begin{align}
        \begin{split}
            g_{tt} ={}& -1 - 2a_{i}x^{i} - (R_{titj} + a_{i}a_{j})x^{i}x^{j} \\
            & - \frac{1}{3}(4 R_{titj}a_{k} + R_{titj;k})x^{i}x^{j}x^{k} + \order{r^4}, \label{eq:metric_background_tt}
        \end{split}\\
        \begin{split}
            g_{ta} ={}& -\frac{2}{3}R_{tiaj}x^{i}x^{j} - \frac{1}{3}R_{tiaj}a_{k}x^{i}x^{j}x^{k} \\
            & - \frac{1}{4}R_{tiaj;k}x^{i}x^{j}x^{k} + \order{r^4}, \label{eq:metric_background_ta}
        \end{split}\\
        g_{ab} ={}& \delta_{ab} - \frac{1}{3}R_{aibj}x^{i}x^{j} - \frac{1}{6}R_{aibj;k}x^{i}x^{j}x^{k} + \order{r^4}, \label{eq:metric_background_ab}
    \end{align}
\end{subequations}
which explicitly depends on $\gamma$'s $\e$-dependent covariant acceleration \(a^{\mu} \coloneqq {D}^{2} z^\mu/\dd{t^2} = (0,a^i)\).

All Riemann terms in the metric are evaluated on the worldline and are therefore also implicitly dependent on $\e$. As \(g_{\mu\nu}\) is a vacuum spacetime, we can use the identities in Appendix~D3 from Ref.~\cite{Poisson2010} to write these Riemann quantities in terms of tidal moments:
\begin{subequations}
    \label{eq:R_tidal}
    \begin{align}
        R_{tatb} ={}& \E_{ab}, \label{eq:R2t_tidal} \\
        R_{abct} ={}& \epsilon_{abi}\B^{i}{}_{c}, \label{eq:R1t_tidal} \\
        R_{abcd} ={}& -\epsilon_{abi}\epsilon_{cdj}\E^{ij}, \label{eq:R0t_tidal}
    \end{align}
\end{subequations}
and 
\begin{subequations}
    \label{eq:R_deriv_tidal}
    \begin{align}
        R_{tatb;c} ={}& \E_{abc} + \frac{2}{3}\epsilon_{ci(a}\Bdot_{b)}{}^{i}, \label{eq:R2t_deriv_tidal} \\
        R_{abct;d} ={}& \epsilon_{ab}{}^{i}\pqty{\frac{4}{3}\B_{icd} - \frac{2}{3}\epsilon_{dj(i}\Edot^{j}{}_{c)}}, \label{eq:R1t_deriv_tidal} \\
        R_{abcd;e} ={}& -\epsilon_{abi}\epsilon_{cdj}\pqty{\E^{ij}{}_{e} + \frac{2}{3}\epsilon_{ek}{}^{(i}\Bdot^{j)k}}, \label{eq:R0t_deriv_tidal}
    \end{align}
\end{subequations}
where \(\E_{ab}\), \(\B_{ab}\), \(\E_{abc}\), \(\B_{abc}\) are the external background's quadrupolar and octupolar  tidal moments, which the small object feels as it travels along the worldline. They are STF over all indices and only depend on $t$. A dot denotes a derivative with respect to $t$.

As discussed in the introduction, the local form of the perturbations $h^n_{\mu\nu}$ in these coordinates is determined using matched asymptotic expansions.
Sufficiently close to the small object, at a distance \(r\sim m\), any terms \({\sim}(\flatfrac{m}{r})^n\) in the perturbations reduce in order and become the same `size' as the background spacetime; moreover, such terms have much larger gradients than the external background metric, implying that their gravitational effects dominate over background ones. This causes the expansion in Eq.~\eqref{eq:pert_expansion} to break down. To account for this, we introduce a second asymptotic expansion that uses a scaled distance,
\begin{equation}
    \rtilde \coloneqq \frac{r}{\epsilon}. \label{eq:distanceRescale}
\end{equation}
Now when we take the limit as \(\epsilon \to 0\) at fixed \(\rtilde\), we keep the scale of the small object fixed and send the external universe to infinity.
This is in contrast to our original \(\epsilon\to 0\) limit, which fixes the external universe and sends the size of the small object to zero.

In our new expansion near the small object, we rewrite our full spacetime metric as
\begin{equation}
    \fullg_{\mu\nu}(r,\epsilon) = \gobj_{\mu\nu}(\rtilde) + \epsilon H^1_{\mu\nu}(\rtilde) + \epsilon^2 H^2_{\mu\nu}(\rtilde) + \order{\epsilon^3}, \label{eq:metricnearsmallobj}
\end{equation}
where \(\gobj_{\mu\nu}\) is the metric the small object would have if it were isolated in spacetime, and where the components refer to an unscaled coordinate basis. Implicit in this form are two key assumptions: (i)  $\e$ is the only relevant length scale near the object, meaning the object must be compact, with a spatial extension comparable to its mass; and (ii) there is no small time scale $\sim \e$ in the spacetime, meaning the object is approximately in equilibrium with its surroundings and not undergoing any internal dynamics on the scale of its light-crossing time.

Both Eqs.~\eqref{eq:fullmetric}--\eqref{eq:pert_expansion} and Eq.~\eqref{eq:metricnearsmallobj} are expansions of the same metric, which we refer to as the outer and inner expansions, respectively. So for a sufficiently well-behaved metric, they must agree when appropriately compared; that is, they must satisfy a ``matching" condition. There are various formulations of such conditions, and various assumptions that imply them. Ref.~\cite{Gralla2008} assumes a strong set of smoothness conditions. A more common, weaker assumption~\cite{Kevorkian1996} is that the two expansions agree in an overlap region on some length scale between $r\sim \e$ and $r\sim \e^0$; this will fall somewhere within the ``buffer region" $\e\ll r\ll \e^0$, in which $\e$, $r$, and $\e/r$ are all small (and in which $\tilde r$ is large compared to $m$). However, there exist functions violating the above assumptions that still satisfy the explicit matching condition that is generally used in practice~\cite{Eckhaus:79}. That condition is that the two expansions must commute, in the sense that if the outer expansion is re-expanded for small $r$, and the inner expansion is re-expanded for large $\tilde r$, then (after re-expressing the inner expansion result in terms of $r$ and $\e$) the result in both cases is a double expansion for small $r$ and $\e$, and the coefficients in these double expansions must agree with one another term by term. Here we merely assume that this matching condition holds, without adopting any stronger set of assumptions.

Using this condition and requiring that the outer and inner expansion are well behaved (i.e., that there are no negative powers of \(\epsilon\) in either expansion), we constrain the powers of \(r\) and \(\rtilde\) that can appear in our expansions.
Following the argument in Paper I, if the outer perturbations are expanded for small $r$, as in \(h^n_{\mu\nu} = \sum_p r^p h^{n,p}_{\mu\nu}\), then terms with \(p<-n\) would have to match terms in the inner expansion with inverse powers of \(\e\). To see this, simply note that an outer-expansion term of the form $\e^n/r^{n+1}$ corresponds to $1/(\e\tilde r^{n+1})$ in the inner expansion. Hence, such terms are ruled out.
This argument also applies to the inner expansions but with \(r^p\) replaced by \(1/\rtilde^p\).
Therefore, the expansions for small \(r\) and large \(\rtilde\) must have the form
\begin{align}
    g_{\mu\nu} &= \sum_{p\geq 0} r^{p}g^p_{\mu\nu}, \label{eq:metricconstraintsg} \\
    h^n_{\mu\nu} &= \sum_{p\geq -n} r^p h^{n,p}_{\mu\nu}, \label{eq:metricconstraintsh} \\
    \gobj_{\mu\nu} &= \sum_{p\geq 0} \frac{1}{\rtilde^p}\gobjp{p}_{\mu\nu}, \label{eq:metricconstraintsgobj} \\
    H^n_{\mu\nu} &= \sum_{p\geq -n} \frac{1}{\rtilde^p}H^{n,p}_{\mu\nu}, \label{eq:metricconstraintsH}
\end{align}
where \(\ln(r)\) terms may appear but have been absorbed into the coefficients for visual clarity. When we express \(\rtilde\) as \(r/\e\), each term in Eqs.~\eqref{eq:metricconstraintsgobj} and~\eqref{eq:metricconstraintsH} must be in one-to-one correspondence with, and agree identically with, a term in Eqs.~\eqref{eq:metricconstraintsg} and~\eqref{eq:metricconstraintsh}.

We have already given the explicit form~\eqref{eq:metric_background} of the external background~\eqref{eq:metricconstraintsg} in Fermi--Walker coordinates. It is determined by the acceleration of the worldline and the external tidal moments. The inner background, conversely, is determined by the multipole moments of the small object. We see from Eq.~\eqref{eq:metricconstraintsgobj} that \(\gobj_{\mu\nu}\) is asymptotically flat. Furthermore, it is quasistationary, as it only varies on the time scale $t\sim \e^0$. Its expansion~\eqref{eq:metricconstraintsgobj} can therefore be written completely in terms of the Geroch--Hansen multipole moments~\cite{Geroch1970,Hansen1974}.\footnote{The Geroch--Hansen moments are defined for strictly stationary spacetimes. We can define them for our quasistationary spacetime by fixing $t$ in the coefficients $\gobjp{p}_{\mu\nu}$; that fixed-$t$ spacetime approximates $\gobj_{\mu\nu}$. The result is that the multipole moments depend on $t$.}
Broadly, this means that in the buffer region, it has the form~\cite{Pound2017}
\begin{align}
    \gobj_{\mu\nu} &\sim 1 + \frac{\epsilon}{r}m + \frac{\epsilon^2}{r^2}\pqty{m^2 + M_{i}n^i + \epsilon_{ijk}S^{j}n^k} \nonumber\\
    &\quad + \order{\frac{\epsilon^3}{r^3}}, \label{eq:schematicobject}
\end{align}
where \(m\) and \(S^i\) are the Arnowitt--Deser--Misner (ADM) mass and angular momentum of \(\gobj_{\mu\nu}\). $M_i$ is its mass dipole moment relative to $\gamma$, discussed in the introduction.

The $n$th-order term in the expansion~\eqref{eq:schematicobject} must match the leading-order term in $h^{n}_{\mu\nu}$: 
\beq
\frac{\e^n h^{n,-n}_{\mu\nu}}{r^n} = \frac{\e^n g^{\rm obj,n}_{\mu\nu}}{r^n}.
\eeq
In words, the most singular (at $r=0$) term in $h^n_{\mu\nu}$ is uniquely determined by the large-$\tilde r$ expansion of the object's metric $\gobj_{\mu\nu}$. This is the essential fact that implies the existence of a highly regular gauge.

To see why this implication follows, consider a spherically symmetric, nonspinning object. Its metric $\gobj_{\mu\nu}$ is uniquely given by the Schwarzschild metric, which can be written in ingoing Eddington--Finkelstein coordinates as
\begin{equation}
    \dd{s}^{2}_{\text{obj}} = -\left(1 - \frac{2m}{r}\right)\dd{v}^{2} + 2\dd{v}\dd{r} + r^{2}\dd{\Omega}^{2}. \label{eq:schwarz_ds2}
\end{equation}
We see immediately that this is linear in \(m/r\); the series~\eqref{eq:metricconstraintsgobj} terminates at $p=1$. Hence, for such an object, we have $h^{n,-n}_{\mu\nu}=0$ for all $n>1$. Even if the object is spinning and nonspherical, the collection of all terms that are independent of higher moments in Eq.~\eqref{eq:schematicobject} together form an expansion of the Schwarzschild metric, and we can impose the Eddington--Finkelstein gauge on them to set them all to zero (for $n>1$).  

\subsection{Metric perturbations in a lightcone gauge}

To obtain the metric perturbations $h^n_{\mu\nu}$ in a highly regular gauge, we follow the approach detailed in Paper I; all results in this section are taken from that reference.

We begin by calculating the form of the inner expansion's metric in a highly regular gauge that is also a {\em rest gauge}. In such a gauge, the object is manifestly at rest on $\gamma$, in the sense that no acceleration or mass dipole terms appear in the metric in the buffer region.
Such gauges exist because we can always find an effective metric in which our small object's center-of-mass worldline is a geodesic~\cite{Pound2015Small}.
We then translate this inner expansion into a small-$r$ expansion of the outer expansion.
This will lead to perturbations $h^n_{\mu\nu}$ in a gauge that is impractical for numerical implementation. In the next section, we transform it to a less restrictive, practical gauge that maintains the high regularity of our initial gauge.

We immediately specialize to a nonspinning, approximately spherical small object.
It follows that even if the object is a material body rather than a black hole, for our purposes we can take the inner expansion to be the metric of a tidally perturbed, nonspinning black hole as presented in Ref.~\cite{Poisson2005}. Because the composition of the small object is fully encoded in its multipole moments, the difference between this metric and that of a material body will not manifest itself in the outer perturbations until order $\e^3$, at which order the object's quadrupole moment would appear.
The metric of Ref.~\cite{Poisson2005} is written with $\gobj_{\mu\nu}$ in the Eddington--Finkelstein form~\eqref{eq:schwarz_ds2}, and with the perturbations $H^n_{\mu\nu}$ in a lightcone gauge. In terms of Cartesian Eddington--Finkelstein coordinates $(\mathsf{v},\mathsf{x^a})$ defined on the manifold of $\gobj_{\mu\nu}$, this gauge condition reads
\begin{equation}
    H^n_{\mu \mathsf{a}} n^{\mathsf{a}} = 0. \label{eq:lightconegauge}
\end{equation}
Note that the coordinates we use here for the inner expansion, denoted with sans serif fonts, differ from the Fermi--Walker coordinates we use for the outer expansion. Ref.~\cite{Poisson2005} additionally refines the gauge to enforce \(H^{1}_{\mu\nu} = 0\). The Eddington--Finkelstein form of $\gobj_{\mu\nu}$ contains no leading-order mass dipole moment, ensuring that the coordinates are mass-centered at leading order, and \(H^{1}_{\mu\nu} = 0\) ensures that they are mass-centered at the first subleading order.

The resulting inner expansion is given explicitly by Eqs.~(61)--(64) in Ref.~\cite{Pound2017}. It is naturally written in terms of $\tilde{\mathsf{r}}:=\mathsf{r}/\e$. We then re-expand it for small \(\e\) at fixed $\mathsf{r}$ (or equivalently, re-expand it for large $\tilde{\mathsf{r}}$ and then re-express it in terms of $\mathsf{r}$ and $\e$) and perform a small-$r$, $\e^0$ transformation from local advanced coordinates to Fermi--Walker coordinates; this transformation is given by Eq.~(65) in Ref.~\cite{Pound2017}.
This gives an expansion valid in the buffer region which we write as
\beq
\mathsf{g}_{\mu\nu} = \gzero_{\mu\nu} + \e\mathring{h}^{1'}_{\mu\nu} + \e^2\mathring{h}^{2'}_{\mu\nu} + O(\e^3).\label{eq:outer expansion v2}
\eeq
By the assumptions of matched asymptotic expansions, this is necessarily the local form of the outer expansion, which will ultimately provide punctures for equations of the form of Eqs.~\eqref{eq:punctureP1} and~\eqref{eq:punctureP2}.
The primes indicate that the perturbations are in the lightcone rest gauge. The overset rings indicate that the expansion is organized slightly differently than Eqs.~\eqref{eq:fullmetric}--\eqref{eq:pert_expansion}, in a manner described momentarily. 

The leading term in Eq.~\eqref{eq:outer expansion v2} is 
\begin{subequations}
    \label{eq:metric_no_acc}
\begin{align}
    \gzero_{tt} &= - 1 - r^2\E_{ab}\nhat^{ab} - \frac{1}{3}r^3\E_{abc}\nhat^{abc} + \order{r^4}, \label{eq:metric_no_acc_tt} \\
    \gzero_{ta} &= -\frac{2}{3}r^{2}\B^{bc}\epsilon_{acd}\nhat_{b}{}^{d} + \frac{r^3}{60}\big(3\Edot_{ab}\nhat^{b} - 5\Edot_{bc}\nhat_{a}{}^{bc} \nonumber \\
        &\quad - 20\B^{bcd}\epsilon_{ab}{}^{i}\nhat_{cdi}\big) + \order{r^4}, \label{eq:metric_no_acc_ta} \\
    \gzero_{ab} &= \delta_{ab} - \frac{r^2}{9}\big(\E_{ab} - 6\E_{(a}{}^{c}\nhat_{b)c} + 3\delta_{ab}\E_{cd}\nhat^{cd}\big) \nonumber \\
        &\quad + \frac{r^3}{90}\big(30\E_{(a}{}^{cd}\nhat_{b)cd} - 3\E_{abc}\nhat^{c} - 8\Bdot_{(a}{}^{d}\epsilon_{b)cd}\nhat^{c} \nonumber \\
        &\quad + 10\Bdot^{cd}\epsilon_{c(a}{}^{i}\nhat_{b)di} - 15\delta_{ab}\E_{cdi}\nhat^{cdi}\big) + \order{r^4}, \label{eq:metric_no_acc_ab}
\end{align}
\end{subequations}
where we have introduced \(\nhat^{L} \equiv n^{\langle i_1}\cdots n^{i_l\rangle}\). This metric is {\em not} identical to the external background~\eqref{eq:metric_background}; instead, it is Eq.~\eqref{eq:metric_background} with the acceleration terms set to zero. The reason is that the coordinates are tethered to an $\e$-dependent worldline with a small acceleration. The inner expansion has implicitly included an expansion of that acceleration,
\begin{equation}
    a^\mu = \sum_{n>0} \epsilon^n f^\mu_n, \label{eq:accexpand}
\end{equation}
so that any acceleration terms have implicitly been moved to the first- or second-order outer perturbations. The full external background metric then reads
\begin{equation}
    g_{\mu\nu} = \prescript{0}{}{g}_{\mu\nu} + \epsilon \prescript{1}{}{g}_{\mu\nu} + \epsilon^2 \prescript{2}{}{g}_{\mu\nu} + \order{\epsilon^3}, \label{eq:metric_acc_exp}
\end{equation}
where
\begin{align}
    \prescript{0}{}{g}_{\mu\nu} &= \gzero_{\mu\nu},\\
    \prescript{1}{}{g}_{\mu\nu} &= -2 f^{1}_{i}x^{i}\delta^{t}_{\mu}\delta^{t}_{\nu} + \order{r^3}, \label{eq:g1} \\
    \prescript{2}{}{g}_{\mu\nu} &= -2 f^{2}_{i}x^{i}\delta^{t}_{\mu}\delta^{t}_{\nu} + \order{r^2}. \label{eq:g2}
\end{align}

We adopt analogous notation for other quantities, denoting their re-expansion for small acceleration with the notation
\begin{equation}
    A = \prescript{0}{}{\!\!A} + \epsilon\prescript{1}{}{\!\!A} + \epsilon^{2}\prescript{2}{}{\!\!A} + \order{\epsilon^3}, \label{eq:tensoracc}
\end{equation}
where the prescript is the order of the acceleration term. We then have
\begin{align}
\mathring{h}^{1'}_{\mu\nu} &= \prescript{0}{}{h}^{1'}_{\mu\nu} +{}^1 g_{\mu\nu} , \\
\mathring{h}^{2'}_{\mu\nu} &= \prescript{0}{}{h}^{2'}_{\mu\nu} + \prescript{1}{}{h}^{1'}_{\mu\nu} +\prescript{2}{}{g}_{\mu\nu}.  
\end{align}
This notation differs from that of Paper I, where a dagger was used in place of the overset ring.

The first-order term in Eq.~\eqref{eq:outer expansion v2} reads
\beq
\mathring{h}^{1'}_{\mu\nu} = \mathring{h}^{{\rm R}1'}_{\mu\nu} + \mathring{h}^{{\rm S}1'}_{\mu\nu}.
\eeq
The regular field 
\beq
\mathring{h}^{{\rm R}1'}_{\mu\nu} = \prescript{0}{}{h}^{\R 1'}_{\mu\nu} +{}^1 g_{\mu\nu}\label{hR1'}
\eeq
is given by
\begin{subequations}
\label{eq:hR1Prime}
\begin{align}
    \prescript{0}{}{h}^{\R 1'}_{tt} &= - r^{2}\delta\E_{ab}\nhat^{ab} + \order{r^3}, \label{eq:hR1Primett} \\
    \prescript{0}{}{h}^{\R 1'}_{ta} &= - \frac{2}{3}r^{2}\delta\B^{bc}\epsilon_{acd}\nhat_{b}{}^{d} + \order{r^3}, \label{eq:hR1Primeta} \\
    \prescript{0}{}{h}^{\R 1'}_{ab} &= -\frac{1}{9}r^2\big(\delta\E_{ab} - 6\delta\E_{(a}{}^{c}\nhat_{b)c} + 3\delta_{ab}\delta\E_{cd}\nhat^{cd}\big) \nonumber \\
        &\quad + \order{r^3}, \label{eq:hR1Primeab}
\end{align}
\end{subequations}
and Eq.~\eqref{eq:g1} where \(\delta\E_{ab}\) and \(\delta\B_{ab}\) are corrections to the respective tidal moments; this is identical in form to the tidal terms in Eq.~\eqref{eq:metric_no_acc}, and it is hence a smooth vacuum perturbation at $r=0$. The singular field 
\beq
\mathring{h}^{\S1'}_{\mu\nu} = {}^0h^{\S1'}_{\mu\nu}
\eeq
is given by
\begin{subequations}
\label{eq:hS1Prime}
\begin{align}
    \mathring{h}^{\S1'}_{tt} ={}& \frac{2m}{r} + \frac{11}{3}mr\E_{ab}\nhat^{ab} + \frac{1}{12}mr^2\big(8\Edot_{ab}\nhat^{ab} \big[5 \nonumber \\
        & - 3\log(\tfrac{2m}{r})\big] + 19\Edot_{abc}\nhat^{abc}\big) + \order{r^3}, \label{eq:hS1Primett} \displaybreak[0] \\
    \mathring{h}^{\S1'}_{ta} ={}& \frac{2m}{r}\nhat_a + \frac{2}{15}mr\big(11\E_{ab}\nhat^{b} + 10\B^{bc}\epsilon_{acd}\nhat_{b}{}^{d} \nonumber \\
        & + 15\E_{bc}\nhat_{a}{}^{bc}\big) + \frac{1}{1260}mr^2\big(126\Edot_{ab}\nhat^{b}\big[25 \nonumber \\
        &- 16\log(\tfrac{2m}{r})\big] + 140\Bdot^{bc}\epsilon_{acd}\nhat_{b}{}^{d} \big[13 - 12\log(\tfrac{2m}{r})\big] \nonumber \\
        & + 1095\E_{abc}\nhat^{bc} + 70\Edot^{bc}\nhat_{abc}\big[25 - 12\log(\tfrac{2m}{r})\big] \nonumber \\
        & + 1400\B^{bcd}\epsilon_{ab}{}^{i}\nhat_{cdi} + 840\E_{bcd}\nhat_{a}{}^{bcd}\big) + \order{r^3}, \label{eq:hS1Primeta} \displaybreak[0] \\
    \mathring{h}^{\S1'}_{ab} ={}&  \frac{2m}{3r}\big(\delta_{ab} + 3\nhat_{ab}\big) + \frac{1}{315}mr\big(154\E_{ab} \nonumber \\
        & - 168\B^{d}_{(a}\epsilon_{b)cd}\nhat^{c} + 580\E^{c}{}_{(a}\nhat_{b)c} + 15\E_{cd}\delta_{ab}\nhat^{cd} \nonumber \\
        & + 840\B^{cd}\epsilon_{c}{}^{i}{}_{(a}\nhat_{b)di} + 105\E_{cd}\nhat_{ab}{}^{cd}\big) + \frac{1}{3780}mr^2 \nonumber \\
        & \times \big(252\Edot_{ab}\big[29 - 20\log(\tfrac{2m}{r})\big] + 2322\E_{abc}\nhat^{c} \nonumber \\
        & - 504\Bdot^{d}{}_{(a}\epsilon_{b)cd}\nhat^{c}\big[11 - 12\log(\tfrac{2m}{r})\big] \nonumber \\
        & + 1980\Edot^{c}{}_{(a}\nhat_{b)c} + 60\Edot_{cd}\delta_{ab}\nhat^{cd}\big[59 - 42\log(\tfrac{2m}{r})\big] \nonumber \\
        & - 4800\B_{(a|c|}{}^{i}\epsilon_{b)di}\nhat^{cd} - 420\E_{(a}{}^{cd}\nhat_{b)cd} \nonumber \\
        & + 1680\Bdot^{cd}\epsilon_{c}{}^{i}{}_{(a}\nhat_{b)di}\big[4 - 3\log(\tfrac{2m}{r})\big] \nonumber \\
        & + 1295\Edot_{cdi}\delta_{ab}\nhat^{cdi} + 1260\Edot_{cd}\nhat_{ab}{}^{cd} \nonumber \\
        & + 5040\B^{cdi}\epsilon_{c}{}^{j}{}_{(a}\nhat_{b)dij} + 315\E_{cdi}\nhat_{ab}{}^{cdi}\big) \nonumber\\
        &+ \order{r^3}. \label{eq:hS1Primeab}
\end{align}%
\end{subequations}

The second-order term in Eq.~\eqref{eq:outer expansion v2} reads
\beq
\mathring{h}^{2'}_{\mu\nu} = \mathring{h}^{\R 2'}_{\mu\nu} + \mathring{h}^{\S 2'}_{\mu\nu}.
\eeq
The regular field 
\beq\label{hR2'}
\mathring{h}^{\R 2'}_{\mu\nu} = \prescript{0}{}{h}^{\R 2'}_{\mu\nu} + \prescript{1}{}{h}^{\R 1'}_{\mu\nu} +\prescript{2}{}{g}_{\mu\nu}
\eeq
is given by
\begin{equation}
    \mathring{h}^{\R 2'}_{\mu\nu}  = \order{r^2}. \label{eq:hR2Prime}
\end{equation}
The singular field 
\beq\label{hS2'}
\mathring{h}^{\S 2'}_{\mu\nu} = \prescript{0}{}{h}^{\S 2'}_{\mu\nu} + \prescript{1}{}{h}^{\S 1'}_{\mu\nu}
\eeq
is split into two pieces,
\begin{equation}
    \mathring{h}^{\S 2'}_{\mu\nu} = \mathring{h}^{\S\S'}_{\mu\nu} + \mathring{h}^{\S\R'}_{\mu\nu}, \label{eq:hS2PrimeSplit}
\end{equation}
where \(\mathring{h}^{\S\S'}_{\mu\nu}\) is the ``singular times singular" piece containing all terms proportional to \(m^2\), and \(\mathring{h}^{\S\R'}_{\mu\nu}\) is the ``singular times regular" piece featuring all terms with the form \(m\delta\E\) and \(m\delta\B\).
Individually, these are
\begin{subequations}
\label{eq:hSRPrime}
\begin{align}
    \mathring{h}^{\S\R'}_{tt} ={}& \frac{11}{3}mr\delta\E_{ab}\nhat^{ab}, \label{eq:hSRPrimett} \\
   \mathring{h}^{\S\R'}_{ta} ={}& \frac{2}{15}mr\big(11\delta\E_{ab}\nhat^{b} + 10\delta\B^{bc}\epsilon_{acd}\nhat_{b}{}^{d} \nonumber \\
        & + 15\delta\E_{bc}\nhat_{a}{}^{bc}\big), \label{eq:hSRPrimeta} \\
    \mathring{h}^{\S\R'}_{ab} ={}& \frac{1}{315}mr\big(154\delta\E_{ab} - 168\delta\B^{d}{}_{(a}\epsilon_{b)cd}\nhat^{c} \nonumber \\
        & + 480\delta\E^{c}_{(a}\nhat_{b)c} + 15\delta\E_{cd}\delta_{ab}\nhat^{cd} \nonumber \\
        & + 840\delta\B^{cd}\epsilon_{c}{}^{i}{}_{(a}\nhat_{b)di} + 105\delta\E_{cd}\nhat_{ab}{}^{cd}\big), \label{eq:hSRPrimeab}
\end{align}
\end{subequations}
and\footnote{All \(\order{m^0}\) terms in Eq.~(116) of Paper I have been corrected to include the factor $m^2$.}
\begin{subequations}
\label{eq:hSSPrime}
    \begin{align}
    \mathring{h}^{\S\S'}_{tt} ={}& -4m^2\big[\mathcal{E}_{ab} \nhat^{ab} + r\big(\tfrac{1}{3} \dot{\mathcal{E}}_{ab} \nhat^{ab} \big\{11 - 6 \log(\tfrac{2 m}{r})\big\} \nonumber \\
        & + \tfrac{2}{3} \mathcal{E}_{abc} \nhat^{abc}\big)\big] + \order{r^2}, \label{eq:hrgaugeSSPrimett} \\
    \mathring{h}^{\S\S'}_{ta} ={}& - 4m^2\big[\tfrac{2}{5}\E_{ab}\nhat^{b} + \E_{bc}\nhat_{a}{}^{bc} + r\bigl(\tfrac{6}{5}\Edot_{ab}\nhat^{b} \nonumber \\
        & \times \bigl\{2 - \log(\tfrac{2m}{r})\bigr\} + \tfrac{2}{9}\Bdot^{bc}\e_{acd}\nhat_{b}{}^{d}\bigl\{4 - \log(\tfrac{2m}{r})\bigr\} \nonumber \\
            & + \tfrac{8}{21}\E_{abc}\nhat^{bc} + \tfrac{1}{9}\Edot_{bc}\nhat_{a}{}^{bc}\bigl\{19 - 12\log(\tfrac{2m}{r})\bigr\} \nonumber \\
            & + \tfrac{2}{9}\B^{bcd}\e_{ab}{}^{i}\nhat_{cdi} + \tfrac{1}{2}\E_{bcd}\nhat_{a}{}^{bcd}\bigr)\big] + \order{r^2}, \label{eq:hrgaugeSSPrimeta} \\
   \mathring{h}^{\S\S'}_{ab} ={}& - 4m^2\bigl[\tfrac{4}{5}\B^{d}{}_{(a}\e_{b)cd}\nhat^{c} + \tfrac{8}{7}\E^{c}{}_{(a}\nhat_{b)c} \nonumber \\
        & - \tfrac{1}{21}\E_{cd}\delta_{ab}\nhat^{cd} + \B^{cd}\e_{c}{}^{i}{}_{(a}\nhat_{b)di} + \tfrac{5}{6}\E_{cd}\nhat_{ab}{}^{cd} \nonumber \\
        & + r\bigl(\tfrac{2}{45}\Edot_{ab}\bigl\{31 - 12\log(\tfrac{2m}{r})\bigr\} + \tfrac{4}{21}\E_{abc}\nhat^{c} \nonumber \\
        & - \tfrac{4}{45}\Bdot^{d}{}_{(a}\e_{b)cd}\nhat^{c}\bigl\{4 - 3\log(\tfrac{2m}{r})\bigr\} + \tfrac{4}{7}\Edot^{c}{}_{(a}\nhat_{b)c} \nonumber \\
        & \times \bigl\{4 - 3\log(\tfrac{2m}{r})\bigr\} + \tfrac{1}{63}\Edot_{cd}\delta_{ab}\nhat^{cd} \nonumber \\
        & \times \bigl\{29 - 6\log(\tfrac{2m}{r})\bigr\} - \tfrac{8}{63}\B_{c}{}^{i}{}_{(a}\e_{b)di}\nhat^{cd} \nonumber \\
        & + \tfrac{5}{9}\E_{cd(a}\nhat_{b)}{}^{cd} + \tfrac{1}{27}\E_{cdi}\delta_{ab}\nhat^{cdi} + \tfrac{4}{9}\Bdot^{cd}\e_{c}{}^{i}{}_{(a}\nhat_{b)di} \nonumber \\
        & \times \bigl\{4 - 3\log(\tfrac{2m}{r})\bigr\} + \tfrac{2}{9}\Edot^{cd}\nhat_{abcd}\bigl\{4 - 3\log(\tfrac{2m}{r})\bigr\} \nonumber \\
        & + \tfrac{4}{9}\B^{cdi}\e_{c}{}^{j}{}_{(a}\nhat_{b)dij} + \tfrac{1}{3}\E^{cdi}\nhat_{abcdi}\bigr)\bigr] + \order{r^2}. \label{eq:hrgaugeSSPrimeab}
    \end{align}%
\end{subequations}

We can extend this split into singular and regular fields to arbitrary order in $r$ by including all explicitly $m$-dependent terms in the singular fields, leaving the regular fields to include all terms that depend only on tidal moments, with no explicit $m$ dependence. The regular field is then manifestly a smooth solution to the vacuum Einstein equations,
\begin{align}
   \mathring{\deltaG}_{\mu\nu}[\mathring{h}^{\R1'}] ={}& 0, \label{eq:vacuumEFE1} \\
    \mathring{\deltaG}_{\mu\nu}[\mathring{h}^{\R2'}] ={}& -\mathring{\delta^2 G}_{\mu\nu}[\mathring{h}^{R1'}], \label{eq:vacuumEFE2}
\end{align}
where the overset ring indicates that these are the linearized and second-order Einstein operators defined from $\gzero_{\mu\nu}$. When combined with $\gzero_{\mu\nu}$, the regular field forms an effective metric
\beq
\tilde g'_{\mu\nu} = \gzero_{\mu\nu} +\e \mathring{h}^{\R1'}_{\mu\nu} + \e^2\mathring{h}^{\R2'}_{\mu\nu} + \ldots,
\eeq
as in Eq.~\eqref{eq:eff_metric}, which is a vacuum metric, and in which the small object follows a geodesic. To see that $\gamma$ is a geodesic in this metric, simply note that $\tilde g'_{\mu\nu}=\eta_{\mu\nu}+\order{r^2}$, where $\eta_{\mu\nu}$ is the Minkowski metric; the coordinates are therefore inertial along $\gamma$, which would be impossible of $\gamma$ were accelerated. To see that the object moves on $\gamma$, simply recall that there is no mass dipole moment in $\gobj_{\mu\nu}$ or in $H^{1}_{\mu\nu}$ (nor does the transformation from local lightcone coordinates to Fermi coordinates induce a mass dipole moment).

By inspection of the metric perturbations, we  see that in this gauge we have \(h^{2'}_{\mu\nu} \sim r^0\) instead of the generic behavior \(\sim\flatfrac{1}{r^2}\). This achieves the goal of eliminating the most singular, most problematic term in the second-order metric perturbation. One final step remains, however: to transform the perturbations into a practical gauge suitable for use in concrete implementations.

\section{Transformation to a generic highly regular gauge}\label{sec:HRgaugetransfo}

\subsection{Outline of method}

Although the lightcone rest gauge eliminates the \({\sim}\flatfrac{1}{r^2}\) pieces of \(h^2_{\mu\nu}\) that appears in a generic gauge, its ``rest gauge" aspect forces the regular field to behave as \(\sim r^2\), meaning that $h^\R_{\mu\nu}$ and its first derivative vanish on the worldline. In practice, we wish to be able to adopt a gauge that is convenient in the external background spacetime; in an EMRI, this is typically a radiation gauge~\cite{Chrzanowski1975}, the Regge--Wheeler--Zerilli gauge~\cite{Regge1957,Zerilli1970}, or the Lorenz gauge. In all these cases,  the choice is motivated at least in part by the fact that it leads to hyperbolic field equations for the metric perturbation or for some related variable. However, imposing these gauge conditions does not simultaneously allow one to enforce that the regular field vanishes on the worldline; for a given set of hyperbolic field equations, the regular field on the worldline is fully determined by global boundary conditions. To allow the end user to adopt a convenient gauge such as the radiation or Lorenz gauge, in this section we perform a smooth gauge transformation that puts $h^{\R}_{\mu\nu}$ in {\em any} desired gauge, while preserving the highly regular form of $h^{\S}_{\mu\nu}$. Our method again closely follows Paper~I.

Under a gauge transformation induced by a smooth vector field \(\xi^\mu = \epsilon\xi^\mu_1 + \epsilon^{2}\xi^\mu_2 + \order{\epsilon^3}\), perturbations of the metric \(g_{\mu\nu}\) transform as~\cite{Bruni1997}
\begin{align}
    h^{1}_{\mu\nu} \to{}& h^{1}_{\mu\nu} + \lie_{\xi_1} g_{\mu\nu}, \label{eq:1stordergauge} \\
    h^{2}_{\mu\nu} \to{}& h^{2}_{\mu\nu} +\lie_{\xi_2}g_{\mu\nu}+ \frac{1}{2}\lie^2_{\xi_1}g_{\mu\nu} + \lie_{\xi_1}h^{1}_{\mu\nu}. \label{eq:2ndordergauge}
\end{align}
However, we must divide these transformations into singular and regular pieces, and we must account for the fact that we have written our perturbations as perturbations of $\gzero_{\mu\nu}$.
An appropriate division of the gauge transformation is
\begin{align}
    \mathring{h}^{\text{R1}}_{\mu\nu} ={}& \mathring{h}^{\text{R1}'}_{\mu\nu} + \lie_{\xi_1}\gzero_{\mu\nu}, \label{eq:hR1transform} \\
    \mathring{h}^{\text{S1}}_{\mu\nu} ={}& \mathring{h}^{\text{S1}'}_{\mu\nu},  \label{eq:hS1transform} \\
    \mathring{h}^{\text{R2}}_{\mu\nu} ={}& \mathring{h}^{\text{R2}'}_{\mu\nu} + \lie_{\xi_2}\gzero_{\mu\nu} + \frac{1}{2}\lie^2_{\xi_1}\gzero_{\mu\nu} + \lie_{\xi_1}\mathring{h}^{\text{R1}'}_{\mu\nu}, \label{eq:hR2transform} \\
    \mathring{h}^{\text{S2}}_{\mu\nu} ={}& \mathring{h}^{\text{S2}'}_{\mu\nu} + \lie_{\xi_1}\mathring{h}^{\text{S1}'}_{\mu\nu}. \label{eq:hS2transform}
\end{align}
This ensures that $\tilde g_{\mu\nu}$ transforms as any smooth vacuum metric would under the gauge transformation, meaning that 
\beq
\tilde g_{\mu\nu} = \gzero_{\mu\nu} + \e \mathring{h}^{\R 1}_{\mu\nu}+\e^2\mathring{h}^{\R2}_{\mu\nu}+\ldots
\eeq
remains a vacuum metric and that geodesics in it, such as $\gamma$, remain geodesics. 
Apart from smoothness, we only impose one other condition on the transformation: that it is \emph{worldline preserving}, satisfying
\begin{equation}
    \xi^a_n\big|_{\gamma} = 0. \label{eq:worldlinepreserving}
\end{equation}
This ensures that the worldline in the practical gauge is identical to the worldline in the rest gauge.
An equivalent way to say this is that no mass dipole moment is introduced as a result of the transformation.

Still following Paper I, we now use the approach in Ref.~\cite{Gralla2012}: rather than choosing a gauge condition and finding a vector $\xi^\mu$ that enforces that condition, we allow the regular fields \(h^{\R n}_{\mu\nu}\) to be in an arbitrary gauge, and we solve Eqs.~\eqref{eq:hR1transform} and Eq.~\eqref{eq:hR2transform} for \(\xi_n^\mu\) in terms of \(h^{\R n}_{\mu\nu}\). The gauge of \(h^{\R n}_{\mu\nu}\) can then be freely chosen to put the field equations in any convenient form in the external background.

After finding $\xi^\mu_1$, we can calculate the second-order singular field in the new gauge via Eq.~\eqref{eq:hS2transform}. Despite the gauge vector being smooth, it introduces an unbounded term into \(h^{\mathrm{S}2}_{\mu\nu}\): \(\lie_{\xi_1}h^{\mathrm{S}1'}_{\mu\nu}\), which behaves as \({\sim}\flatfrac{1}{r}\). This is more divergent than the singular field \(\mathring{h}^{\mathrm{S}2'}_{\mu\nu}\) in the rest gauge, which was bounded at \(r=0\). However, as we will discuss in Sec.~\ref{sec:application}, ``singular times regular" terms like $\lie_{\xi_1}h^{\mathrm{S}1'}_{\mu\nu}$ are actually more benign than ``singular times singular" terms like \(\mathring{h}^{\mathrm{S}2'}_{\mu\nu}\) even if their divergence is superficially stronger.

In addition to determining the gauge vectors $\xi^\mu_n$ in terms of the regular field, Eqs.~\eqref{eq:hR1transform} and \eqref{eq:hR2transform} also determine the other functions in the effective metric in terms of the regular field: the accelerations $f^\mu_n$ and the tidal moments $\delta{\cal E}_{ab}$ and $\delta{\cal B}_{ab}$. To better bring out the structure of the equations, we note that after the gauge transformation, our full metric has the form
\begin{multline}
    \fullg_{\mu\nu} = \gzero_{\mu\nu} + \e\overbrace{\bigl(\prescript{0}{}{h}^{1}_{\mu\nu} + \prescript{1}{}{g}_{\mu\nu}\bigr)}^{\mathring{h}^{1}_{\mu\nu}} +\e^{2}\overbrace{\bigl(\prescript{0}{}{h}^{2}_{\mu\nu} + \prescript{1}{}{h}^{1}_{\mu\nu} + \prescript{2}{}{g}_{\mu\nu}\bigr)}^{\mathring{h}^{2}_{\mu\nu}} \\
        + \order{\e^3}, \label{eq:fullgdagger}
\end{multline}
and the regular and singular fields can be divided in analogy with Eqs.~\eqref{hR1'}, \eqref{hR2'}, and \eqref{hS2'}:\footnote{The \(\prescript{0}{}{g}_{\mu\nu}\) term in Eq.~(125) in Paper I should read \(\prescript{1}{}{g}_{\mu\nu}\).}
\begin{align}
    \mathring{h}^{\text{R1}}_{\mu\nu} \coloneqq{}& {}^0\hrtext{R1}_{\mu\nu} + \prescript{1}{}{g}_{\mu\nu}, \label{eq:hR1acc} \\
    \mathring{h}^{\text{S1}}_{\mu\nu} \coloneqq{}& {}^0\hrtext{S1}_{\mu\nu}, \label{eq:hS1acc} \\
    \mathring{h}^{\text{R2}}_{\mu\nu} \coloneqq{}& {}^0\hrtext{R2}_{\mu\nu} + {}^1\hrtext{R1}_{\mu\nu} + {}^2g_{\mu\nu}, \label{eq:hR2acc} \\
    \mathring{h}^{\text{S2}}_{\mu\nu} \coloneqq{}& {}^0\hrtext{S2}_{\mu\nu} + {}^1\hrtext{S1}_{\mu\nu}. \label{eq:hS2acc}
\end{align}
When solving the perturbative field equations in the external spacetime, the variables of interest are the perturbations $h^n_{\mu\nu}$ of the external background \(g_{\mu\nu}\), \emph{not} those of the background \(\gzero_{\mu\nu}\). Hence, we wish to express $\xi^\mu_n$ in terms of $h^{\R n}_{\mu\nu}$, not in terms of $\mathring{h}^{\R n}_{\mu\nu}$.
Using this decomposition, we rewrite Eqs.~\eqref{eq:hR1transform} and \eqref{eq:hR2transform} as
\begin{align}
     {}^0\hrtext{R1}_{\mu\nu} &=  \mathring{h}^{\text{R1}'}_{\mu\nu}+\lie_{\xi_1}\gzero_{\mu\nu} - {}^1g_{\mu\nu}, \label{eq:hR10}\\
    {}^0 h^{\text{R2}}_{\mu\nu}+{}^1 h^{\text{R1}}_{\mu\nu} &= \mathring{h}^{\text{R2}'}_{\mu\nu}+\lie_{\xi_2}\gzero_{\mu\nu} - {}^2 g_{\mu\nu}  \nonumber\\
    &\quad + \frac{1}{2}\lie^2_{\xi_1}\gzero_{\mu\nu} + \lie_{\xi_1}\mathring{h}^{\text{R1}'}_{\mu\nu}.\label{eq:hR20}
\end{align}
Here we have grouped all the unknowns ($\xi^\mu_n$, $f^\mu_n$, $\delta{\cal E}_{ab}$, and $\delta{\cal B}_{ab}$) on the right side of the equations.

Paper I solved Eq.~\eqref{eq:hR10} for portions of $\xi^\mu_1$ through order $r^2$  and used it to calculate the leading, $\sim 1/r$ term in $h^{\S2}_{\mu\nu}$. To make practical use of the highly regular gauge, we must know  two additional orders of the singular field: following Eq.~\eqref{eq:pun_sing_deriv}, we require  \(h^{\mathrm{S2}}_{\mu\nu}\) through order \(r\) to be able to correctly calculate the second-order self-force.
We already have \(\hrtext{S2'}_{\mu\nu}\) from Eqs.~\eqref{eq:hS2PrimeSplit}--\eqref{eq:hSSPrime}.
In the remaining parts of this section we calculate the complete $\xi^\mu_1$ through order $r^2$ and use it to calculate \(\lie_{\xi_1}h^{\text{S1}'}_{\mu\nu}\) through the necessary order. We also briefly discuss the solution for $\xi^\mu_2$.

\subsection{STF decomposition of the gauge vector and the regular field}\label{sec:STF_gauge_vector}

To solve Eq.~\eqref{eq:hR10} for the gauge vector, we begin by expanding both \(\xi^{1}_{\mu}\) (with index down) and \(\hronehr_{\mu\nu}\) in irreducible STF form using  Appendix~A of Ref.~\cite{Blanchet1986} and Appendix~B of Ref.~\cite{Poisson2011}.

\subsubsection{Gauge vector}\label{sec:STF_gauge}

The gauge vector is decomposed as
\begin{equation}
    \xi^1_{\mu} = \sum_{p,l \geq 0} r^{p} \xi^{(p,l)}_{\mu L}(t) \nhat^{L}, \label{eq:xiDecomp}
\end{equation}
where the \(t\) and \(a\) components are, respectively, given by
\begin{subequations}
\label{eq:xiDecompComponents}
\begin{align}
    \xi^{(p,l)}_{t\langle L\rangle} ={}& \That{p,l}_{L}, \label{eq:xiDecompT} \\
    \xi^{(p,l)}_{a\langle L\rangle} ={}& \Xhat{p,l}_{aL} + \epsilon^{j}{}_{a\langle i_{l}} \Yhat{p,l}_{L-1\rangle j} + \delta_{a\langle i_{l}}\Zhat{p,l}_{L-1\rangle}, \label{eq:xiDecompA}
\end{align}
\end{subequations}
with the hat indicating that these are STF tensors. Each term in this decomposition is linearly independent from the others. The quantities $\nhat^L$ form a complete basis, equivalent to scalar spherical harmonics, for scalar fields on the unit sphere, and the further decomposition of Cartesian 3-vectors and 3-tensors into irreducible STF pieces is equivalent to a decomposition into spin-weighted or tensor spherical harmonics.

As mentioned we only impose two conditions on \(\xi^1_{\mu}\): Firstly, that \(\xi^1_{\mu}\) is smooth so that our two gauges are smoothly related and secondly, that \(\xi^1_{\mu}\) is \emph{worldline preserving}, satisfying Eq.~\eqref{eq:worldlinepreserving}. These conditions imply that the expansion~\eqref{eq:xiDecomp} must be equivalent to a Taylor series
\beq
\xi^1_{\mu} = \sum_{k\geq 0} \frac{1}{k!}\partial_K\xi^1_{\mu}(t,0)x^K
\eeq
with $\xi^1_{a}(t,0)=0$. Here $x^K = x^{i_1}\cdots x^{i_k}$. When written as a sum of STF quantities, 
\beq
x^K = r^k[\nhat^K+c_1\delta^{(a_1 a_2}\nhat^{K-2)}+ c_2\delta^{(a_1 a_2}\delta^{a_3 a_4}\nhat^{K-4)}+\ldots]
\eeq
for some some numerical coefficients $c_n$. Hence, our conditions on the gauge vector impose
\begin{subequations}
\label{eq:gaugevectorSTF}
\begin{align}
    \xi^1_t ={}& \That{0,0} + r \nhat^{a}\,\That{1,1}_{a} + r^{2}\big(\That{2,0} + \That{2,2}_{ab}\nhat^{ab}\big) \nonumber \\
        & + r^{3}\big(\That{3,1}_{a}\nhat^{a} + \That{3,3}_{abc}\nhat^{abc}\big) + \order{r^4}, \label{eq:gaugevectorSTFt} \displaybreak[0] \\
    \xi^1_a ={}& r\nhat^{b}\big(\Xhat{1,1}_{ab} + \epsilon^{j}{}_{ab}\Yhat{1,1}_{j} + \delta_{ab}\Zhat{1,1}\big) + r^{2}\big[\Xhat{2,0}_{a} \nonumber \\
        & + \nhat^{bc}\big(\Xhat{2,2}_{abc} + \epsilon^{j}{}_{ab}\Yhat{2,2}_{cj} + \delta_{ab}\Zhat{2,2}_{c}\big)\big] \nonumber \\
        & + r^{3}\big[\nhat^{b}\big(\Xhat{3,1}_{ab} + \epsilon^{j}{}_{ab}\Yhat{3,1}_{j} + \delta_{ab}\Zhat{3,1}\big) \nonumber \\
        & + \nhat^{bcd}\big(\Xhat{3,3}_{abcd} + \epsilon^{j}{}_{ab}\Yhat{3,3}_{cdj} + \delta_{ab}\Zhat{3,3}_{cd}\big)\big] \nonumber \\
        & + \order{r^4}. \label{eq:gaugevectorSTFa}
\end{align}
\end{subequations}

It is necessary to carry this expansion to order \(r^3\) because the Lie derivative and the singular form of \(\mathring{h}^{\S1'}_{\mu\nu}\) in Eq.~\eqref{eq:hS2transform} each reduce the order in \(r\) by one.
Thus, order \(r^3\) in the gauge vector is required for accuracy through order \(r\) in \(\mathring{h}^{\text{S2}}_{\mu\nu}\).

\subsubsection{Regular field}\label{sec:STF_reg_field}

We perform a similar decomposition for the regular field, so that
\begin{equation}
    \hronehr_{\mu\nu} = \sum_{p,l \geq 0} r^{p} \hronehrpl{p,l}_{\mu\nu L}(t)\nhat^{L}. \label{eq:hR1Decomp}
\end{equation}
The \(tt\), \(ta\), and \(ab\) components are given by
\begin{subequations}
\label{eq:hR1DecompComponents}
\begin{align}
    \hronehrpl{p,l}_{tt\langle L\rangle} ={}& \Ahat{p,l}_{L}, \label{eq:hR1DecompTT} \\
    \hronehrpl{p,l}_{ta\langle L\rangle} ={}& \Bhat{p,l}_{aL} + \epsilon^{j}{}_{a\langle i_{l}} \Chat{p,l}_{L-1\rangle j} + \delta_{a\langle i_{l}}\Dhat{p,l}_{L-1\rangle}, \label{eq:hR1DecompTA} \displaybreak[0] \\
    \hronehrpl{p,l}_{ab\langle L\rangle} ={}& \Ehat{p,l}_{abL} + \delta_{ab}\Khat{p,l}_{L} + \STF_{L}\STF_{ab}\bigl(\e^{j}{}_{a{i}_{l}}\Fhat{p,l}_{bjL-1} \nonumber \\
        & + \delta_{a{i}_{l}}\Ghat{p,l}_{bL-1} + \delta_{a{i}_{l}}\e^{j}{}_{b{i}_{l-1}}\Hhat{p,l}_{jL-2} \nonumber \\
        & + \delta_{a{i}_{l}}\delta_{b{i}_{l-1}}\Ihat{p,l}_{L-2}\bigr). \label{eq:hR1DecompAB}
\end{align}
\end{subequations}

Since \(\hronehr_{\mu\nu}\) is smooth, we require this expansion to be equivalent to a Taylor series in $x^a$.
This leaves us with the expansion
\begin{subequations}
\label{eq:hR1STF}
\begin{align}
    \hronehr_{tt} ={}& \Ahat{0,0} + r\Ahat{1,1}_{i}\nhat^{i} + r^2\bigl(\Ahat{2,0} + \nhat^{ij}\Ahat{2,2}_{ij}\bigr) \nonumber \\
        & + \order{r^3}, \label{eq:hR1STFtt} \displaybreak[0] \\
    \hronehr_{ta} ={}& \Bhat{0,0}_{a} + r\nhat^{i}\bigl(\Bhat{1,1}_{ai} + \e^{j}{}_{ai}\Chat{1,1}_{j} + \delta_{ai}\Dhat{1,1}\bigr) \nonumber \\
        & + r^2\bigl[\Bhat{2,0}_{a} + \nhat^{ij}\bigl(\Bhat{2,2}_{aij} + \e^{k}{}_{ai}\Chat{2,2}_{jk} \nonumber \\
        & + \delta_{ai}\Dhat{2,2}_{j}\bigr)\bigr] + \order{r^3}, \label{eq:hR1STFta} \displaybreak[0] \\
    \hronehr_{ab} ={}& \Ehat{0,0}_{ab} + \delta_{ab}\Khat{0,0} + r\nhat^i\bigl[\Ehat{1,1}_{abi} + \delta_{ab}\Khat{1,1}_{i} \nonumber \\
        & + \STF_{ab}\bigl(\e^{k}{}_{ai}\Fhat{1,1}_{bk} + \delta_{ai}\Ghat{1,1}_{b}\bigr)\bigr] + r^2\bigl[\Ehat{2,0}_{ab} \nonumber \\
        & + \delta_{ab}\Khat{2,0} + \nhat^{ij}\bigl(\Ehat{2,2}_{abij} + \delta_{ab}\Khat{2,2}_{ij} \nonumber \\
        & + \STF_{ab}\bigl\{\e^{k}{}_{ai}\Fhat{2,2}_{bkj} + \delta_{ai}\Ghat{2,2}_{bj} + \delta_{ai}\e^{k}{}_{bj}\Hhat{2,2}_{k} \nonumber \\
        & + \delta_{ai}\delta_{bj}\Ihat{2,2}\bigr\}\bigr)\bigr] + \order{r^3}. \label{eq:hR1STFab}
\end{align}
\end{subequations}
Appendix~\ref{app:reg_field} gives the relation between the individual STF tensors and derivatives of the regular field evaluated on the worldline.

Additionally, we use constraints from the linearised vacuum Einstein equations
\begin{equation}
    \mathring{\deltaG}_{\mu\nu}[\,^{0}{}h^{\mathrm{R1}}] = 0. \label{eq:pertEFE1}
\end{equation}
Note that \(\mathring{\deltaG}_{\mu\nu}[\mathring{h}^{\mathrm{R1}}] = \mathring{\deltaG}_{\mu\nu}[\,^{0}{}h^{\mathrm{R1}}]\) because ${}^1 g_{\mu\nu}$ is a linear vacuum perturbation of $\gzero_{\mu\nu}$.

The \(tt\) and \(ta\) components of Eq.~\eqref{eq:pertEFE1} give
\begin{subequations}
\label{eq:EFEconstraintsTTTA}
\begin{align}
    \Ihat{2,2} ={}& \frac{1}{5}\E^{ab}\Ehat{0,0}_{ab} + \frac{6}{5}\Khat{2,0}, \label{eq:EFEconstraintsTT} \\
    \Dhat{2,2}_{a} ={}& \frac{6}{5}\Bhat{2,0}_{a} + \frac{3}{5}\Bhat{0,0}_{b}\E_{a}{}^{b} + \frac{3}{5}\B^{bc}\e_{ac}{}^{d}\Ehat{0,0}_{bd} \nonumber \\
        & - \frac{1}{2}\dv{t}\Ghat{1,1}_{a} + \frac{3}{5}\dv{t}\Khat{1,1}_{a}. \label{eq:EFEconstraintsTA}
\end{align}
\end{subequations}
We use these equations to eliminate \(\Ihat{2,2}\) and \(\Dhat{2,2}_{a}\), but the choice is arbitrary; we could have easily chosen  two other STF tensors to remove.

From the \(ab\) component of Eq.~\eqref{eq:pertEFE1} we get two restrictions, one at \(l=0\) and one at \(l=2\). These are
\begin{subequations}
\label{eq:EFEconstraintsAB}
\begin{align}
    \Ahat{2,0} ={}& -\frac{1}{3}\E^{ab}\Ehat{0,0}_{ab} + \dv{t}\Dhat{1,1} - \frac{1}{2}\dv[2]{t}\Khat{0,0}, \label{eq:EFEconstraintsAB0} \\
    \Ahat{2,2}_{ab} ={}& \Ahat{0,0}\E_{ab} - 2\Bhat{0,0}_{c}\B_{d(a}\e_{b)}{}^{cd} + \Ehat{2,0}_{ab} \nonumber \\
        & - 2\E^{c}{}_{\langle a}\Ehat{0,0}_{b\rangle c} - \frac{7}{6}\Ghat{2,2}_{ab} + \E_{ab}\Khat{0,0} \nonumber \\
        & + \Khat{2,2}_{ab} + \dv{t}\Bhat{1,1}_{ab} - \frac{1}{2}\dv[2]{t}\Ehat{0,0}_{ab}, \label{eq:EFEconstraintsAB2}
\end{align}
\end{subequations}
where the constraints from the \(tt\) and \(ta\) components have been used to simplify these expressions.

Combining Eqs.~\eqref{eq:hR1STF},~\eqref{eq:EFEconstraintsTTTA}, and~\eqref{eq:EFEconstraintsAB} gives us the final expression for the components of \(\hronehr_{\mu\nu}\):
\begin{subequations}
\label{eq:hR1STFExpand}
\begin{align}
    \hronehr_{tt} ={}& \Ahat{0,0} + r\Ahat{1,1}_{i}\nhat^{i} + r^2\Bigl[-\frac{1}{3}\E^{ab}\Ehat{0,0}_{ab} + \dv{t}\Dhat{1,1} \nonumber \\
        & - \frac{1}{2}\dv[2]{t}\Khat{0,0} + \nhat^{ij}\Bigl(\Ahat{0,0}\E_{ij} - 2\Bhat{0,0}_{c}\B_{di}\e_{j}{}^{cd} \nonumber \\
        & + \Ehat{2,0}_{ij} - 2\E^{c}{}_{i}\Ehat{0,0}_{jc} - \frac{7}{6}\Ghat{2,2}_{ij} + \E_{ij}\Khat{0,0} \nonumber \\
        & + \Khat{2,2}_{ij} + \dv{t}\Bhat{1,1}_{ij} - \frac{1}{2}\dv[2]{t}\Ehat{0,0}_{ij}\Bigr)\Bigr] + \order{r^3}, \label{eq:hR1STFExpandtt} \displaybreak[0] \\
    \hronehr_{ta} ={}& \Bhat{0,0}_{a} + r\nhat^{i}\bigl(\Bhat{1,1}_{ai} + \e^{j}{}_{ai}\Chat{1,1}_{j} + \delta_{ai}\Dhat{1,1}\bigr) \nonumber \\
        & + r^2\Bigl[\Bhat{2,0}_{a} + \nhat^{ij}\Bigl(\Bhat{2,2}_{aij} + \e^{k}{}_{ai}\Chat{2,2}_{jk} \nonumber \\
        & + \delta_{ai}\Bigl\{\frac{6}{5}\Bhat{2,0}_{j} + \frac{3}{5}\Bhat{0,0}_{b}\E_{j}{}^{b} + \frac{3}{5}\B^{bc}\e_{jc}{}^{d}\Ehat{0,0}_{bd} \nonumber \\
        & - \frac{1}{2}\dv{t}\Ghat{1,1}_{j} + \frac{3}{5}\dv{t}\Khat{1,1}_{j}\Bigr\}\Bigr)\Bigr] + \order{r^3}, \label{eq:hR1STFExpandta} \displaybreak[0] \\
    \hronehr_{ab} ={}& \Ehat{0,0}_{ab} + \delta_{ab}\Khat{0,0} + r\nhat^i\bigl[\Ehat{1,1}_{abi} + \delta_{ab}\Khat{1,1}_{i} \nonumber \\
        & + \STF_{ab}\bigl(\e^{k}{}_{ai}\Fhat{1,1}_{bk} + \delta_{ai}\Ghat{1,1}_{b}\bigr)\bigr] + r^2\Bigl[\Ehat{2,0}_{ab} \nonumber \\
        & + \delta_{ab}\Khat{2,0} + \nhat^{ij}\Bigl(\Ehat{2,2}_{abij} + \delta_{ab}\Khat{2,2}_{ij} \nonumber \\
        & + \STF_{ab}\Bigl\{\e^{k}{}_{ai}\Fhat{2,2}_{bkj} + \delta_{ai}\Ghat{2,2}_{bj} + \delta_{ai}\e^{k}{}_{bj}\Hhat{2,2}_{k} \nonumber \\
        & + \delta_{ai}\delta_{bj}\Bigl(\frac{1}{5}\E^{cd}\Ehat{0,0}_{cd} + \frac{6}{5}\Khat{2,0}\Bigr)\Bigr\}\Bigr)\Bigr] + \order{r^3}. \label{eq:hR1STFExpandab}
\end{align}
\end{subequations}
This form is particularly advantageous as it automatically includes any constraints that would be imposed by the Einstein equations onto the form of our regular field.

\subsection{Solving for $\xi^\mu_1$}\label{sec:gauge_matching}

We now return to Eq.~\eqref{eq:hR10}, where, recall, \(\gzero_{\mu\nu}\) is given by Eq.~\eqref{eq:metric_no_acc}, \(\prescript{1}{}{g}_{\mu\nu}\) by Eq.~\eqref{eq:g1}, and \(\mathring{h}^{\R 1'}_{\mu\nu}\) by Eq.~\eqref{eq:hR1Prime}. To solve for the gauge vector, we substitute the expansions~\eqref{eq:gaugevectorSTF} and \eqref{eq:hR1STFExpand} and then work order by order in \(r\) and \(\nhat^L\). 
This is possible because \(\nhat^L\) forms an orthogonal basis, implying \(A_{P\langle L\rangle}\nhat^{L} = B_{P\langle L\rangle}\nhat^{L} \implies A_{P\langle L\rangle} = B_{P\langle L\rangle}\). As a result, Eq.~\eqref{eq:hR10} reduces to a hierarchical set of equations for the STF tensors $\hat T^{(p,l)}_L$, $\hat X^{(p,l)}_{L+1}$, $\hat Y^{(p,l)}_{L}$, and $\hat Z^{(p,l)}_{L-1}$.

Rather than belabouring the technical details of the calculation, which are largely mechanical, we state the results that follow from each order in $r$ in  Eq.~\eqref{eq:hR10}.

Note that in the equations that follow, \(\hronehr_{\mu\nu}\) and its derivatives are always evaluated on the worldline, but we omit the notation ${}|_{\gamma}$ for brevity.
Additionally, we define \(\hronehr := \hronehr_a{}^a \equiv \delta^{ab} \,\hronehr_{ab}\).

\subsubsection{Order \(r^0\)}\label{sec:HR_order0}

Starting at the lowest order in the expansion of Eq.~\eqref{eq:hR10}, we immediately discover rules for four of our gauge vector components.
These are
\begin{align}
    \That{0,0} ={}& \frac{1}{2} \int\Ahat{0,0}\dd{t}, \label{eq:T00} \\
    \That{1,1}_a ={}& \Bhat{0,0}_{a}, \label{eq:T11} \\
    \Xhat{1,1}_{ab} ={}& \frac{1}{2}\Ehat{0,0}_{ab}, \label{eq:X11} \\
    \Zhat{1,1} ={}& \frac{1}{2}\Khat{0,0}. \label{eq:Z11}
\end{align}
In Paper I, the relations in Eqs.~(129)--(131) were given in terms of the full gauge vector, \(\xi^1_\mu\).
To compare, we perform equivalent operations but now on our expansion of \(\xi^1_\mu\), substituting our values for the STF tensors from Eqs.~\eqref{eq:T00}--\eqref{eq:Z11} and using Appendix~\ref{app:reg_field} to relate the STF tensors to derivatives of the regular field.
The results are
\begin{align}
    \dv{t}\xi^1_t ={}& \dv{t}\That{0,0} = \frac{1}{2}\Ahat{0,0} \nonumber \\
        ={}& \frac{1}{2}\hronehr_{tt}, \label{eq:xi1tt} \displaybreak[0] \\
    \xi^1_{t,a} ={}& \That{1,1}_a = \Bhat{0,0}_{a} \nonumber \\
        ={}& \hronehr_{ta}, \label{eq:xi1ta} \displaybreak[0] \\
    \xi^1_{(a,b)} ={}& \Xhat{1,1}_{ab} + \delta_{ab}\Zhat{1,1} \nonumber \\
        ={}& \frac{1}{2}\Ehat{0,0}_{ab} + \frac{1}{2}\delta_{ab}\Khat{0,0} \nonumber \\
        ={}& \frac{1}{2}\hronehr_{ab}, \label{eq:xi1abSym}
\end{align}
which exactly match the expressions in Paper I, as expected.
The value of \(\xi^1_{[a,b]}\) is also given in Paper I but relies on \(\Yhat{1,1}_{c}\), which is found at order \(r\).

\subsubsection{Order \(r\)}\label{sec:HR_order1}

Having correctly reproduced the leading expressions from Paper I, we can confidently move on to higher orders.
We continue our procedure, but now we find our higher-order STF tensors in terms of not just the STF tensors in \(\hronehr_{\mu\nu}\) but also the tidal moments.

From the \(tt\) component of Eq.~\eqref{eq:hR10}, we obtain an expression for the first-order self-force,
\begin{equation}
    f^{1}_{a} = \frac{1}{2}\Ahat{1,1}_{a} - \dv{t}\Bhat{0,0}_{a}. \label{eq:f1HRSTF}
\end{equation}
When rewritten in terms of \(\hronehr_{\mu\nu}\), this gives,
\begin{equation}
    f^{1}_{a} = \frac{1}{2}\hronehr_{tt,a} - \dv{t}\hronehr_{ta}, \label{eq:f1HR}
\end{equation}
which is the standard result for the first-order self-force when written in component form~\cite{Pound2017}.

The \(ta\) component gives
\begin{align}
    \That{2,0} ={}& \frac{1}{2}\Dhat{1,1} - \frac{1}{4}\dv{t}\Khat{0,0}, \label{eq:T20} \\
    \That{2,2}_{ab} ={}& \frac{1}{2}\Bhat{2,2}_{ab} + \frac{1}{2}\E_{ab}\int\Ahat{0,0}\dd{t} - \frac{1}{4}\dv{t}\Ehat{0,0}_{ab}, \label{eq:T22} \\
    \Yhat{1,1}_{a} ={}& \int\Chat{1,1}_{a}\dd{t}. \label{eq:Y11}
\end{align}
Using the value of \(\Yhat{1,1}_{a}\), we can now compare to Paper I's result for the antisymmetric part of the spatial derivative of the gauge vector.
This gives
\begin{align}
    \xi^{1}_{[a,b]} ={}& \e_{ab}{}^{c}\Yhat{1,1}_{c} = \e_{ab}{}^{c}\int\Chat{1,1}_{c}\dd{t} \nonumber \\
        ={}& \int\hronehr_{t[a,b]}\dd{t}, \label{eq:xi1abAnti}
\end{align}
which matches Eq.~(133) from Paper I.

Finally, the \(ab\) component  of Eq.~\eqref{eq:hR10} gives
\begin{align}
    \Xhat{2,0}_{a} ={}& \frac{5}{18}\Ghat{1,1}_{a} - \frac{1}{12}\Khat{1,1}_{a}, \label{eq:X20} \\
    \Xhat{2,2}_{abc} ={}& \frac{1}{4}\Ehat{1,1}_{abc}, \label{eq:X22} \\
    \Yhat{2,2}_{ab} ={}& \frac{1}{2}\Fhat{1,1}_{ab} - \frac{1}{3}\B_{ab}\int\Ahat{0,0}\dd{t}, \label{eq:Y22} \\
    \Zhat{2,2}_{a} ={}& \frac{1}{2}\Khat{1,1}_{a} - \frac{1}{6}\Ghat{1,1}_{a}. \label{eq:Z22}
\end{align}

\subsubsection{Order \(r^2\)}\label{sec:HR_order2}

At the final order, not only do we find the last components of the gauge vector but we also fix the forms of \(\delta\E_{ab}\) and \(\delta\B_{ab}\) that appear in Eq.~\eqref{eq:hR1Prime}.
They are
\begin{align}
    \delta\E_{ab} ={}& 2\B^{d}{}_{(a}\Bhat{0,0}_{|c|}\e_{b)}{}^{c}{}_{d} + \E^{c}{}_{\langle a}\Ehat{0,0}_{b\rangle c} - \Ehat{2,0}_{ab} + \frac{7}{6}\Ghat{2,2}_{ab} \nonumber \\
        & - 2\E_{ab}\Khat{0,0} - \Khat{2,2}_{ab} + \frac{1}{2}\Edot_{ab}\int\Ahat{0,0}\dd{t} \nonumber \\
        & - 2\E^{d}{}_{(a}\e_{b)}{}^{c}{}_{d}\int\Chat{1,1}_{c}\dd{t}, \label{eq:deltaEgauge} \displaybreak[0] \\
    \delta\B_{ab} ={}& \frac{1}{2}\Ahat{0,0}\B_{ab} + \frac{3}{2}\Chat{2,2}_{ab} - \Bhat{0,0}_{c}\E^{d}{}_{(a}\e_{b)}{}^{c}{}_{d} \nonumber \\
        & + \frac{1}{2}\Bdot_{ab}\int\Ahat{0,0}\dd{t} - \frac{3}{2}\B_{ab}\Khat{0,0} - \frac{3}{4}\dv{t}\Fhat{1,1}_{ab} \nonumber \\
        & - 2\B^{d}{}_{(a}\e_{b)}{}^{c}{}_{d}\int\Chat{1,1}_{c}\dd{t}. \label{eq:deltaBgauge}
\end{align}

When the values of the STF tensors are substituted, however, these become\footnote{While we do not manipulate \(\delta\B_{ab}\) after substitution, we do manipulate \(\delta\E_{ab}\). Arriving at our second expression for \(\delta\E_{ab}\) necessitates rewriting the Einstein field equation's condition for \(\Ahat{2,2}_{ab}\) from Eq.~\eqref{eq:EFEconstraintsAB2} in terms of \(\Bhat{0,0}_{c}\B_{d(a}\e_{b)}{}^{cd}\) and then substituting it into our initial expression for \(\delta\E_{ab}\).}
\begin{align}
    \delta\E_{ab} ={}& \E_{ab}\hronehr_{tt} - \E_{\langle a}{}^{c}\hronehr_{b\rangle c} - \frac{1}{2}\hronehr_{tt,\langle ab\rangle} + \dv{t}\hronehr_{t\langle a,b\rangle} \nonumber \\
        & + \frac{1}{2}\Edot_{ab}\inthrhr{_{tt}} - \frac{1}{2}\dv[2]{t}\hronehr_{\langle ab\rangle} \nonumber \\
        & + 2\STF_{ab}\E_{a}{}^{c}\inthrhr{_{t[b,c]}}, \label{eq:deltaE} \displaybreak[0] \\
    \delta\B_{ab} ={}& - \frac{1}{2}\B_{ab}\hronehr + \E_{(a}{}^{c}\epsilon_{b)c}{}^{d}\hronehr_{td} + \frac{1}{2}\B_{ab}\hronehr_{tt} \nonumber \\
        & + \frac{1}{2}\epsilon^{cd}{}_{(a}\hronehr_{|tc|,b)d} + \frac{1}{2}\Bdot_{ab}\inthrhr{_{tt}} \nonumber \\
        & + 2\STF_{ab}\B_{a}{}^{c}\inthrhr{_{t[b,c]}} - \frac{1}{2}\epsilon^{cd}{}_{(a}\dv{t}\hronehr_{b)c,d}. \label{eq:deltaB}
\end{align}
These expressions match those found for the transformation from the rest gauge to the Lorenz gauge in Paper I but with the omission of the term \(\propto m\).
As in Paper I, we can also write the perturbations of the tidal moments as
\begin{align}
    \delta\E_{ab} ={}& \mathring{\delta R}_{tatb}[\hronehr - \lie_{\xi_{1}}\mathring{g}], \label{eq:deltaERiemann} \\
    \delta\B_{ab} ={}& \frac{1}{2}\epsilon^{pq}{}_{(a}\mathring{\delta R}_{b)tpq}[\hronehr - \lie_{\xi_{1}}\mathring{g}], \label{eq:deltaBRiemann}
\end{align}
in agreement with analogous results in Ref.~\cite{Gralla2012}. These forms of $\delta\E_{ab}$ and $\delta\B_{ab}$ let us interpret them as the tidal moments of \(\hronehr_{\mu\nu}\) (up to a gauge transformation).

The rest of the STF tensors are found to be
\begin{align}
    \That{3,1}_{a} ={}& \frac{3}{5}\Bhat{2,0}_{a} + \frac{2}{5}\E_{a}{}^{b}\Bhat{0,0}_{b} + \frac{1}{5}\B^{bc}\e_{ac}{}^{d}\Ehat{0,0}_{bd} \nonumber \\
        & + \frac{2}{5}\B_{a}{}^{b}\int\Chat{1,1}_{b}\dd{t} - \frac{1}{6}\dv{t}\Ghat{1,1}_{a} + \frac{1}{20} \dv{t}\Khat{1,1}_{a}, \label{eq:T31} \displaybreak[0] \\
    \That{3,3}_{abc} ={}& \Bhat{2,2}_{abc} + \frac{1}{6}\E_{abc}\int\Ahat{0,0}\dd{t} - \frac{1}{12}\dv{t}\Ehat{1,1}_{abc} \nonumber \\
        & + \STF_{abc}\biggl(\frac{2}{3}\Bhat{0,0}_{a}\E_{bc} + \frac{1}{3}\B^{d}{}_{a}\e_{bd}{}^{i}\Ehat{0,0}_{ci} \nonumber \\
        & - \frac{2}{9}\B_{ab}\int\Chat{1,1}_{c}\dd{t}\biggr), \label{eq:T33} \displaybreak[0] \\
    \Xhat{3,1}_{ab} ={}& \frac{4}{15}\Ehat{2,0}_{ab} + \frac{7}{180}\Ghat{2,2}_{ab} - \frac{1}{30}\Khat{2,2}_{ab} \nonumber \\
        & - \frac{1}{15}\B^{d}{}_{(a}\Bhat{0,0}_{|c|}\e_{b)}{}^{c}{}_{d} - \frac{1}{10}\E^{c}{}_{\langle a}\Ehat{0,0}_{b\rangle c} \nonumber \\
        & - \frac{1}{40}\Edot_{ab}\int\Ahat{0,0}\dd{t} - \frac{1}{15}\E^{d}{}_{(a}\e_{b)}{}^{c}{}_{d}\int\Chat{1,1}_{c}\dd{t}, \label{eq:X31} \displaybreak[0] \\
    \Xhat{3,3}_{abcd} ={}& \frac{1}{6}\Ehat{2,2}_{abcd}, \label{eq:X33} \displaybreak[0] \\
    \Yhat{3,1}_{a} ={}& - \frac{1}{10}\B_{a}{}^{b}\Bhat{0,0}_{b} - \frac{1}{60}\E^{bc}\e_{ac}{}^{d}\Ehat{0,0}_{bd} + \frac{1}{4}\Hhat{2,2}_{a} \nonumber \\
        & + \frac{7}{30}\E_{a}{}^{b}\int\Chat{1,1}_{b}\dd{t}, \label{eq:Y31} \displaybreak[0] \\
    \Yhat{3,3}_{abc} ={}& \frac{1}{4}\Fhat{2,2}_{abc} - \frac{1}{6}\B_{abc}\int\Ahat{0,0}\dd{t} \nonumber \\
        & + \STF_{abc}\biggl(\frac{1}{6}\E^{d}{}_{a}\e_{bd}{}^{i}\Ehat{0,0}_{ci} - \frac{1}{3}\B_{ab}\Bhat{0,0}_{c} \nonumber \\
        & - \frac{1}{3}\E_{ab}\int\Chat{1,1}_{c}\dd{t}\biggr), \label{eq:Y33} \displaybreak[0] \\
    \Zhat{3,1} ={}& \frac{1}{45}\E^{ab}\Ehat{0,0}_{ab} + \frac{3}{10}\Khat{2,0}, \label{eq:Z31} \displaybreak[0] \\
    \Zhat{3,3}_{ab} ={}& \frac{1}{9}\B^{d}{}_{(a}\e_{b)}{}^{c}{}_{d}\Bhat{0,0}_{c} - \frac{1}{6}\Ehat{2,0}_{ab} + \frac{1}{36}\Ghat{2,2}_{ab} \nonumber \\
        & + \frac{1}{6}\E^{c}{}_{\langle a}\Ehat{0,0}_{b\rangle c} + \frac{1}{3}\Khat{2,2}_{ab} + \frac{1}{24}\Edot_{ab}\int\Ahat{0,0}\dd{t} \nonumber \\
        & + \frac{1}{9}\E^{d}{}_{(a}\e_{b)}{}^{c}{}_{d}\int\Chat{1,1}_{c}\dd{t}. \label{eq:Z33}
\end{align}

\subsubsection{Final result for $\xi^\mu_1$}\label{app:HRgauge}

Substituting the above results for the STF tensors into Eq.~\eqref{eq:gaugevectorSTF}, we obtain our final form of the gauge vector required to transform from the rest gauge into the practical highly regular gauge.
The components are given  by\footnote{\(\xi^{1}_{a}\) was additionally simplified using the constraint in Eq.~\eqref{eq:EFEconstraintsTT} from the Einstein field equations in terms of \(\hronehr_{\mu\nu}\).}
\begingroup
\allowdisplaybreaks
\begin{subequations}
\label{eq:HRgaugevector}
\begin{align}
    \xi^1_t ={}& \frac{1}{2}\inthrhr{_{tt}} + r\nhat^a\hronehr_{ta} + \frac{r^2}{12}\bigg[2\hronehr_{ta,}{}^a - \dv{t}\hronehr \nonumber \\
        & + 3\nhat^{ab}\bigg(2\hronehr_{ta,b} + 2\E_{ab}\inthrhr{_{tt}} - \dv{t}\hronehr_{ab}\bigg)\bigg] \nonumber \\
        & + \frac{r^3}{60}\bigg[3\nhat^a\bigg(4\B^{bc}\epsilon_{ac}{}^d\hronehr_{bd} + 8\E_{a}{}^{b}\hronehr_{tb} + 2\hronehr_{ta,b}{}^b \nonumber \\
        & + 4\B_{ab}\epsilon^{bcd}\inthrhr{_{tc,d}} - 2\dv{t}\hronehr_{ab,}{}^b + \dv{t}\hronehr_{,a}\bigg) \nonumber \\
        & + 5\nhat^{abc}\bigg(8\E_{ab}\hronehr_{tc} + 2\E_{abc}\inthrhr{_{tt}} + 2\hronehr_{ta,bc} \nonumber \\
        & + 4\B_{a}{}^{d}\e_{cd}{}^{i}\hronehr_{bi} - 4\B_{ab}\e_{c}{}^{di}\inthrhr{_{td,i}} \nonumber \\
        & - \dv{t}\hronehr_{ab,c}\bigg)\bigg] + \order{r^4}, \label{eq:HRgaugevectort} \displaybreak[0] \\
    \xi^1_a ={}& \frac{r\nhat^b}{2}\bigg[\hronehr_{ab} + 2\inthrhr{_{t[a,b]}}\bigg] + \frac{r^2}{12}\bigg[2\hronehr_{ab,}{}^b - \hronehr_{,a} \nonumber \\
        & + \nhat^{bc}\bigg(6\hronehr_{ab,c} - 4\B_{b}{}^{d}\epsilon_{acd}\inthrhr{_{tt}} - 3\hronehr_{bc,a} \bigg)\bigg] \nonumber \displaybreak[0] \\
        & + \frac{r^3}{360}\bigg[12\E^{bc}\hronehr_{bc}\nhat_a - 3\nhat^b\bigg(4\E_{b}{}^{c}\hronehr_{ac} \nonumber \\
        & + 8\E_{a}{}^{c}\hronehr_{bc} + 12\B^{cd}\epsilon_{abd}\hronehr_{tc} - 4\E_{ab}\hronehr \nonumber \\
        & - 8\B^{c}{}_{(a}\epsilon_{b)c}{}^{d}\hronehr_{td} - 8\hronehr_{b[a,c]}{}^c - 8\hronehr_{ac,b}{}^c \nonumber \\
        & + 2\hronehr_{,ab} + 3\Edot_{ab}\inthrhr{_{tt}} + 24\E_{b}{}^{c}\inthrhr{_{t[a,c]}} \nonumber \\
        & - 32\E_{a}{}^{c}\inthrhr{_{t[b,c]}}\bigg) + 5\nhat_{a}{}^{bc}\bigg(12\E_{b}{}^{d}\hronehr_{cd} \nonumber \\
        & - 8\B_{b}{}^{d}\epsilon_{cd}{}^{i}\hronehr_{ti} - 24\E_{b}{}^{d}\inthrhr{_{t[c,d]}} \nonumber \\
        & + 3\Edot_{bc}\hronehr_{tt}\bigg) - 10\nhat^{bcd}\bigg(12\E_{b[c}\hronehr_{a]d} - 6\hronehr_{ab,cd} \nonumber \\
        & + 8\B_{b}{}^{i}\epsilon_{adi}\hronehr_{tc} + 4\B_{bc}\epsilon_{ad}{}^{i}\hronehr_{ti} + 3\hronehr_{bc,ad} \nonumber \\
        & + 12\E_{bc}\inthrhr{_{t[a,d]}} + 6\B_{bc}{}^{i}\epsilon_{adi}\inthrhr{_{tt}}\bigg)\bigg] \nonumber \\
        & + \order{r^4}. \label{eq:HRgaugevectora}
\end{align}%
\end{subequations}
\endgroup

The order-\(r^0\) and -\(r\) terms match those found previously in Eqs.~(129)--(131) and~(133) of Paper I.

\subsection{Solving for $\xi^\mu_2$}\label{sec:gauge_matching2}

We can solve Eq.~\eqref{eq:hR20} for $\xi^\mu_2$ exactly as we solved Eq.~\eqref{eq:hR10} for $\xi^\mu_1$. The only change is that the STF tensors $\hat A^{(p,l)}_{L}$ through $\hat K^{(p,l)}_L$ now refer to terms in the irreducible STF decomposition of the quantity
$ {}^0 h^{\text{R2}}_{\mu\nu}+{}^1 h^{\text{R1}}_{\mu\nu} - \frac{1}{2}\lie^2_{\xi_1}\gzero_{\mu\nu} - \lie_{\xi_1}\mathring{h}^{\text{R1}'}_{\mu\nu}$. A smaller change is that we cannot eliminate any of these coefficients using the linearized vacuum Einstein equation.

In addition to determining $\xi^\mu_2$, Eq.~\eqref{eq:hR20} determines the second-order term in the acceleration, $f^\mu_2$. The calculation and its outcome were given in Sec. VIB4 of Paper I. The total acceleration $a^\mu=\e f_1^\mu+\e^2 f^\mu_2 +\order{\e^3}$ is given by Eq.~\eqref{eq:2nd_eom} with 
\beq
h^{\R}_{\mu\nu} = \e\ {}^0 h^{\R1}_{\mu\nu} + \e^2\left({}^0 h^{\R2}_{\mu\nu} + {}^1 h^{\R1}_{\mu\nu}\right) +\order{\e^3}.
\eeq

Again following Paper I, we do not present explicit results for $\xi^\mu_{2}$. The reason is that we can simply leave the regular field to be implicitly defined from the full and singular fields: $h^R_{\mu\nu} = h_{\mu\nu}-h^S_{\mu\nu}$. As input for a numerical scheme, all that is required is the singular field.

\subsection{Second-order singular field}\label{sec:hS2inHR}

With the gauge vector determined through order $r^3$, we can now take the Lie derivative of \(\mathring{h}^{\S1'}_{\mu\nu}\) as required to determine \(\hrtext{S2}_{\mu\nu}\).
Following Paper I, to more explicitly reveal the structure of the singular field, we perform an SS/SR split as in Eq.~\eqref{eq:hS2PrimeSplit} so that
\beq
\mathring{h}^{\S2}_{\mu\nu} = \mathring{h}^{\S\S}_{\mu\nu} + \mathring{h}^{\S\R}_{\mu\nu}\label{eq:hS2 HR}
\eeq
with
\begin{align}
    \mathring{h}^{\mathrm{SR}}_{\mu\nu} ={}& \mathring{h}^{\S\R'}_{\mu\nu} + \lie_{\xi_{1}} \mathring{h}^{\S1'}_{\mu\nu}, \label{eq:hSR_HR} \\
    \mathring{h}^{\mathrm{SS}}_{\mu\nu} ={}& \mathring{h}^{\S\S'}_{\mu\nu}. \label{eq:hSS_HR}
\end{align}
This means that \(\mathring{h}^{\mathrm{SR}}_{\mu\nu}\) comprises terms \(\sim m\,\hronehr_{\mu\nu}\), and  \(\mathring{h}^{\mathrm{SS}}_{\mu\nu}\) features all terms \(\propto m^2\).

Calculating $\lie_{\xi_{1}}\mathring{h}^{\S1'}_{\mu\nu}$ and combining it with \(\mathring{h}^{\text{SR}'}_{\mu\nu}\) in Eqs.~\eqref{eq:hSRPrime}, we find the first three orders of \(\mathring{h}^{\mathrm{SR}}_{\mu\nu}\) are
\begin{widetext}
\begin{subequations}
\label{eq:hrgaugeSR}
\begin{align}
    \mathring{h}^{\mathrm{SR}}_{tt} ={}& -2m\biggl[\frac{1}{r}\bigg({}^{0}{}h^{\mathrm{R1}}_{tt} + \frac{1}{2} \,^{0}{}h^{\mathrm{R1}}_{ab} n^{ab}\bigg) + \biggl(\frac{1}{4} \, ^{0}{}h^{\mathrm{R1}}_{ab,c} n^{abc} - n^{ab} \dv{t}\, ^{0}{}h^{\mathrm{R1}}_{ab} + 2 n^{a} \dv{t}\, ^{0}{}h^{\mathrm{R1}}_{ta}\biggr) + r\biggl(\frac{11}{6}\E^{ab}\hronehr_{ab} \nonumber \\
        & + n^{ab}\biggl\{-\frac{11}{3}\E_{a}{}^{c}\hronehr_{bc}+\frac{11}{6}\E_{ab}\delta^{ij}\hronehr_{ij} + \frac{11}{3}\B_{a}{}^{c}\e_{bc}{}^{d}\hronehr_{td} + \frac{11}{6}\E_{ab}\hronehr_{tt} + \frac{11}{12}\hronehr_{ab,c}{}^{c} - \frac{11}{6}\hronehr_{ac,b}{}^{c} \nonumber \\
        & + \frac{11}{12}\delta^{ij}\hronehr_{ij,ab} + \dv{t}\hronehr_{ta,b} - \frac{1}{2}\dv[2]{t}\hronehr_{ab}\biggr\} - \frac{1}{2}n^{abc}\dv{t}\hronehr_{ab,c} + \frac{1}{12}n^{abcd}\biggl\{11\E_{ab}\hronehr_{cd} + \hronehr_{ab,cd}\biggr\}\biggr)\biggr] + \order{r^2}, \label{eq:hrgaugeSRtt} \displaybreak[0] \\
    \mathring{h}^{\mathrm{SR}}_{ta} ={}& -2m\biggl[\frac{1}{r}\biggl({}^{0}{}h^{\mathrm{R1}}_{ta} + \frac{1}{2} \,^{0}{}h^{\mathrm{R1}}_{tt} n_{a} -  \,^{0}{}h^{\mathrm{R1}}_{ab} n^{b} + {}^{0}{}h^{\mathrm{R1}}_{bc} n_{a}{}^{bc}\biggr) + \bigg(n^b\bigg\{{}^{0}{}h^{\mathrm{R1}}_{t[a,b]} -  \frac{1}{2} \dv{t}\,^{0}{}h^{\mathrm{R1}}_{ab}\bigg\} + n_a{}^{b}\dv{t}\,^{0}{}h^{\mathrm{R1}}_{tb} \nonumber \\
        & - \frac{1}{4}n^{bc}\bigg\{2 \,^{0}{}h^{\mathrm{R1}}_{ab,c} + {}^{0}{}h^{\mathrm{R1}}_{bc,a}\bigg\} - \frac{1}{2}n_a{}^{bc}\dv{t}\,^{0}{}h^{\mathrm{R1}}_{bc}\biggr) + r\biggl(\frac{1}{3}\B^{bc}\e_{ac}{}^{d}\hronehr_{bd} + \frac{4}{3}n_{a}\E^{bc}\hronehr_{bc} + \frac{1}{6}n^{b}\biggl\{-4\E^{c}{}_{(a}\hronehr_{b)c} \nonumber \\
        & + 2\E_{ab}\delta^{ij}\hronehr_{ij} + 4\B^{c}{}_{(a}\e_{b)c}{}^{d}\hronehr_{td} + \E_{ab}\hronehr_{tt} + 2\hronehr_{a[b,c]}{}^{c} - 2\hronehr_{[b}{}^{c}{}_{,c]a}\biggr\} - \frac{1}{3}n_{a}{}^{b}\B^{cd}\e_{bd}{}^{i}\hronehr_{ci} \nonumber \\
        & + \frac{1}{12}n^{bc}\biggl\{-8\B_{b}{}^{d}\e_{d}{}^{i}{}_{[a}\hronehr_{c]i} + 10\E_{bc}\hronehr_{ta} + \B^{di}\e_{aci}\hronehr_{bd} + 6\hronehr_{ta,bc} - 6\dv{t}\hronehr_{ab,c} + 3\dv{t}\hronehr_{bc,a}\biggr\} \nonumber \\
        & + n_{a}{}^{bc}\biggl\{2\E_{b}{}^{d}\hronehr_{cd} + \E_{bc}\delta^{ij}\hronehr_{ij} + 2\B_{b}{}^{d}\e_{cd}{}^{i}\hronehr_{ti} + \frac{1}{2}\E_{bc}\hronehr_{tt} + \frac{1}{2}\hronehr_{bc,d}{}^{d} - \hronehr_{bd,c}{}^{d} + \frac{1}{2}\delta^{ij}\hronehr_{ij,bc} \nonumber \\
        & + \frac{1}{2}\dv{t}\hronehr_{tb,c} - \frac{1}{4}\dv[2]{t}\hronehr_{bc}\biggr\} - \frac{1}{6}n^{bcd}\biggl\{6\E_{bc}\hronehr_{ad} + 2\hronehr_{b(a,c)d}\biggr\} - \frac{1}{4}n_{a}{}^{bcd}\dv{t}\hronehr_{bc,d} - \frac{1}{3}n^{bcdi}\B_{b}{}^{j}\e_{aij}\hronehr_{cd} \nonumber \\
        & + n_{a}{}^{bcdi}\biggl\{\E_{bc}\hronehr_{di} + \frac{1}{6}\hronehr_{bc,di}\biggr\}\biggr)\biggr] + \order{r^2}, \label{eq:hrgaugeSRta} \displaybreak[0] \\
    \mathring{h}^{\mathrm{SR}}_{ab} ={}& -2m\biggl[\frac{1}{r}\biggl(2\,^{0}{}h^{\mathrm{R1}}_{t(a} n_{b)} -  2\,^{0}{}h^{\mathrm{R1}}_{c(a} n_{b)}{}^{c} + \frac{3}{2} \,^{0}{}h^{\mathrm{R1}}_{cd} n_{ab}{}^{cd}\biggr) + \biggl({}^{0}{}h^{\mathrm{R1}}_{t(a,|c|} n_{b)}{}^{c} + \,^{0}{}h^{\mathrm{R1}}_{tc,(a} n_{b)}{}^{c} - \,^{0}{}h^{\mathrm{R1}}_{c(a,|d|} n_{b)}{}^{cd} \nonumber \\
        & -  \frac{1}{2} \,^{0}{}h^{\mathrm{R1}}_{cd,(a} n_{b)}{}^{cd} + \frac{3}{4} \,^{0}{}h^{\mathrm{R1}}_{cd,i} n_{ab}{}^{cdi} - n^{c}{}_{(a} \dv{t}\,^{0}{}h^{\mathrm{R1}}_{b)c}\biggr) + r\biggl(\frac{2}{3}n_{(a}\e_{b)d}{}^{i}\B^{cd}\hronehr_{ci} + \frac{2}{3}n^{c}\E_{c(a}\hronehr_{b)t} + \frac{5}{6}n_{ab}\E^{cd}\hronehr_{cd} \nonumber \\
        & + \frac{2}{3}\sym_{ab}n_{a}{}^{c}\biggl\{\E_{bc}\delta^{ij}\hronehr_{ij} - 2\E_{(b}{}^{d}\hronehr_{c)d} + 2\B_{(b}{}^{d}\e_{c)d}{}^{i}\hronehr_{ti} + \hronehr_{b[c,d]}{}^{d} - \hronehr_{[c}{}^{d}{}_{,d]b}\biggr\} - \frac{2}{3}n^{cd}\biggl\{\E_{c(a}\hronehr_{b)d} \nonumber \displaybreak[0] \\
        & + 2\B_{c}{}^{i}\e_{di(a}\hronehr_{b)t}\biggr\} - \frac{2}{3}n_{ab}{}^{c}\B^{di}\e_{ci}{}^{j}\hronehr_{dj} + \frac{1}{6}\sym_{ab}n_{a}{}^{cd}\biggl\{8\B_{c}{}^{i}\e_{i}{}^{j}{}_{[b}\hronehr_{d]j} + 4\B^{ij}\e_{bdj}\hronehr_{cj} - \B_{c}{}^{i}\e_{bdi}\delta^{jk}\hronehr_{jk} \nonumber\\
        & + 4\B_{c}{}^{i}\e_{bdi}\hronehr_{tt} + 6\hronehr_{tb,cd} - 6\dv{t}\hronehr_{bc,d} + 3\dv{t}\hronehr_{cd,b}\biggr\} + \frac{4}{3}n^{cdi}\B_{c}{}^{j}\e_{ij(a}\hronehr_{b)d} + \frac{1}{12}n_{ab}{}^{cd}\biggl\{2\E_{cd}\delta^{ij}\hronehr_{ij} \nonumber \\
        & - 4\E_{c}{}^{i}\hronehr_{di} + 4\B_{c}{}^{i}\e_{di}{}^{j}\hronehr_{tj} + \hronehr_{cd,i}{}^{i} - 2\hronehr_{ci,d}{}^{i} + \delta^{ij}\hronehr_{ij,cd}\biggr\} - \frac{2}{3}\sym_{ab}n_{a}{}^{cdi}\biggl\{\E_{c[d}\hronehr_{b]i} + \hronehr_{c(b,d)i}\biggr\} \nonumber \\
        & - \frac{4}{3}n^{cdij}{}_{(a}\e_{b)jk}\B_{c}{}^{k}\hronehr_{di} + \frac{1}{4}n_{ab}{}^{cdij}\biggl\{\E_{cd}\hronehr_{ij} + \hronehr_{cd,ij}\biggr\}\biggr)\biggr] + \order{r^2}. \label{eq:hrgaugeSRab}
\end{align}%
\end{subequations}
\end{widetext}
Here the first two orders, $\sim 1/r$ and $\sim r^0$, arise purely from $\lie_{\xi_{1}} \mathring{h}^{\S1'}_{\mu\nu}$, while the linear-in-$r$ terms contain contributions from both $\lie_{\xi_{1}} \mathring{h}^{\S1'}_{\mu\nu}$ and $\mathring{h}^{\S\R'}_{\mu\nu}$.

The ``singular times singular" piece of the perturbation is given in Eq.~\eqref{eq:hSSPrime}, which we rewrite here as
\begin{subequations}
\label{eq:hrgaugeSS}
\begin{align}
    \mathring{h}^{\mathrm{SS}}_{tt} ={}& -4m^2\big[\mathcal{E}_{ab} n^{ab} + r\big(\tfrac{1}{3} \dot{\mathcal{E}}_{ab} n^{ab} \big\{11 - 6 \log(\tfrac{2 m}{r})\big\} \nonumber \\
        & + \tfrac{2}{3} \mathcal{E}_{abc} n^{abc}\big)\big] + \order{r^2}, \label{eq:hrgaugeSStt} \displaybreak[0] \\
    \mathring{h}^{\mathrm{SS}}_{ta} ={}& - 4m^2\big[\mathcal{E}_{bc} n_{a}{}^{bc} + r\big(\tfrac{2}{9} \dot{\mathcal{E}}_{ab} n^{b} \big\{7 - 3 \log(\tfrac{2 m}{r})\big\} \nonumber \\
        & + \tfrac{1}{6} \mathcal{E}_{abc} n^{bc} - \tfrac{2}{9} \dot{\mathcal{B}}_{b}{}^{d} \epsilon_{acd} n^{bc} \big\{4 - 3 \log(\tfrac{2 m}{r})\big\} \nonumber \\
        & + \tfrac{1}{9} \dot{\mathcal{E}}_{bc} n_{a}{}^{bc} \big\{19 - 12 \log(\tfrac{2 m}{r})\big\} + \tfrac{1}{2} \mathcal{E}_{bcd} n_{a}{}^{bcd} \nonumber \\
        & - \tfrac{2}{9} \mathcal{B}_{bc}{}^{i} \epsilon_{adi} n^{bcd}\big)\big] + \order{r^2}, \label{eq:hrgaugeSSta} \displaybreak[0] \\
    \mathring{h}^{\mathrm{SS}}_{ab} ={}& - 4m^2\bigl[-\tfrac{1}{3} \mathcal{E}_{ab} + \mathcal{B}_{(a}{}^{d} \epsilon_{b)cd} n^{c} + \tfrac{2}{3} \mathcal{E}_{c(a} n_{b)}{}^{c} \nonumber \\
        & -  \tfrac{1}{6} \mathcal{E}_{cd} \delta_{ab} n^{cd} -  \mathcal{B}_{c}{}^{i} \epsilon_{di(a} n_{b)}{}^{cd} + \tfrac{5}{6} \mathcal{E}_{cd} n_{ab}{}^{cd} \nonumber \\
        & + r\big(\tfrac{2}{3} \dot{\mathcal{E}}_{ab} + \tfrac{4}{9} \dot{\mathcal{E}}_{c(a} n_{b)}{}^{c} \big\{4 - 3 \log(\tfrac{2 m}{r})\big\} \nonumber \\
        & + \tfrac{1}{3} \dot{\mathcal{E}}_{cd} \delta_{ab} n^{cd} +\tfrac{1}{9}n^{cd}{}_{(a}\big\{3 \mathcal{E}_{b)cd}  - 4 \dot{\mathcal{B}}_{|c|}{}^{i} \epsilon_{b)di} \nonumber \\
        & \times \big[4 - 3 \log(\tfrac{2 m}{r})\big]\big\} - \tfrac{4}{9} \mathcal{B}_{cd}{}^{j} \epsilon_{ij(a} n_{b)}{}^{cdi} \nonumber \\
        & + \tfrac{2}{9} \dot{\mathcal{E}}_{cd} n_{ab}{}^{cd} \big\{4 - 3 \log(\tfrac{2 m}{r})\big\} + \tfrac{1}{3} \mathcal{E}_{cdi} n_{ab}{}^{cdi} \big)\bigr] \nonumber \\
        & + \order{r^2}. \label{eq:hrgaugeSSab}
\end{align}%
\end{subequations}
In $\mathring{h}^{\S\S}_{\mu\nu}$ we have simply rewritten $\mathring{h}^{\S\S'}_{\mu\nu}$, as given in Eq.~\eqref{eq:hSSPrime}, in terms of \(n^L:=n^{i_1}\cdots n^{i_l}\) instead of \(\nhat^L=n^{\langle L\rangle}\). This will simplify the conversion to fully covariant form, as required for use in a puncture scheme; such a conversion can be done following the method in Ref.~\cite{Pound2014}.

$\mathring{h}^{\S\S}_{\mu\nu}$ and the leading, $1/r$ terms in $\mathring{h}^{\S\R}_{\mu\nu}$ were given previously in Paper I.\footnote{The \(\hronehr_{ab}\) term in Eq.~(134c) of Paper I has a typo and has been corrected in Eq.~\eqref{eq:hrgaugeSRab} to \(\hronehr_{cd} n_{ab}{}^{cd}\).} The $\sim r^0$ and linear-in-$r$ terms in $\mathring{h}^{\S\R}_{\mu\nu}$ appear here for the first time. We also provide our full results for the singular field in a user-ready Mathematica form in the supplementary material~\cite{SuppMat}.

This completes our calculation of the second-order singular field. In the next section, we turn to the skeleton stress-energy that this field is associated with.

\section{The Detweiler stress-energy: derivation and properties in highly regular gauges}\label{sec:o2stress}

To find the form of \(T_2^{\mu\nu}\) in a practical highly regular gauge, we first find it in the rest gauge. We then find its transformation to the generic highly regular gauge and derive some of its useful properties. However, before doing so, we discuss how the stress-energy is defined in self-force theory. 

\subsection{Stress-energy tensors in self-force theory}\label{sec:stress-energy in SF} 

We begin by reiterating our comment in the introduction: in self-force theory founded on matched asymptotic expansions, we cannot freely prescribe a stress-energy based on some desired physical characteristics. Instead, we can only prescribe the values of the multipole moments in the local metric perturbation. The stress-energy tensor, when it is well defined at all, is defined by the Einstein curvature tensor of the local perturbations.

The construction is more easily explained if we revert to an ordinary Taylor series $h_{\mu\nu}(x,\e) = \e h^1_{\mu\nu}(x) + \e^2 h^2_{\mu\nu}+\order{\e^2}$ rather than the self-consistent expansion. The stress-energy would then also have the form of a Taylor series,
\beq\label{eq:T Taylor}
    T^{\mu\nu}(x,\e) = \e T^{\mu\nu}_{1}(x) + \e^2 T^{\mu\nu}_{2}(x) +\order{\e^3},
\eeq
where
\begin{align}
    8\pi T^{\mu\nu}_1 &:= \delta G^{\mu\nu}[h^1],\label{eq:T1 Taylor}\\
    8\pi T^{\mu\nu}_2 &:= \delta G^{\mu\nu}[h^2] + \delta^2 G^{\mu\nu}[h^1,h^1].\label{eq:T2 Taylor}
\end{align}
To obtain such a series from our results in the previous section, we could expand the worldline $\gamma$ around a background geodesic, following Ref.~\cite{Pound2015}; this would introduce a mass dipole moment into $h^2_{\mu\nu}$, which would contribute to $T^{\mu\nu}_2$.

Even in that simpler approach, we should note that Eq.~\eqref{eq:T Taylor} need not be an actual Taylor expansion of an extended stress-energy distribution describing an extended material body; the quantities $T^{\mu\nu}_n$ are the same for a black hole with some multipole structure as for a material body with the same multipole structure, even though the finite-sized black hole does not have a well-defined stress-energy.  $T^{\mu\nu}$ in Eq.~\eqref{eq:T Taylor} should instead be thought of as an {\em effective} stress-energy that encodes the object's multipole structure. The field equations translate this encoded information in both directions, from the metric perturbations to $T^{\mu\nu}$ and from $T^{\mu\nu}$ to the metric perturbations.

In the self-consistent scheme, additional, more tangible subtleties enter. First, we have not actually calculated the perturbations $h^n_{\mu\nu}$ that appear in the self-consistent expansion; these are defined as the coefficients in an expansion at fixed $(z^\mu,u^\mu)$ in an external, $\e$-independent coordinate system. We have instead calculated the fields $\mathring{h}^n_{\mu\nu}$, which are the coefficients in such an expansion in an $\e$-dependent coordinate system. Second, even if we had obtained $h^n_{\mu\nu}$ rather than $\mathring{h}^n_{\mu\nu}$, we would not define \(T^{\mu\nu }_1\) using Eq.~\eqref{eq:T1 Taylor}. If we did,  we would find that $T^{\mu\nu}_1$ is {\em not} the stress-energy of a point mass; to see this, note that if $\delta G^{\mu\nu}[h^1]$ were equal to a point-mass stress-energy, then $\nabla_\nu T^{\mu\nu}_1=\nabla_\nu\delta G^{\mu\nu}[h^1]=0$ would imply that \(\gamma\) is a geodesic of the background \(g_{\mu\nu}\). Since our point mass is accelerated, we instead have
\begin{equation}
    \e\deltaG^{\mu\nu}[h^1] = 8\pi\e T_{1}^{\mu\nu} + \order{\epsilon^2},\label{eq:dGscT}
\end{equation}
where $T^{\mu\nu}_1$ is the stress-energy of a point mass on the accelerated curve $\gamma$, and the $\order{\e^2}$ term is a spatially noncompact source proportional to the acceleration.\footnote{The fact that the extra term is noncompact rather than confined to $\gamma$ can be easily confirmed with an explicit calculation in the Lorenz gauge, where we know $h^1_{\mu\nu}$ and not just ${}^0 h^1_{\mu\nu}$.}

Given these subtleties, there are at least two paths we can follow (in Sec.~\ref{sec:T_lorenz}, we discuss a third path that one of us followed in previous papers). One option is to work with total quantities, defining
\begin{multline}
    8\pi T^{\mu\nu} \coloneqq \deltaG^{\mu\nu}[\e h^1 + \e^2 h^2] + \e^2 \delta^2 G^{\mu\nu}[h^1,h^1]\\+ \order{\e^3}\label{eq:EFEsc}
\end{multline}
or, expanding around $\gzero_{\mu\nu}$,
\begin{multline}
    8\pi T^{\mu\nu} = \mathring{\deltaG}{}^{\mu\nu}[\e\, \mathring{h}^1 + \e^2 \mathring{h}^2] + \e^2\, \mathring{\delta^2 G}{}^{\mu\nu}[\mathring{h}^1,\mathring{h}^1]\\
    + \order{\e^3}, \label{eq:EFEsc expanded}
\end{multline}
where, recall, $\mathring{\deltaG}{}^{\mu\nu}$ and $\mathring{\delta^2G}{}^{\mu\nu}$ are the linearized and quadratic Einstein tensors defined on the background $\gzero_{\mu\nu}$. Alternatively, we can use $\gzero_{\mu\nu}$ and $\mathring{h}^n_{\mu\nu}$ to define $n$th-order stress-energies:
\begin{align}
    8\pi T^{\mu\nu}_1 &:= \mathring{\deltaG}{}^{\mu\nu}[\mathring{h}^1],\label{eq:T1 SC} \\
    8\pi T^{\mu\nu}_2 &:= \mathring{\deltaG}{}^{\mu\nu}[\mathring{h}^2] + \mathring{\delta^2G}{}^{\mu\nu}[\mathring{h}^1,\mathring{h}^1].\label{eq:T2 SC}
\end{align}
We will use both the form~\eqref{eq:T1 SC}--\eqref{eq:T2 SC} and the summed forms~\eqref{eq:EFEsc}--\eqref{eq:EFEsc expanded}.

With the definition~\eqref{eq:T1 SC}, $T^{\mu\nu}_1$ is precisely invariant under the transformation
\beq
\mathring{h}^1_{\mu\nu} \to \mathring{h}^1_{\mu\nu} + \lie_{\xi}\gzero_{\mu\nu}
\eeq
because $\mathring{\deltaG}{}^{\mu\nu}$ is invariant under that transformation. We can therefore calculate $T^{\mu\nu}_1$ in any convenient gauge; the result will be that $T^{\mu\nu}_1$ is the stress-energy of a point mass on $\gamma$, given in Eq.~\eqref{eq:stress1}. In Sec.~\ref{sec:T_lorenz} we review that calculation, and its extension to second order, in the Lorenz gauge.

In contrast, the quantity $T^{\mu\nu}_2$ is gauge dependent. As discussed in the introduction, in a generic gauge compatible with the assumptions of matched asymptotic expansions, it is not obvious whether $T^{\mu\nu}_2$ is well defined because $\mathring{\delta^2G}{}^{\mu\nu}[\mathring{h}^1,\mathring{h}^1]$ is a product of distributions.\footnote{Note that a quantity of the form $\partial^2 h^1\sim 1/r^3$, which appears in $\mathring{\delta^2G}{}^{\mu\nu}[\mathring{h}^1,\mathring{h}^1]$, is well defined as a distribution but not as a locally integrable function.} By construction, the total quantity $\mathring{\deltaG}{}^{\mu\nu}[\mathring{h}^2] + \mathring{\delta^2G}{}^{\mu\nu}[\mathring{h}^1,\mathring{h}^1]$ vanishes at all points $r>0$, which might suggest that we can promote it to a distribution on $r\geq0$ even if we cannot promote $\mathring{\delta^2G}{}^{\mu\nu}[\mathring{h}^1,\mathring{h}^1]$ on its own. But (i) there is no unique choice of this promotion, and (ii) even if we choose some way to promote the total quantity, there's no pragmatic purpose to doing so unless we can write a well-defined field equation for $\mathring{h}^2_{\mu\nu}$ (or $h^2_{\mu\nu}$). If all three quantities in Eq.~\eqref{eq:T2 SC} are individually well defined as distributions on $r\geq0$, then we can rearrange it to write
\beq
\mathring{\deltaG}{}^{\mu\nu}[\mathring{h}^2] = 8\pi T^{\mu\nu}_2-\mathring{\delta^2G}{}^{\mu\nu}[\mathring{h}^1,\mathring{h}^1].
\eeq
This (or an analogue of it for $h^2_{\mu\nu}$) would allow us to solve for the physical field $h^2_{\mu\nu}$, just as one can solve for the physical first-order field. But if we have only defined the total quantity $\mathring{\deltaG}{}^{\mu\nu}[\mathring{h}^2] + \mathring{\delta^2G}{}^{\mu\nu}[\mathring{h}^1,\mathring{h}^1]$, then we have not given $\mathring{\delta^2G}{}^{\mu\nu}[\mathring{h}^1,\mathring{h}^1]$ by itself a distributional definition, meaning we do not have a meaningful equation for $h^2_{\mu\nu}$ on the domain $r\geq0$. 

We can glean more about the nature of this problem by splitting $\mathring{h}^1_{\mu\nu}$ into singular and regular fields, such that
\begin{multline}
   \mathring{\delta^2G}{}^{\mu\nu}[\mathring{h}^1,\mathring{h}^1] = \mathring{\delta^2G}{}^{\mu\nu}[\mathring{h}^{\S1},\mathring{h}^{\S1}] + 2\mathring{\delta^2G}{}^{\mu\nu}[\mathring{h}^{\R1},\mathring{h}^{\S1}] \\
    + \mathring{\delta^2G}{}^{\mu\nu}[\mathring{h}^{\R1},\mathring{h}^{\R1}]. \label{eq:d2G_HR}
\end{multline}
The final term, \(\mathring{\delta^2G}{}^{\mu\nu}[\mathring{h}^{\R1},\mathring{h}^{\R1}]\), is a smooth field defined at all points in the spacetime and, as such, is well-defined distributionally. The second term behaves like $\mathring{\delta^2G}{}^{\mu\nu}[\mathring{h}^{\R1},\mathring{h}^{\S1}]\sim \mathring{h}^{\R1}\partial^2\mathring{h}^{\S1}\sim \mathring{h}^{\R1}/r^3$. This is not locally integrable, but because $\mathring{h}^{\R1}_{\mu\nu}$ is smooth, $\mathring{\delta^2G}{}^{\mu\nu}[\mathring{h}^{\R1},h]$ (for any $h_{\mu\nu}$) is a smooth linear operator acting on $h_{\mu\nu}$, which we write as \begin{equation}
    \mathring{Q}_{\R}^{\mu\nu}[h] \coloneqq \mathring{\delta^2G}{}^{\mu\nu}[\mathring{h}^{\R1},h] \label{eq:Qdef}
\end{equation}
[a special case of Eq.~\eqref{eq:Qdef general}].
This is then well defined in the distributional sense when acting on the integrable function \(\mathring{h}^{\S1}_{\mu\nu}\).

Therefore the problems arise entirely from the ``singular times singular" piece of the second-order Einstein tensor, $\mathring{\delta^2G}{}^{\mu\nu}[\mathring{h}^{\S1},\mathring{h}^{\S1}]$. In the next two sections, following an argument in Paper I, we show that in a highly regular gauge, this generically problematic quantity is an integrable function on $r\geq0$. We then use that fact to derive the second-order stress-energy tensor in these gauges. 

\subsection{Stress-energy in the lightcone rest gauge}

Before specializing to the rest gauge, we first consider the distributional nature of the individual terms in the class of highly regular gauges.
While it is not immediately obvious that $\mathring{\delta^2G}{}^{\mu\nu}[\mathring{h}^{\S1},\mathring{h}^{\S1}]$ is well defined as a distribution in these gauges, we note that because \(\mathring{h}^{\R1}_{\mu\nu}\) features no terms with explicit factors of \(m\), and \(\mathring{h}^{\S1}_{\mu\nu}\) only features terms with an explicit factor of \(m\),  \(\mathring{\delta^2G}{}^{\mu\nu}[\mathring{h}^{\S1}]\) must be the source for \(\mathring{h}^{\S\S}_{\mu\nu}\) in \(\mathring{h}^{2}_{\mu\nu}\) as it features all the terms with the factor \(m^2\).
This implies that
\begin{equation}
    \mathring{\delta^2G}{}^{\mu\nu}[\mathring{h}^{\S1}] = -\mathring{\deltaG}{}^{\mu\nu}[\mathring{h}^{\S\S}] \qc{r > 0}. \label{eq:d2GS1_gen}
\end{equation}
The previous relation is of course true in any gauge as we are free to choose the split of \(h^2_{\mu\nu}\) so that is satisfies this equality.
However, in the class of highly regular gauges, the right-hand side of Eq.~\eqref{eq:d2GS1_gen} behaves as \(\sim 1/r^2\) because \(\mathring{h}^{\S\S}_{\mu\nu} \sim r^0\).
As such, it is a locally integrable function across the entire space $r\geq0$, so we can write
\begin{equation}
    \mathring{\delta^2G}{}^{\mu\nu}[\mathring{h}^{\S1}] = -\mathring{\deltaG}{}^{\mu\nu}[\mathring{h}^{\S\S}] \qc{\forall r}. \label{eq:d2GS1_HR}
\end{equation}

We can now specialize to the rest gauge and evaluate the definition of $T^{\mu\nu}_2$ in Eq.~\eqref{eq:T2 SC}. Because the regular field is a vacuum solution, only terms involving the singular parts of the perturbations contribute to the stress-energy tensor. Hence, Eq.~\eqref{eq:T2 SC} can be simplified to
\begin{align}
    8\pi T^{\mu\nu}_{2'} &= \mathring{\deltaG}{}^{\mu\nu}[\mathring{h}^{\S\R'}] + \mathring{\deltaG}{}^{\mu\nu}[\mathring{h}^{\S\S'}]  \nonumber \\
        &\quad + 2\mathring{\delta^2G}{}^{\mu\nu}[\mathring{h}^{\S1'},\mathring{h}^{\R1'}] \nonumber \\
        &\quad + \mathring{\delta^2G}{}^{\mu\nu}[\mathring {h}^{\S1'},\mathring{h}^{\S1'}]. \label{eq:Tprime_source}
\end{align}
In the rest gauge, the quantities on the right sum to zero for \(r>0\) and are all ordinary integrable functions. Therefore their sum vanishes when integrated against a test function, and we can write
\begin{equation}
    T^{\mu\nu}_{2'} = 0. \label{eq:T2rest}
\end{equation}

This means, in physical terms, that in a nonspinning object's local rest gauge, its stress-energy is effectively that of a point mass in the external background (up to possible corrections of order $\e^3$).

\subsection{Stress-energy in a generic highly regular gauge}\label{sec:T_HR}

We now find $T^{\mu\nu}_2$ by finding how the stress-energy transforms under a gauge transformation from the rest gauge. We use Eqs.~\eqref{eq:hR1transform}--\eqref{eq:hS2transform} to write the \(\mathring{h}^n_{\mu\nu}\)'s in terms of the rest gauge quantities. Additionally, we require the identities (C2)--(C4) from Ref.~\cite{Pound2015}, which are
\begin{align}
    \lie_{\xi} A[g] ={}& \delta A[\lie_{\xi}g], \label{eq:gmc2} \\
    \lie^2_\xi A[g] ={}& \delta A[\lie^2_\xi g] + 2\delta^2 A[\lie_{\xi}g,\lie_{\xi}g], \label{eq:gmc3} \\
    \lie_{\xi}\delta A[h] ={}& \delta A[\lie_{\xi}h] + 2\delta^2 A[\lie_{\xi}g,h], \label{eq:gmc4}
\end{align}
where \(A\) is a tensor of arbitrary rank which is constructed from a metric \(g\). The first of these reduces to the invariance of the linearized Einstein tensor, $\delta G^{\mu\nu}[\lie_\xi g]=0$, when the background is vacuum.

Together, the above replacements and identities gives
\begin{align}
8\pi T_2^{\mu\nu} &=
        \mathring{\deltaG}{}^{\mu\nu}[\mathring{h}^{2'} + \lie_{\xi_{1}}\mathring{h}^{1'} + \tfrac{1}{2}\lie^2_{\xi_{1}}\gzero + \lie_{\xi_{2}}\gzero] \nonumber \\
        &\quad + \mathring{\delta^2G}{}^{\mu\nu}[\mathring{h}^{1'} + \lie_{\xi_{1}}g, \mathring{h}^{1'} + \lie_{\xi_{1}}\gzero] \nonumber \\
    &= \mathring{\deltaG}{}^{\mu\nu}[\mathring{h}^{2'}] + \mathring{\delta^2G}{}^{\mu\nu}[\mathring{h}^{1'},\mathring{h}^{1'}] \nonumber \\
        &\quad +  \mathring{\deltaG}{}^{\mu\nu}[\lie_{\xi_{1}}\mathring{h}^{1'}] + \tfrac{1}{2}\mathring{\deltaG}{}^{\mu\nu}[\lie^2_{\xi_{1}}\gzero] \nonumber \\
        &\quad + 2\mathring{\delta^2G}{}^{\mu\nu}[\mathring{h}^{1'},\lie_{\xi_{1}}\gzero] + \mathring{\delta^2G}{}^{\mu\nu}[\lie_{\xi_{1}}\gzero,\lie_{\xi_{1}}\gzero] \nonumber \\
    &= 8\pi T_{2'}^{\mu\nu} + \lie_{\xi_{1}}\mathring{\deltaG}{}^{\mu\nu}[\mathring{h}^{1'}] + \tfrac{1}{2} \lie^2_{\xi_{1}}G^{\mu\nu}[\gzero] \nonumber \\
    &= 8\pi T_{2'}^{\mu\nu} + 8\pi\lie_{\xi_{1}}T_{1}^{\mu\nu}. \label{eq:T_rest_to_HR}
\end{align}
In the first line, we have substituted Eqs.~\eqref{eq:hR1transform}--\eqref{eq:hS2transform} into the right-hand side of the definition~\eqref{eq:T2 SC}. In the third equality we have appealed to Eqs.~\eqref{eq:gmc3} and \eqref{eq:gmc4}. In the fourth, we have appealed to  $G^{\mu\nu}[\gzero]=0$ and $\mathring{\deltaG}{}[\mathring{h}^{1'}]=8\pi T^{\mu\nu}_{1'}=8\pi T^{\mu\nu}_{1}$.

Equation~\eqref{eq:T_rest_to_HR} tells us we can write
\begin{equation}
    T^{\mu\nu}_{2} = T^{\mu\nu}_{2'} + \lie_{\xi_{1}}T^{\mu\nu}_{1} \label{eq:T2_transform}.
\end{equation}
This is not too surprising as it is just the transformation law for a second-order tensor when the background tensor vanishes~\cite{Bruni1997}. In our case, we have effectively defined \(T^{\mu\nu}_{2}\) as the second-order term in an expansion of the Einstein tensor. However, note that the steps involved in Eq.~\eqref{eq:T_rest_to_HR} rely on the properties of the highly regular gauges; we have not established Eq.~\eqref{eq:T2_transform} for the transformation between any two generic gauges.

Next, since $T^{\mu\nu}_{2'}=0$, Eq.~\eqref{eq:T2_transform} becomes
\begin{equation}
    T^{\mu\nu}_{2} = \lie_{\xi_1}T^{\mu\nu}_{1}. \label{eq:T2HRlie}
\end{equation}
The right-hand side was previously calculated in Eq.~(D1) of Ref.~\cite{Pound2015} and is rederived in Eq.~\eqref{eq:LieXT1Final}.\footnote{The \(\tau\) derivative term has a missing minus sign in Ref.~\cite{Pound2015}, which has been added here.\label{footnote:LieDT1}} It reads
\begin{equation}
    \lie_{\xi_1}T^{\mu\nu}_{1} = -m\int_{\gamma} g^{\mu}_{\mu'} g^{\nu}_{\nu'} u^{\mu'} u^{\nu'} \bigg(\xi^\rho_{1;\rho} - \dv{\xi^{1}_{\parallel}}{\tau}\bigg) \delta^4(x,z) \dd{\tau}, \label{eq:LieXi1T1}
\end{equation}
where \(\xi^{1}_{\parallel} \coloneqq u_\rho \xi^\rho_1\) and we have removed the orthogonal parts of the gauge vector, \(\xi^{\mu}_{1\perp} \coloneqq P^{\mu}{}_{\nu}\xi^{\nu}_{1}\), as the worldline-preserving condition sets them to zero. We detail the derivation of Eq.~\eqref{eq:LieXi1T1} and various related results in Appendix~\ref{app:LieDT1}.

By taking \(\xi^{\rho}_1\) to be the gauge vector from Eq.~\eqref{eq:HRgaugevector} and the proper time to be \(t\), we find
\begin{equation}
    \dv{\xi^{1}_{\parallel}}{\tau}\bigg|_{\gamma} = - \dv{\xi^{t}_{1}}{t}\bigg|_{\gamma} = \frac{1}{2} \hronehr_{tt}\egamma = \frac{1}{2}u^\mu u^\nu {}^0 h^{\R1}_{\mu\nu}\egamma \label{eq:tDerivXiPar}
\end{equation}
and
\begin{align}
    \xi^{\rho}_{1;\rho}\big|_{\gamma} ={}& (\partial_{t} \xi^{t}_{1} + \partial_{a} \xi^{a}_{1})\big|_{\gamma} \nonumber\\
    ={}& \frac{1}{2}(-\hronehr_{tt} + \delta^{ab}\hronehr_{ab})\big|_{\gamma} \nonumber \\
    ={}& \frac{1}{2}g^{\alpha\beta}\hronehr_{\alpha\beta}\big|_{\gamma}. \label{eq:xiDivergence}
\end{align}
Thus, the second-order stress-energy tensor in the highly regular gauge is given by
\begin{equation}
    T^{\mu\nu}_{2} = - \frac{m}{2}\int u^\mu u^\nu \left(g^{\alpha\beta} - u^{\alpha}u^{\beta}\right)\hronehr_{\alpha\beta}  \delta^4(x,z) \dd{\tau}. \label{eq:T2HRcov}
\end{equation}

\subsection{Point mass in the effective spacetime}\label{sec:effT}

With a short calculation, we can show the total stress-energy $\e T^{\mu\nu}_1+\e^2 T^{\mu\nu}_2$ derived above is exactly equal, through order $\e^2$, to the stress-energy tensor of a point mass in the effective spacetime \(\tilde{g}_{\mu\nu} = g_{\mu\nu} + \hrtext{R}_{\mu\nu}\).
That stress-energy tensor is given by
\begin{equation}
    \tilde{T}^{\mu\nu} = \epsilon m\int_{\gamma}\tilde{u}^{\mu}\tilde{u}^{\nu} \frac{\delta^4(x-z)}{\sqrt{-\tilde{g}}} \dd{\tilde{\tau}}. \label{eq:effstressenergy}
\end{equation}
Expanding this for small $h^\R_{\mu\nu}$, we see that
\begin{align}
    \tilde{T}^{\mu\nu} ={}& \epsilon m \int_{\gamma} \dv{\tau}{\tilde{\tau}}u^{\mu}u^{\nu} \delta^4(x,z) \bigg(1 - \frac{1}{2}\epsilon g^{\alpha\beta}\hrtext{R1}_{\alpha\beta}\bigg)\dd{\tau} \nonumber \\
        & + \order{\epsilon^3} \nonumber \\
    ={}& \e m \int_\gamma u^{\mu}u^{\nu}\delta^{4}(x,z)\biggl(1 - \frac{1}{2}\e\biggl[g^{\alpha\beta} - u^{\alpha}u^{\beta}\biggr]h^{\rone}_{\alpha\beta}\biggr)\dd{\tau} \nonumber \\
        & + \order{\e^3}, \label{eq:Teff}
\end{align}
where we have used the standard expansion of a determinant and expanded \(\dv*{\tau}{\tilde{\tau}}\) using
\begin{equation}
    \dv{\tau}{\tilde{\tau}} = \frac{1}{\sqrt{1 - \hrtext{R}_{\mu\nu}u^\mu u^\nu}} = 1 + \frac{1}{2}\epsilon\hrtext{R1}_{\mu\nu}u^\mu u^\nu + \order{\epsilon^2}, \label{eq:dtdttilde}
\end{equation}
which follows from 
\beq
-1 = \tilde{g}_{\mu\nu}\tilde{u}^{\mu}\tilde{u}^{\nu} = (g_{\mu\nu} + \hrtext{R}_{\mu\nu})\left(\dv{\tau}{\tilde{\tau}}\right)^{2}u^{\mu}u^{\nu}.
\eeq

Comparing Eqs.~\eqref{eq:T2HRcov} and~\eqref{eq:Teff}, we see that
\begin{equation}
    \e T^{\mu\nu}_{1} + \e^2 T^{\mu\nu}_{2} = \tilde{T}^{\mu\nu} + \order{\e^3}. \label{eq:TeffT1T2}
\end{equation}
This confirms Detweiler's postulate in Ref.~\cite{Detweiler2012}. 

As Detweiler also noted, we can use this to write the field equations in a more transparent form. Eq.~\eqref{eq:d2GS1_HR}, together with $G_{\mu\nu}[\tilde g]=0$, implies that
\begin{align}
G^{\mu\nu}[{\sf g}] &= \e\delta G^{\mu\nu}[h^{\S1}] + \e^2\delta G^{\mu\nu}[h^{\S\R}] \nonumber\\
&\quad + 2\e^2\delta^2G^{\mu\nu}[h^{\S1},h^{\R1}]+\order{\e^3}\nonumber\\
&= \delta\tilde G^{\mu\nu}[\e h^{\S1}+\e^2 h^{\S\R}],\label{eq:Gexact=Geff}
\end{align}
where ``$G^{\mu\nu}[{\sf g}]$" is to be understood as the expansion of that quantity through order $\e^2$, and $\delta\tilde G^{\mu\nu}$ is the linearized Einstein tensor constructed from $\tilde g_{\mu\nu}$. In words, the Einstein curvature of the physical spacetime (extended to all $r>0$ from outside the body) is identical to the linearized Einstein curvature of the perturbation $\e h^{\S1}_{\mu\nu}+\e^2 h^{\S\R}_{\mu\nu}$ atop the effective background $\tilde g_{\mu\nu}$. Combining this with Eq.~\eqref{eq:TeffT1T2} allows us to write the field equations in the form of a point mass sourcing a linear perturbation of an effective background:
\beq
\delta \tilde G^{\mu\nu}[\e h^{\S1}+\e^2 h^{\S\R}] = 8\pi \tilde T^{\mu\nu} +\order{\e^3}.\label{eq:Detweiler EFE}
\eeq

In the remainder of the section, we derive several useful properties of this stress-energy. In all cases, the properties further show that the Detweiler stress-energy behaves as an ordinary stress-energy tensor in the effective metric, even as it behaves strikingly {\em unlike} an ordinary stress-energy in the physical spacetime.

\subsection{Raising and lowering indices}\label{sec:T2 raising and lowering}

Suppose our stress-energy were an ordinary tensor ${\sf T}^{\mu\nu}(x,\e)$ with the expansion ${\sf T}^{\mu\nu}=\e T^{\mu\nu}_1+\e^2 T^{\mu\nu}_2+\order{\e^3}$. Its indices would be raised and lowered with ${\sf g}_{\mu\nu}$, such that 
\begin{align}
{\sf T}_{\mu}{}^\nu &= {\sf g}_{\mu\rho}{\sf T}^{\rho\nu} \\
&= \e T^1_\mu{}^\nu +\e^2(T_\mu^{2\ \nu} +h^1_{\mu\rho}T^{\rho\nu}_1) +\order{\e^3}.
\end{align}
Clearly our stress-energy cannot behave in this way. If it did, then the second-order stress-energy with one index down would contain the term $h^1_{\mu\rho}T^{\rho\nu}_1$, which has the form $\sim \frac{\delta^3(x^a)}{r}$. This is manifestly ill defined.

Instead, we show that the stress-energy's indices are raised and lowered with the effective metric \(\tilde{g}_{\mu\nu} = g_{\mu\nu} + \hrtext{R}_{\mu\nu}\). That is, if we define
\begin{align}
    8\pi\tilde T_{\mu}{}^{\nu} &:= \delta G_{\mu}{}^{\nu}[\e h^1 + \e^2 h^2] \nonumber\\
    &\quad + \e^2 \delta^2(g_{\mu\rho} G^{\rho\nu})[h^1,h^1] + \order{\e^3},\label{eq:T mixed  def}\\
    8\pi\tilde T_{\mu\nu} &:= \delta G_{\mu\nu}[\e h^1 + \e^2 h^2] \nonumber\\
    &\quad + \e^2 \delta^2(g_{\mu\rho}g_{\nu\sigma} G^{\rho\sigma})[h^1,h^1] + \order{\e^3}\label{eq:T down def} 
\end{align}
in analogy with Eq.~\eqref{eq:EFEsc}, then
\begin{align}
    \tilde{T}_{\mu}{}^{\nu} &= \tilde{g}_{\mu\alpha}\tilde{T}^{\alpha\nu} \nonumber\\
    &= \e T^1_\mu{}^{\nu} + \e^2(T^2_\mu{}^\nu+{}^0h^{\R1}_{\mu\alpha}T_1^{\alpha\nu})+\order{\e^3}, \label{eq:TeffDU} \\
    \tilde{T}_{\mu\nu} &= \tilde{g}_{\mu\alpha}\tilde{g}_{\nu\beta}\tilde{T}^{\alpha\beta}\nonumber\\
    &= \e T^1_{\mu\nu} + \e^2(T^2_{\mu\nu}+2\,{}^0h^{\R1}_{\alpha(\mu}g_{\nu)\beta}T_1^{\alpha\beta})+\order{\e^3}. \label{eq:TeffDD}
\end{align}
The right-hand sides of Eqs.~\eqref{eq:T mixed def} and \eqref{eq:T down def} are the expansions of the Einstein tensor with mixed indices and both indices down, as given in Eqs.~\eqref{eq:G mixed}--\eqref{eq:ddG down}.

We establish these results following the same method we used to derive $T^{\mu\nu}_2$. Repeating the steps in Eq.~\eqref{eq:T_rest_to_HR}, we find the analogues of Eq.~\eqref{eq:T2HRlie},
\begin{align}
\tilde T_\mu{}^\nu &= \e T^1_\mu{}^\nu + \e^2 \lie_{\xi_1}T^1_{\mu}{}^{\nu},\\
T_{\mu\nu} &= \e T^1_\mu{}^\nu + \e^2 \lie_{\xi_1}T_{\mu\nu}^{1}.
\end{align}
The Lie derivatives are given in Eqs.~\eqref{eq:LieXi1T1DU} and \eqref{eq:LieXi1T1DD}. By substituting the values of the gauge vector from Eq.~\eqref{eq:HRgaugevector} and converting to Fermi--Walker coordinates, we see that the individual components for mixed indices are given by
\begin{subequations}
\label{eq:LieXi1T1DUComps}
\begin{align}
    \lie_{\xi_{1}}T^{1}_{t}{}^{t} ={}& \frac{m}{2}\int \delta^{ab}\hronehr_{ab}\delta^4(x,z) \dd{t}, \label{eq:LieXi1T1DUTT} \\
    \lie_{\xi_{1}}T^{1}_{t}{}^{a} ={}& 0, \label{eq:LieXi1T1DUTA} \\
    \lie_{\xi_{1}}T^{1}_{a}{}^{t} ={}& m\int\! \hronehr_{ta} \delta^4(x,z) \dd{t}, \label{eq:LieXi1T1DUAT} \\
    \lie_{\xi_{1}}T^{1}_{a}{}^{b} ={}& 0, \label{eq:LieXi1T1DUAB}
\end{align}
\end{subequations}
and for both indices down by
\begin{subequations}
\label{eq:LieXi1T1DDComps}
\begin{align}
    \lie_{\xi_{1}}T^{1}_{tt} &= - \frac{m}{2}\int\! \bigl(2\hronehr_{tt} + \delta^{ab}\hronehr_{ab}\bigr)\delta^4(x,z) \dd{t}, \label{eq:LieXi1T1DDTT} \\
    \lie_{\xi_{1}}T^{1}_{ta} &= - m\int\! \hronehr_{ta} \delta^4(x,z) \dd{t}, \label{eq:LieXi1T1DDTA} \\
    \lie_{\xi_{1}}T^{1}_{ab} &= 0. \label{eq:LieXi1T1DDAB}
\end{align}
\end{subequations}
In covariant form, these become
\begin{align}
    \lie_{\xi_{1}}T^{1}_{\mu}{}^{\nu} ={}& - \frac{m}{2}\int_{\gamma}\big[\bigl(g^{\alpha\beta} - u^{\alpha}u^{\beta}\bigr)\hronehr_{\alpha\beta} u_{\mu} \nonumber \\
        & - 2\hronehr_{\mu\alpha} u^{\alpha}\big]u^{\nu}\delta^{4}(x,z)\dd{\tau}, \label{eq:LieT1DU}\\
    \lie_{\xi_{1}}T^{1}_{\mu\nu} ={}& -\frac{m}{2}\int_{\gamma}\bigl[\bigl(g^{\alpha\beta} - u^{\alpha}u^{\beta}\bigr)\hronehr_{\alpha\beta} u_{\mu}u_{\nu} \nonumber \\
        & - 4 u^{\alpha}u_{(\mu}\hronehr_{\nu)\alpha}\bigr] \delta^{4}(x,z) \dd{\tau}. \label{eq:LieT1DD} 
\end{align}
We see by comparison with Eq.~\eqref{eq:T2HRcov} that these agree with the order-$\e^2$ terms in Eqs.~\eqref{eq:TeffDU} and \eqref{eq:TeffDD}.

\subsection{Conservation of stress-energy}

Again suppose our stress-energy were an ordinary tensor ${\sf T}^{\mu\nu}(x,\e)$ with the expansion ${\sf T}^{\mu\nu}=\e T^{\mu\nu}_1+\e^2 T^{\mu\nu}_2+\order{\e^3}$. It would then be conserved in ${\sf g}_{\mu\nu}$: ${}^{\sf g}\nabla_\nu{\sf T}^{\mu\nu}=0$, which would imply
\begin{multline}
\e\nabla_\nu T^{\mu\nu}_1 + \e^2\left(\nabla_\nu T^{\mu\nu}_2 + \delta\Gamma^\mu_{\rho\nu}T_1^{\rho\nu} + \delta\Gamma^\nu_{\rho\nu}T_1^{\mu\rho}\right)\\ = \order{\e^3}.\label{eq:conservation in gexact}
\end{multline}
Here ${}^{\sf g}\nabla_\nu$ is the covariant derivative compatible with ${\sf g}_{\mu\nu}$, and $\delta\Gamma^\mu_{\rho\nu}:=\frac{1}{2}g^{\mu\sigma}(2h^1_{\rho(\nu;\sigma)}-h^1_{\rho\nu;\sigma})$ is the linear correction to the Christoffel symbol associated with ${\sf g}_{\mu\nu}$. Clearly our stress-energy cannot satisfy Eq.~\eqref{eq:conservation in gexact}, as it involves ill-defined terms of the form $(\partial_\alpha h^1_{\beta\gamma})T^{\mu\nu}_1\sim \frac{\delta^3(x^a)}{r^2}$.

Instead, the Detweiler stress-energy is conserved in the effective spacetime, meaning
\beq
\tilde\nabla_\nu \tilde T^{\mu\nu} = 0.\label{eq:conservation in geff}
\eeq
This follows from the textbook result that a point-mass stress-energy in a metric $\tilde g_{\mu\nu}$ is conserved if and only if the mass moves on a geodesic of that metric.

This result may seem at odds with the Bianchi identity, which tells us that the left-hand side of the field equations has zero divergence in ${\sf g}_{\mu\nu}$: ${}^{\sf g}\nabla_\nu G^{\mu\nu}[{\sf g}]=0$, which implies
\begin{align}
\nabla_\nu \delta G^{\mu\nu}[h^1] &= 0,
\end{align}
and
\begin{multline}
\nabla_\nu \delta^2 G^{\mu\nu}[h^1,h^1] + \delta\Gamma^\mu_{\rho\nu}\delta G^{\rho\nu}[h^1] \\+ \delta\Gamma^\nu_{\rho\nu}\delta G^{\mu\rho}[h^1] = 0.\label{eq:Bianchi id in gexact}
\end{multline}
These identities hold for any smooth rank-2 symmetric tensor $h^1_{\mu\nu}$. In our case, they hold for all $r>0$, but Eq.~\eqref{eq:Bianchi id in gexact} is ill defined on the domain $r\geq0$ because it involves products of distributions. The equality that {\em does} hold is the expansion of $\tilde\nabla_\nu G^{\mu\nu}[{\sf g}]=\order{\e^3}$,
\begin{multline}
\nabla_\nu \delta^2 G^{\mu\nu}[h^1,h^1] + \delta\tilde{\Gamma}^\mu_{\rho\nu}\delta G^{\rho\nu}[h^1] \\+ \delta\tilde{\Gamma}^\nu_{\rho\nu}\delta G^{\mu\rho}[h^1] = 0,\label{eq:Bianchi id in geff}
\end{multline}
which can be reduced to
\begin{multline}
2\nabla_\nu \delta^2 G^{\mu\nu}[h^{\S1},h^{\R1}] + \delta\tilde{\Gamma}^\mu_{\rho\nu}\delta G^{\rho\nu}[h^{\S1}] \\+ \delta\tilde{\Gamma}^\nu_{\rho\nu}\delta G^{\mu\rho}[h^{\S1}] = 0.\label{eq:Bianchi id in geff v2}
\end{multline}
Here $\delta\tilde\Gamma^\mu_{\rho\nu}:=\frac{1}{2}g^{\mu\sigma}(2h^{\R1}_{\rho(\nu;\sigma)}-h^{\R1}_{\rho\nu;\sigma})$. We obtain the reduction~\eqref{eq:Bianchi id in geff v2} using $\tilde\nabla_\nu G^{\mu\nu}[ \tilde g]=0$ and noting that because of Eq.~\eqref{eq:d2GS1_HR}, we have  
\beq
\nabla_\nu \delta^2 G^{\mu\nu}[h^{\S1},h^{\S1}] = 0.
\eeq

To see why (the expansion of) $\tilde\nabla_\nu G^{\mu\nu}[{\sf g}]=\order{\e^3}$ holds true, recall Eq.~\eqref{eq:Gexact=Geff}: as a distribution in a neighbourhood of $\gamma$, $G^{\mu\nu}[{\sf g}]=\delta\tilde G^{\mu\nu}[\e h^{\S1}+\e^2h^{\S\R}]+O(\e^3)$ (again interpreting the left-hand side as its expansion through order $\e^2$). The Bianchi identity $\tilde\nabla_\nu\delta\tilde G^{\mu\nu}[\e h^{\S1}+\e^2 h^{\S\R}]=0$ then trivially implies $\tilde\nabla_\nu G^{\mu\nu}[{\sf g}]=\order{\e^3}$.

\subsection{Gauge invariance under smooth transformations}\label{sec:T2gaugeInvSmooth}

All of our results are valid for any member of our class of highly regular gauges. However, our derivation relied on the notion of a worldline-preserving transformation: for each highly regular gauge, we have started from an associated rest gauge in which the worldline is identical. In this section, as a consistency check, we show that under an arbitrary smooth transformation between two highly regular gauges with differing, gauge-related worldlines, the functional form of the Detweiler stress-energy is invariant.

Under such a transformation, we have~\cite{Pound2015}
\begin{align}
h^{\R1}_{\mu\nu} &\to h^{\R1}_{\mu\nu} + \lie_{\xi_1} g_{\mu\nu},\\
z^\mu &\to z^\mu -\e \xi^\mu_1 +\order{\e^2}.\label{eq:Dz}
\end{align}
Following through the calculation in Eq.~\eqref{eq:T_rest_to_HR} once again, but now accounting for the shift~\eqref{eq:Dz} in the worldline, we obtain 
$T^{\mu\nu}_{2\ddagger} = T^{\mu\nu}_{2} + \left(\lie_{\xi_{1}} + \lieZ_{\xi_{1}}\right)T^{\mu\nu}_{1}$, where we denote our new gauge with a double-ended dagger. Here $\lieZ_{\xi_{1}}$ acts on $T^{\mu\nu}_{1}$'s dependence on $z^\mu$; see Ref.~\cite{Pound2015} for a thorough description of this type of transformation. 

Equation~\eqref{eq:LieXZT1} gives the action of the Lie derivatives on $T^{\mu\nu}_1$. The gauge vector in that equation can be expressed in terms of $\Delta h^{\R 1}_{\mu\nu}$ by solving $\lie_{\xi_1} g_{\mu\nu}=\Delta h^{\R 1}_{\mu\nu}$ (no longer subject to $\xi^a_1|_\gamma=0$). The result is that the terms involving gauge vectors are again given by Eqs.~\eqref{eq:tDerivXiPar} and~\eqref{eq:xiDivergence} but with \(\hrtext{R1}_{\mu\nu}\) replaced by \(\Delta h^{\mathrm{R1}}_{\mu\nu}\). Making those substitutions, we obtain
\begin{multline}
    \left(\lie_{\xi_{1}} + \lieZ_{\xi_{1}}\right)T^{\mu\nu}_{1} = - \frac{m}{2}\int u^\mu u^\nu \left(g^{\alpha\beta} - u^{\alpha}u^{\beta}\right)\Delta\hrtext{R1}_{\alpha\beta} \\
        \times \delta^4(x,z) \dd{\tau}. \label{eq:LieXZT1HR1}
\end{multline}
This generalizes our previous result for the special case of a transformation from a rest gauge. In that case, \(\hrtext{R1'}_{\mu\nu} = 0\) on the worldline, resulting in \(\Delta h^{\mathrm{R1}}_{\mu\nu} = h^{\mathrm{R1}}_{\mu\nu}\).

The stress-energy in the new gauge is therefore
\begin{align}
    T^{\mu\nu}_{2\ddagger} &= T^{\mu\nu}_{2} + \left(\lie_{\xi_{1}} + \lieZ_{\xi_{1}}\right)T^{\mu\nu}_{1} \nonumber \\
        &= - \frac{m}{2}\int u^\mu u^\nu \left(g^{\alpha\beta} - u^{\alpha}u^{\beta}\right){h}^{\mathrm{R1}\ddagger}_{\alpha\beta}  \delta^4(x,z) \dd{\tau}, \label{eq:T2GeneralGauge}
\end{align}
which confirms that the functional form of Eq.~\eqref{eq:T2HRcov} is always valid for smoothly related highly regular gauges but with a regular field specific to the particular gauge. Note that this is also consistent with the value of zero in the rest gauge.
In the rest gauge, \(h^{\mathrm{R}}_{\mu\nu}\egamma = 0\), leading to a vanishing \(T^{\mu\nu}_{2'}\).

\section{The Detweiler stress-energy in the Lorenz gauge}\label{sec:T_lorenz}

In the last section we established the validity of the Detweiler stress-energy in the class of highly regular gauges. In this section, we investigate whether it remains valid in less regular gauges. We do not consider commonly used gauges with pathological singularities away from $\gamma$, such as the Regge-Wheeler-Zerilli~\cite{Thompson2018} or radiation gauges~\cite{Pound2013}. We focus on the Lorenz gauge, the most commonly used gauge possessing the generic level of regularity assumed in matched asymptotic expansions: $h^n_{\mu\nu}\sim m^n/r^n$ and $h^n_{\mu\nu}$ is smooth away from the particle. The Lorenz gauge has been central to many foundational derivations in self-force theory and has been used in numerous practical calculations~\cite{Barack2018}. More relevantly here, it has been the basis for the concrete development of second order numerical schemes~\cite{Pound2014,WardellWarburton2015,Pound2015LargeScale,Miller2016,Pound2020,Miller2020}. 

Unfortunately, it is not possible to perform the same treatment in the Lorenz gauge as in the highly regular gauge because the transformation from a highly regular gauge to the Lorenz gauge is singular on $\gamma$, making $\lie_\xi T^{\mu\nu}_1$ ill defined. We instead perform a direct calculation of the right-hand side of Eq.~\eqref{eq:EFEsc}. The calculation is based on a particular distributional definition of $\delta^2 G^{\mu\nu}[h^{1},h^{1}]$, which we call the Detweiler canonical definition. Using this choice, we recover the Detweiler stress-energy.

At the end of the section we discuss whether this result applies in all gauges with generic regularity.

\subsection{Field equations and local form of the metric perturbation}

In the self-consistent Lorenz-gauge scheme~\cite{Pound2010,Pound2012,Pound2012SmallBody}, the gauge condition
\begin{equation}
    \nabla_\nu\bar{h}^{\mu\nu} = 0 \label{eq:lorenz_condition}
\end{equation}
is imposed on the total perturbation $h_{\mu\nu}$, not on each coefficient $h^n_{\mu\nu}$. Here 
\begin{equation}
    \bar{h}_{\mu\nu} \coloneqq h_{\mu\nu} - \tfrac{1}{2}g_{\mu\nu}g^{\alpha\beta}h_{\alpha\beta} \label{eq:trace-reversed}
\end{equation}
is the trace-reversed perturbation. The coefficients $h^n_{\mu\nu}$ satisfy
\begin{align}
    E^{\mu\nu}[\bar{h}^{1*}] &= 0 &\text{for }x\notin\gamma,\label{eq:Eh1 off gamma}\\
    E^{\mu\nu}[\bar{h}^{2*}] &= - \delta^2 G^{\mu\nu}[h^{1*},h^{1*}]&\text{for }x\notin\gamma,\label{eq:Eh2 off gamme} 
\end{align}
where
\begin{equation}
    E^{\mu\nu}[\bar{h}] \coloneqq -\frac{1}{2}(\bar{h}^{\mu\nu;\alpha}{}_{\alpha} + 2R^{\mu}{}_{\alpha}{}^{\nu}{}_{\beta}\bar{h}^{\alpha\beta}) \label{eq:linearG_lorenz}
\end{equation}
is the linearized Einstein tensor in the Lorenz gauge, and where we use an asterisk to denote Lorenz-gauge quantities.

In this gauge we have access to the full perturbations $h^n_{\mu\nu}$ rather than just $\mathring{h}^n_{\mu\nu}$~\cite{Pound2010,Pound2012SmallBody,Pound2014}. The first-order singular field takes the form
\begin{equation}
    h^{\mathrm{S1*}}_{\mu\nu} = \frac{2m}{r}(g_{\mu\nu} + 2 u_{\mu}u_{\nu}) + \order{r^0}, \label{eq:hS1Lorenz}
\end{equation}
where \(u^{\alpha} = (1,0,0,0)\) so that \(u^{\alpha}n_{\alpha} = 0\). The second-order singular field is split into three pieces,
\begin{equation}
    h^{\mathrm{S2*}}_{\mu\nu} = h^{\mathrm{SS*}}_{\mu\nu} + h^{\mathrm{SR*}}_{\mu\nu} + h^{\delta m*}_{\mu\nu}, \label{eq:hS2Lorenz}
\end{equation}
which satisfy
\begin{align}
E^{\mu\nu}[\bar h^{\S\S*}] &= -\delta^2 G^{\mu\nu}[h^{\S1*},h^{\S1*}] \quad \text{for }x\notin\gamma,\label{eq:EhSS r>0}\\
E^{\mu\nu}[\bar h^{\S\R*}] &= -2\delta^2 G^{\mu\nu}[h^{\R1*},h^{\S1*}] \quad \text{for }x\notin\gamma,\label{eq:EhSR r>0}\\
E^{\mu\nu}[\bar h^{\delta m*}] &= 0 \quad \text{for }x\notin\gamma.\label{eq:Ehdm r>0}
\end{align}
\(h^{\mathrm{SS*}}_{\mu\nu}\) contains all local terms explicitly proportional to $m^2$. It has the form $\sim m^2/r^2$, as in a generic gauge compatible with matched expansions, but we will not need explicit expressions for it. The combined quantity \(h^{\mathrm{SR*}}_{\mu\nu}+h^{\delta m*}_{\mu\nu}\) is the analogue of what we have called $h^{\S\R}_{\mu\nu}$ in the highly regular gauge, containing products of $m$ with $h^{\R1}_{\mu\nu}$. The components of the ``singular times regular" pieces are given by
\begin{subequations}
    \label{eq:hSR_Lorenz}
    \begin{align}
    h^{\mathrm{SR*}}_{tt} &= - \frac{m}{r} \hrstar_{ab}\nhat^{ab} + \order{r^0}, \label{eq:hSRttLorenz} \\
    h^{\mathrm{SR*}}_{ta} &= - \frac{m}{r} \hrstar_{tb}\nhat_{a}{}^{b} + \order{r^0}, \label{eq:hSRtaLorenz} \\
    h^{\mathrm{SR*}}_{ab} &= \frac{m}{r} \Big[2\nhat^{c}{}_{(a}\hrstar_{b)a} - \delta_{ab}\hrstar_{cd}\nhat^{cd} \nonumber \\
        &\quad\, - \big(\hrstar_{ij}\delta^{ij} + \hrstar_{tt}\big)\nhat_{ab}\Big] + \order{r^0} \label{eq:hSRabLorenz}
    \end{align}
\end{subequations}
and 
\begin{subequations}
    \label{eq:hDM_Lorenz}
    \begin{align}
    \hdm_{tt} &= - \frac{m}{3r} \Big(\hrstar_{ab}\delta^{ab} + 6\hrstar_{tt}\Big) + \order{r^0}, \label{eq:hDMtt} \\
    \hdm_{ta} &= - \frac{4m}{3r}\hrstar_{ta}+ \order{r^0}, \label{eq:hDMta} \\
    \hdm_{ab} &= \frac{m}{3r} \Big(2\hrstar_{ab} + \delta_{ab}\delta^{cd}\hrstar_{cd} + 2\delta_{ab}\hrstar_{tt}\Big) \nonumber \\
        &\quad\, + \order{r^0}. \label{eq:hDMab}
    \end{align}
\end{subequations}
$\hdm_{\mu\nu}$ consists entirely of $\ell=0$ terms in a decomposition of the form $h^2_{\mu\nu} = \sum_{l\geq0} h^{2(l)}_{\mu\nu}(t,r)\nhat^L$.

Previous papers by one of us (e.g.,~\cite{Pound2012SmallBody,Pound2015Small}) defined effective stress-energy tensors associated with particular pieces of the metric perturbation:
\begin{align}
8\pi T^{\mu\nu}_1 &:= E^{\mu\nu}[\bar h^{1*}],\\
8\pi T^{\mu\nu}_{\delta m} &:= E^{\mu\nu}[\bar h^{\delta m*}],\label{eq:Tdm}
\end{align}
and similar for higher multipole moments. These definitions are well defined to all orders in perturbation theory, and they provide a complete characterization of the object's multipole structure. However, they do not describe the full Einstein curvature. Moreover, the total curvature is obscured by the division of ``singular times regular" pieces into $h^{\S\R*}_{\mu\nu}$ and $h^{\delta m*}_{\mu\nu}$. The clean split into the field equations~\eqref{eq:EhSR r>0} and \eqref{eq:Tdm} does not guarantee that, for example, Eq.~\eqref{eq:EhSR r>0} holds distributionally for $r\geq0$.

In this paper, motivated by the results in highly regular gauges, we deviate from the definitions in Refs.~\cite{Pound2012SmallBody,Pound2015Small} and instead use the definition of $T^{\mu\nu}$ in Eq.~\eqref{eq:EFEsc}. As in the highly regular gauge, the regular fields are defined to be solutions of the vacuum Einstein equations and their Einstein tensor does not contribute to \(T^{\mu\nu}\), leaving us with the analogue of Eq.~\eqref{eq:Tprime_source},
\begin{align}
    8\pi T^{\mu\nu} &= \deltaG^{\mu\nu}[\e h^{\S1*}+\e^2 h^{\S\S*}+\e^2 h^{\S\R*}+\e^2 h^{\delta m*}]  \nonumber \\
        &\quad + 2\e^2\delta^2 G^{\mu\nu}[h^{\S1*},h^{\R1*}]  + \e^2\delta^2 G^{\mu\nu}[h^{\S1*},h^{\S1*}]\nonumber \\
        &\quad +\order{\e^3}. \label{eq:Tstar_source}
\end{align}
To proceed from here we must choose a distributional definition of $\delta^2 G^{\mu\nu}[h^{\S1*},h^{\S1*}]$.
Using the property $\e^2\nabla_\nu \bar h^{\mu\nu}_{\S\S*}=\order{\e^3}$, we can rewrite Eq.~\eqref{eq:EhSS r>0} as
\begin{multline}
\delta G^{\mu\nu}[\bar h^{\S\S*}] = -\delta^2 G^{\mu\nu}[h^{\S1*},h^{\S1*}] +\order{\e^3} \\ \text{for }x\notin\gamma.\label{eq:dGhSS r>0}
\end{multline}
In the highly regular gauge we get ``for free" that the analogue of this equation [Eq.~\eqref{eq:d2GS1_HR}] is true for $r\geq0$. In the Lorenz gauge we only get that free cancellation off of the worldline.
However, we can {\em define} Eq.~\eqref{eq:dGhSS r>0} to be true distributionally on the region $r\geq0$:
\begin{equation}
    \delta^2 G^{\mu\nu}[h^{\S1*},h^{\S1*}] \coloneqq - \deltaG^{\mu\nu}[h^{\mathrm{SS*}}] \qc{\forall r}.\label{eq:d2GS1_Lorenz}
\end{equation}
$\deltaG^{\mu\nu}[h^{\mathrm{SS*}}]$ is a linear operator acting on a locally intregable function, making it (and therefore $\delta^2 G^{\mu\nu}[h^{\S1*},h^{\S1*}]$) well defined as a distribution on $r\geq0$. 

However, the field $h^{\S\S*}_{\mu\nu}$ is only defined in the form of a local expansion around $\gamma$. We can therefore only apply the definition~\eqref{eq:d2GS1_Lorenz} in an infinitesimal neighbourhood of $\gamma$. To localize it to such a neighbourhood, we define $\delta^2G^{\mu\nu}[h^{1*},h^{1*}]$ as the limit $s\to 0$ of a continuous sequence (i.e., a net) of distributions,
\beq
\delta^2G^{\mu\nu}[h^{1*},h^{1*}] := \lim_{s\to0}\delta^2G_s^{\mu\nu}[h^{1*},h^{1*}],\label{eq:ddG lim def}
\eeq
where
\begin{align}
\delta^2G_s^{\mu\nu}[h^{1*},h^{1*}] &:= \left(-\deltaG^{\mu\nu}[h^{\mathrm{SS*}}] +2\delta^2G[h^{\S1*},h^{\R1*}]\right.\nonumber\\
&\qquad\left.+\delta^2G^{\mu\nu}[h^{\R1*},h^{\R1*}]\right)\theta(s-r) \nonumber\\
&\quad\ + \delta^2G^{\mu\nu}[h^{1*},h^{1*}]\theta(r-s).\label{localized definition}
\end{align}
Here $\theta$ is the Heaviside function. With this definition, outside the infinitesimal region $r<s$, $\delta^2G_s^{\mu\nu}[h^{1*},h^{1*}]$ is simply the smooth function $\delta^2G^{\mu\nu}[h^{1*},h^{1*}]$; inside the region $r<s$, we split $h^{1*}_{\alpha\beta}$ into $h^{\S 1*}_{\alpha\beta}$ and $h^{\R 1*}_{\alpha\beta}$ and then replace $\delta^2G_s^{\mu\nu}[h^{\S 1*},h^{\S 1*}]$ using the definition~\eqref{eq:d2GS1_Lorenz}. The definition~\eqref{localized definition} implies that 
as a distribution, $\delta^2G^{\mu\nu}[h^{1*},h^{1*}]$ acts on test fields $\phi_{\mu\nu}$ via\footnote{The integral over the region $r>s$ is an ordinary integral of smooth functions, which diverges as $\sim 1/s$ in the limit $s\to0$. The first integral, on the other hand, is defined in the distributional sense, such that
\beq
\int \phi_{\mu\nu}\deltaG^{\mu\nu}[h^{\mathrm{SS*}}]\theta_s \dd{V} := \int \deltaG^{\mu\nu}[\phi\,  \theta_s]h^{\mathrm{SS*}}_{\mu\nu} \dd{V}.\label{eq:phidGhSS int}
\eeq
Here $\theta_s:=\theta(s-r)$, and we have used that $\delta G^{\mu\nu}$ is self-adjoint; see the next section. Equation~\eqref{eq:phidGhSS int} also diverges as $1/s$, providing a counterterm that  cancels the $1/s$ divergence from the $r>s$ integral.}
\begin{align}
\int \phi_{\mu\nu}&\delta^2G^{\mu\nu}[h^{1*},h^{1*}] \dd{V} \nonumber\\
&:= \lim_{s\to0}\left\{\int\phi_{\mu\nu}\left(-\deltaG^{\mu\nu}[h^{\mathrm{SS*}}]\right.\right.\nonumber\\
&\qquad\qquad +2\delta^2G[h^{\S1*},h^{\R1*}]\nonumber\\
&\qquad\qquad\left.+\delta^2G^{\mu\nu}[h^{\R1*},h^{\R1*}]\right)\theta(s-r)\dd{V}\nonumber\\
&\qquad\qquad + \left.\int_{r>s}\phi_{\mu\nu}\delta^2G^{\mu\nu}[h^{1*},h^{1*}]\dd{V}\right\}\!.\label{eq:ddG definition}
\end{align}
Beyond a certain finite order in the local expansions of $h^{\S\S*}_{\mu\nu}$, $h^{\S1*}_{\mu\nu}$, and $h^{\R1*}_{\mu\nu}$, this definition is insensitive to the truncation order.

In Ref.~\cite{Detweiler2012}, Detweiler takes Eq.~\eqref{eq:dGhSS r>0} to be valid distributionally on the region $r\geq0$, and so we refer to Eq.~\eqref{eq:ddG lim def} as the {\em Detweiler canonical definition} of $\delta^2 G^{\mu\nu}[h^{1*},h^{1*}]$. We return to some of its consequences at the end of the section.

With Eq.~\eqref{eq:d2GS1_Lorenz}, our Einstein equations become
\begin{align}
	8\pi T^{\mu\nu} &= \e \deltaG^{\mu\nu}[\hrtext{S1*}] + \e^2\left(\deltaG^{\mu\nu}[\hrtext{SR*}]  + \deltaG^{\mu\nu}[\hrtext{\delta m*}]\right. \nonumber \\
		&\quad \left.+ 2Q^{\mu\nu}_{\R}[\hrtext{S1*}]\right) + \order{\e^3}, \label{eq:TLorenz_source}
\end{align}
where \(Q^{\mu\nu}_{\R}[h]:=\delta^2G^{\mu\nu}[h^{\R1*},h]\), in analogy with Eq.~\eqref{eq:Qdef}. In Eq.~\eqref{eq:TLorenz_source} we use the expressions in terms of locally defined fields, despite our discussion of localization above, because the total Einstein tensor identically vanishes for $r>0$. If we were to write a field equation to be solved for $h^{2*}_{\mu\nu}$ globally in the external spacetime, with $\deltaG^{\mu\nu}[h^{2*}]$ on the left and $\delta^2G^{\mu\nu}[h^{1*},h^{1*}]$ on the right, we would instead use Eq.~\eqref{eq:ddG lim def}.

\subsection{Distributional analysis}

To determine the distribution $T^{\mu\nu}$, we integrate the right-hand side of Eq.~\eqref{eq:TLorenz_source} against a test function. 

Doing so requires the adjoints of our operators \(\deltaG^{\mu\nu}\) and \(Q_{\R}^{\mu\nu}\). Here the adjoint of a linear operator $D^{\mu\nu}$ is defined by
\begin{equation}
    \phi_{\mu\nu}D^{\mu\nu}[\psi] - D^{\dagger\mu\nu}[\phi]\psi_{\mu\nu} = \nabla_{\mu}K^{\mu}_{D}, \label{eq:adjoint_def}
\end{equation}
where \(\phi_{\mu\nu}\) and \(\psi_{\mu\nu}\) are arbitrary smooth fields and $K^{\mu}_{D}=K^{\mu}_{D}(\phi,\psi)$. If  \(\psi_{\mu\nu}\) is a distribution, then we define the integral of \(D^{\mu\nu}[h]\) against a test field $\phi_{\mu\nu}$ as
\begin{equation}
    \int \phi_{\mu\nu}D^{\mu\nu}[\psi]\dd{V} \coloneqq \int D^{\dagger\mu\nu}[\phi]\psi_{\mu\nu}\dd{V}. \label{eq:Dadj}
\end{equation}
The linearised Einstein operator is self-adjoint~\cite{Wald1978}; that is,  \(\deltaG^{\dagger\mu\nu}[h] = \deltaG^{\mu\nu}[h]\). \(Q_R^{\dagger\mu\nu}\) is given in Eq.~\eqref{eq:Qadjoint} with  $h^\flat_{\mu\nu}= h^{\R1*}_{\mu\nu}$.

We now evaluate the integral of Eq.~\eqref{eq:TLorenz_source} against a test field \(\phi_{\mu\nu}\),
\begin{align}
   8\pi \int\! \phi_{\mu\nu}T^{\mu\nu}dV &=  \int \phi_{\mu\nu}\left\{\e \deltaG^{\mu\nu}[\hrtext{S1*}] + \e^2(\deltaG^{\mu\nu}[\hrtext{SR*}]\right. \nonumber\\
   &\quad + \left.\deltaG^{\mu\nu}[\hdm] 
        + 2Q_{\R}^{\mu\nu}[\hrtext{S1*}])\right\} \dd{V}. \label{eq:int_phi_lorenz}
\end{align}
We then move the operators \(\deltaG^{\mu\nu}\) and \(Q_{\R}^{\mu\nu}\) onto the test tensor using Eq.~\eqref{eq:Dadj}, so that the right-hand side of Eq.~\eqref{eq:int_phi_lorenz} becomes
\begin{align}
    \MoveEqLeft[4] \int \left(\deltaG^{\mu\nu}[\phi]\left\{\e\hrtext{S1*}_{\mu\nu} + \e^2(\hrtext{SR*}_{\mu\nu} + \hrtext{\delta m*}_{\mu\nu})\right\}\right. \nonumber \\
    \MoveEqLeft[3] \left.+ 2\e^{2}Q_{\R}^{\dagger\mu\nu}[\phi]\hrtext{S1*}_{\mu\nu}\right)\dd{V} \nonumber \\
        ={}& \lim_{R\to 0} \int_{r>R} \left(\deltaG^{\mu\nu}[\phi]\left\{\e\hrtext{S1*}_{\mu\nu} + \e^2(\hrtext{SR*}_{\mu\nu} + \hrtext{\delta m*}_{\mu\nu})\right\}\right. \nonumber \\
            & + \left.\e^{2}2Q_{\R}^{\dagger\mu\nu}[\phi]\hrtext{S1*}_{\mu\nu}\right)\dd{V} \nonumber \\
        ={}& \lim_{R\to 0} \biggl[\int_{r>R} \left\{\phi_{\mu\nu}\deltaG^{\mu\nu}[\e\hrtext{S1*} + \e^2(\hrtext{SR*} + \hrtext{\delta m*})]\right. \nonumber \\
            & + \left.2\e^{2}\phi_{\mu\nu}Q_{\R}^{\mu\nu}[\hrtext{S1*}]\right\} \dd{V} - \int_{r=R}\left\{K^{\deltaG}_{\alpha}[\e\hrtext{S1*}\right. \nonumber \\
            & + \left.\e^2(\hrtext{SR*} + \hrtext{\delta m*})] +2\e^{2}K^{Q}_{\alpha}[\hrtext{S1*}]\right\}\dd{S^{\alpha}}\biggr], \label{eq:einstein_lorenz_dist_int}
\end{align}
where \(K^{D}_{\alpha}\) denotes the boundary term for the operator \(D\).
In the first equality we note that as the integral is now over ordinary integrable functions instead of distributions, we can remove the region \(r<R\) and then take the limit as \(R\) goes to \(0\).
Following that, in the second equality, we integrate by parts using Stokes' theorem to move the operators back onto the metric perturbations.
The values of \(K^{D}_{\alpha}\) are given by
\begin{align}
    K^{\delta G}_{\alpha}[h] ={}& \frac{1}{2}\phi^{\beta\mu}h_{\beta\mu;\alpha} - \frac{1}{2}h^{\beta\mu}\phi_{\beta\mu;\alpha} + \phi^{\beta}{}_{\beta}h^{\mu}{}_{[\alpha;\mu]} \nonumber \\
        & + h^{\beta}{}_{\beta}\phi^{\mu}{}_{[\mu;\alpha]} + \frac{1}{2}\phi_{\alpha}{}^{\beta}h^{\mu}{}_{\mu;\beta} - \frac{1}{2}h_{\alpha}{}^{\beta}\phi^{\mu}{}_{\mu;\beta} \nonumber \\
        & + h^{\beta\mu}\phi_{\alpha\beta;\mu} - \phi^{\beta\mu}h_{\alpha\beta;\mu}. \label{eq:dG_source}
\end{align}
and
\begin{align}
    K^{Q}_{\alpha}[h] ={}& \frac{1}{8} \biggl[h^{\beta \gamma } \{\phi^{\zeta }{}_{\zeta } \hrstar_{\beta \gamma ;\alpha } + \phi_{\beta \gamma } \hrstar_{\zeta }{}^{\zeta }{}_{;\alpha } - 4 \phi_{\alpha }{}^{\zeta } \hrstar_{\beta \zeta ;\gamma } \nonumber \\
        & - 2 \phi_{\alpha \beta } \hrstar_{\zeta }{}^{\zeta }{}_{;\gamma }\} -  h^{\beta }{}_{\beta } \bigl\{\phi^{\gamma }{}_{\gamma } \hrstar_{\zeta }{}^{\zeta }{}_{;\alpha } \nonumber \\
        & + \phi^{\gamma \zeta } (\hrstar_{\gamma \zeta ;\alpha } - 2 \hrstar_{\alpha \gamma ;\zeta })\bigr\} + 2 \Bigl\{h_{\alpha }{}^{\beta } \bigl(2 \phi^{\gamma \zeta } \hrstar_{\beta \gamma ;\zeta } \nonumber \\
        & + 2\phi^{\gamma }{}_{\gamma }\hrstar_{[\zeta }{}^{\zeta }{}_{;\beta]}\bigr) + \hrstar{}^{\beta \gamma } \bigl(\phi_{\beta \gamma } h^{\zeta }{}_{\zeta }{}_{;\alpha } \nonumber \\
        & -  h^{\zeta }{}_{\zeta } \phi_{\beta \gamma }{}_{;\alpha } + 2 h_{\beta }{}^{\zeta } \phi_{\gamma \zeta }{}_{;\alpha } -  h_{\beta \gamma } \phi^{\zeta }{}_{\zeta }{}_{;\alpha } \nonumber \\
        & + 2\phi^{\zeta }{}_{\zeta } h_{\beta [\gamma }{}_{;\alpha] } - 2 \phi_{\alpha \beta } h^{\zeta }{}_{\zeta }{}_{;\gamma } - 2 h_{\beta }{}^{\zeta } \phi_{\alpha \zeta }{}_{;\gamma } \nonumber \\
        & + h_{\alpha \beta } \phi^{\zeta }{}_{\zeta }{}_{;\gamma } + 2 \phi_{\beta }{}^{\zeta } [h_{\alpha \zeta }{}_{;\gamma } + h_{\alpha \gamma }{}_{;\zeta } - h_{\gamma \zeta }{}_{;\alpha }] \nonumber \\
        & -  \phi_{\beta \gamma } h_{\alpha }{}^{\zeta }{}_{;\zeta } -  \phi_{\alpha }{}^{\zeta } h_{\beta \gamma }{}_{;\zeta } - 2 h_{\beta }{}^{\zeta } \phi_{\alpha \gamma }{}_{;\zeta } \nonumber \\
        & + h_{\beta \gamma } \phi_{\alpha }{}^{\zeta }{}_{;\zeta } + h_{\alpha }{}^{\zeta } \phi_{\beta \gamma }{}_{;\zeta }\bigr) + \hrstar_{\alpha }{}^{\beta } \bigl(h^{\gamma \zeta } \phi_{\gamma \zeta }{}_{;\beta } \nonumber \\
        & -  h^{\gamma }{}_{\gamma } \phi^{\zeta }{}_{\zeta }{}_{;\beta } + h_{\beta }{}^{\gamma } \phi^{\zeta }{}_{\zeta }{}_{;\gamma } + 2\phi^{\gamma }{}_{\gamma } h^{\zeta }{}_{[\zeta;\beta]} \nonumber \\
        & - \phi^{\gamma \zeta } [h_{\gamma \zeta }{}_{;\beta } - 2 h_{\beta \gamma }{}_{;\zeta }] - 2 h^{\gamma \zeta } \phi_{\beta \gamma }{}_{;\zeta } \nonumber \\
        & + 2 h^{\gamma }{}_{\gamma } \phi_{\beta }{}^{\zeta }{}_{;\zeta }\bigr)\Bigr\}\biggr], \label{eq:d2GhR1_Lorenz}
\end{align}
while the surface element in Fermi--Walker coordinates is given by
\begin{equation}
    \dd{S^\alpha} = - R^2 n^{\alpha}\dd{t}\dd{\Omega} + \order{R^3}, \label{eq:FW_surf_elem}
\end{equation}
where \(n^{\alpha} = (0,n^{i})\) and the minus sign comes from the orientation of the normal vector to the boundary of the region $r>R$.

To evaluate the volume integral, note that the integrand is order \(\e^3\) off the worldline,
\begin{multline}
    \e \deltaG^{\mu\nu}[\hrtext{S1*}] + \e^2(\deltaG^{\mu\nu}[\hrtext{SR*} + \hrtext{\delta m*}] \\
        + 2Q_{\R}^{\mu\nu}[\hrtext{S1*}]) = \order{\e^3} \qc{r > 0}. \label{eq:sing_field_off_worldline}
\end{multline}
So the volume integral contributes nothing to the final result and can be ignored, leaving only the boundary terms:
\begin{align}
\int\! \phi_{\mu\nu}T^{\mu\nu}dV &=
- \frac{1}{8\pi}\lim_{R\to0}\int_{r=R}\left(\e K^{\deltaG}_{\alpha}[\hrtext{S1*}]\right.\nonumber\\
    &\quad +\e^2K^{\deltaG}_{\alpha}[\hrtext{SR*}] + \e^2K^{\deltaG}_{\alpha}[\hrtext{\delta m*}] \nonumber \\
    &\quad \left.+2\e^{2}K^{Q}_{\alpha}[\hrtext{S1*}]\right)\dd{S^{\alpha}}+\order{\e^3}.\label{eq:TLorenz bdry int}
\end{align}

\subsection{Evaluation of boundary terms}\label{sec:lorenz_source}

For the rest of this section, all occurences of \(\hrstar_{\mu\nu}\) and \(\phi_{\mu\nu}\) are evaluated on the worldline, but we omit the notation for visual clarity.
We substitute \(\hrtext{S1*}_{\mu\nu}\) from~\eqref{eq:hS1Lorenz}, \(\hrtext{SR*}_{\mu\nu}\) from~\eqref{eq:hSR_Lorenz} and \(\hdm_{\mu\nu}\) from~\eqref{eq:hDM_Lorenz} into Eq.~\eqref{eq:dG_source}, giving
\begin{align}
    K^{\delta G}_{\alpha}[h^{\mathrm{S1*}}] ={}& - \frac{2mn_{\beta}}{r^2}\Big(\phi^{\mu}{}_{\mu} u_{\alpha}u^{\beta} - 2\phi^{\beta}{}_{\mu} u_{\alpha}u^{\mu} \nonumber \\
        & + g_{\alpha}{}^{\beta}\phi_{\mu\nu} u^{\mu}u^{\nu}\Big) + \order{1/r}, \label{eq:dG_hS1_Lorenz} \displaybreak[0] \\
    K^{\delta G}_{\alpha}[h^{\mathrm{SR*}}] ={}& - \frac{m n_{\alpha}\nhat^{ab}}{2 r^2} \Big(2\hrstar_{ac}\phi_{b}{}^{c} + 2\hrstar_{ta}\phi_{tb} \nonumber \\
        & - \hrstar_{tt}\phi_{ab} - \delta^{ij}\hrstar_{ij}\phi_{ab} - \hrstar_{ab}\phi^{c}{}_{c} \nonumber \\
        & - \hrstar_{ab}\phi_{tt}\Big) + \order{1/r}, \label{eq:dG_hSR_Lorenz} \displaybreak[0] \\
    K^{\delta G}_{\alpha}[h^{\mathrm{\delta m*}}] ={}& - \frac{m}{6r^2}\Big[6\phi_{\alpha a} n^{a}\big(\delta^{ij}\hrstar_{ij} + 2\hrstar_{tt}\big) \nonumber \\
        & + n_{\alpha}\big(2\hrstar_{ab}\phi^{ab} + 8\hrstar_{ta}\phi_{t}{}^{a} - 10\hrstar_{tt}\phi^{a}{}_{a} \nonumber \\
        & - 5\delta^{ij}\hrstar_{ij}\phi^{b}{}_{b} + 5\delta^{ij}\hrstar_{ij}\phi_{tt} \nonumber \\
        & + 6 \hrstar_{tt}\phi_{tt}\big)\Big] + \order{1/r}. \label{eq:dG_hDM_Lorenz}
\end{align}
Note that we only require terms of order \(1/r^n\) where \(n \geq 2\) as all other terms will vanish after taking the limit \(R \to 0\).
We follow the same procedure for \(K^Q_\alpha\), substituting Eq.~\eqref{eq:hS1Lorenz} into Eq.~\eqref{eq:d2GhR1_Lorenz}, to get
\begin{align}
    K^{Q}_{\alpha}[h^{\mathrm{S1*}}] ={}& \frac{m}{r^2} \Bigl[n_{\alpha}\Bigl(\hrstar_{tt}\phi^{a}{}_{a} - \hrstar_{ab}\phi^{ab} - 2\hrstar_{ta}\phi_{t}{}^{a} \nonumber \\
        & + \delta^{ij}\hrstar_{ij}\phi^{b}{}_{b} - \delta^{ij}\hrstar_{ij}\phi_{tt} + 2\hrstar_{tt}\phi_{tt}\Bigr) \nonumber \\
        & + n^{a}\Bigl(4\hrstar_{\alpha b}\phi_{a}{}^{b} - \hrstar_{\alpha a}\phi^{b}{}_{b} - \hrstar_{b}{}^{b}\phi_{\alpha a} \nonumber \\
        & - \hrstar_{tt}\phi_{\alpha a} - 2\hrstar_{ab}\phi_{\alpha}{}^{b} + 2\hrstar_{ta}\phi_{t\alpha} \nonumber \\
        & - \hrstar_{\alpha a}\phi_{tt} + 2u_{\alpha}\bigl(2\hrstar_{tb}\phi_{a}{}^{b} - 2\hrstar_{tt}\phi_{ta} \nonumber \\
        & - \hrstar_{ta}\phi^{b}{}_{b} + 2\hrstar_{ab}\phi^{b}{}_{t} - \hrstar_{ta}\phi_{tt}\bigr)\Bigr)\Bigr] \nonumber \\
        & + \order{1/r}. \label{eq:d2GhR1_hS1_Lorenz}
\end{align}

We then integrate each of these quantities  with the surface element from Eq.~\eqref{eq:FW_surf_elem}, noting that~\cite{Blanchet1986}
\begin{equation}
    \int \nhat_L \dd{\Omega} = 0 \qfor{l \geq 1}. \label{eq:nhat_int}
\end{equation}
The first-order integral is given by
\begin{equation}
    \lim_{R\to 0} \int_{r=R} K^{\delta G}_{\alpha}[h^{\mathrm{S1*}}] \dd{S^\alpha} = -8\pi m \int \phi_{tt} \dd{t}, \label{eq:dG_hS1_Lorenz_surf_int}
\end{equation}
and the second-order ones by
\begin{align}
    \MoveEqLeft[4] \lim_{R\to 0} \int_{r=R} K^{\delta G}_{\alpha}[h^{\mathrm{SR*}}] \dd{S^\alpha} \nonumber \\
        ={}& - \frac{8\pi m}{9}\int \left(\hrstar_{ab}\phi^{ab} + 2\hrstar_{ta}\phi_{t}{}^{a} + \hrstar_{tt}\phi^{a}{}_{a}\right. \nonumber \\
            & - \left.\delta^{ij}\hrstar_{ij}\phi_{tt} - 3\hrstar_{tt}\phi_{tt}\right) \dd{t}, \label{eq:dG_hSR_Lorenz_surf_int} \displaybreak[0] \\
    \MoveEqLeft[4] \lim_{R\to 0} \int_{r=R} K^{\delta G}_{\alpha}[h^{\delta m*}] \dd{S^\alpha} \nonumber \\
        ={}& - \frac{4\pi m}{9}\int \left(\hrstar_{ab}\phi^{ab} + 8\hrstar_{ta}\phi_{t}{}^{a} - 8\hrstar_{tt}\phi^{a}{}_{a}\right. \nonumber \\
            & - \left.3\delta^{ij}\hrstar_{ij}\phi^{b}{}_{b} + 5\delta^{ij}\hrstar_{ij}\phi_{tt} + 6\hrstar_{tt}\phi_{tt}\right) \dd{t}, \label{eq:dG_hDM_Lorenz_surf_int} \displaybreak[0] \\
    \MoveEqLeft[4] \lim_{R\to 0} \int_{r=R} K^{Q}_{\alpha}[h^{\mathrm{S1*}}] \dd{S^\alpha} \nonumber \\
        ={}& \frac{4\pi m}{3}\int \left(\hrstar_{ab}\phi^{ab} + 4\hrstar_{ta}\phi^{a}{}_{t} - 2\hrstar_{tt}\phi^{a}{}_{a}\right. \nonumber \\
            & - \left.\delta^{ij}\hrstar_{ij}\phi^{b}{}_{b} + 4\delta^{ij}\hrstar_{ij}\phi_{tt} - 6\hrstar_{tt}\phi_{tt}\right) \dd{t}. \label{eq:d2GR1_Lorenz_surf_int}
\end{align}

\subsection{Result: recovering the Detweiler stress-energy}\label{sec:boundary_total}

As we explained in Sec.~\ref{sec:stress-energy in SF}, if we were to define $8\pi T^{\mu\nu}_1:= \delta G^{\mu\nu}[h^{1*}]$ in the self-consistent expansion, then we would find $T^{\mu\nu}_1$ contains a subdominant correction that is extended away from $\gamma$. That prompted us to define the total $T^{\mu\nu}$ in Eq.~\eqref{eq:EFEsc}, rather than defining each $T^{\mu\nu}_n$ separately. However, our formula~\eqref{eq:TLorenz bdry int} now provides an unambiguous split:
\begin{align}
\int\! \phi_{\mu\nu}T_1^{\mu\nu}dV &=
- \frac{1}{8\pi}\lim_{R\to0}\int_{r=R}K^{\deltaG}_{\alpha}[\hrtext{S1*}]\dd{S^{\alpha}},\label{eq:T1 bdry def}\\
\int\! \phi_{\mu\nu}T_{2*}^{\mu\nu}dV &= - \frac{1}{8\pi}\lim_{R\to0}\int_{r=R}\left(K^{\deltaG}_{\alpha}[\hrtext{SR*}]\right.\nonumber\\
&\quad \left.+ K^{\deltaG}_{\alpha}[\hrtext{\delta m*}] + 2K^{Q}_{\alpha}[\hrtext{S1*}]\right)\dd{S^{\alpha}}.\label{eq:T2 bdry def}
\end{align}
These are equivalent to the definitions~\eqref{eq:T1 SC} and \eqref{eq:T2 SC}.

At first order, we can immediately see from Eq.~\eqref{eq:dG_hS1_Lorenz_surf_int} that Eq.~\eqref{eq:T1 bdry def} can be written as
\begin{equation}
    \int\phi_{\mu\nu}T^{\mu\nu}_{1}\dd{V} = m\int\!\!\!\int \phi_{\mu\nu}u^\mu u^\nu \delta^4(x,z)\dd{\tau}dV. \label{eq:T1_lorenz}
\end{equation}
Since this holds for an arbitrary test field $\phi_{\mu\nu}$, we infer that  $T^{\mu\nu}_1$ is the point-mass stress-energy in Eq.~\eqref{eq:stress1}, as expected; a nearly identical derivation appears in Ref.~\cite{Gralla2008}.

Moving to second order, we sum the boundary terms to obtain
\begin{multline}
    \lim_{R\to 0} \int_{r=R} \left(K^{\delta G}_{\alpha}[h^{\mathrm{SR*}}] + K^{\delta G}_{\alpha}[h^{\mathrm{\delta m*}}] + K^{Q}_{\alpha}[h^{\mathrm{S1*}}]\right) \dd{S^\alpha} \\
        = 4\pi m \int(2\hrstar_{tt} - \delta^{ij}\hrstar_{ij}) \phi_{tt}\dd{t}. \label{eq:T2_lorenz_boundary}
\end{multline}
We can therefore write Eq.~\eqref{eq:T2 bdry def} as
\begin{equation}
    \int\! \phi_{\mu\nu}T^{\mu\nu}_{2*}\dd{V} = \frac{m}{2} \int\!\!\!\int  \phi_{\mu\nu}u^\mu u^\nu(u^\alpha u^\beta - g^{\alpha\beta})\hrstar_{\alpha\beta}\dd{\tau}dV.
\end{equation}
This implies that, given Detweiler's canonical definition of $\delta^2 G^{\mu\nu}[h^{1*},h^{1*}]$, \(T^{\mu\nu}_{2*}\) in the Lorenz gauge has the same functional form as the \(T^{\mu\nu}_{2}\) found in the highly regular gauge in Eq.~\eqref{eq:T2HRcov}.
Additionally, using the methods and arguments outlined in this section, we can show that the functional forms of \(T^{2}_{\mu}{}^{\nu}\) and \(T^{2}_{\mu\nu}\) in the Lorenz gauge match the ones found in the highly regular gauge, as is to be expected.

\subsection{Generality of Detweiler's canonical definition}\label{sec:T2gaugeInvGen}

In this section we have focused on the Lorenz gauge, but much of the analysis immediately extends to all gauges with a generic level of regularity. Specifically, the canonical definition~\eqref{eq:ddG lim def} suffices to determine a unique $T^{\mu\nu}_{(2)}$ given by an equation of the form~\eqref{eq:T2 bdry def} (though in general we would not split the ``singular times regular" piece of the field into the two pieces $h^{\S\R}_{\mu\nu}$ and $h^{\delta m}_{\mu\nu}$). Moreover, the canonical definition implies that the Einstein equation can be written in the form~\eqref{eq:Detweiler EFE} for some distribution $\tilde T^{\mu\nu}$ supported on $\gamma$. This in turn implies that $\tilde \nabla_\nu \tilde T^{\mu\nu}=0$. We conjecture, based on that fact, that our result in the Lorenz gauge holds true in all gauges with generic regularity: the Detweiler canonical definition of $\delta^2 G^{\mu\nu}[h^1,h^1]$ implies that the Detweiler stress-energy is valid. But we have not attempted to prove this statement.

A separate question is whether the canonical definition has practical utility. One of us made some use of it in Ref.~\cite{Miller2020}, but we defer further discussion of this question to future work.

\section{Applications}\label{sec:application}

In this section, we briefly outline how the highly regular gauge and second-order stress-energy tensor could be utilized in numerical schemes.

\subsection{Puncture scheme}\label{sec:app_punc}

As discussed in the introduction, there is only one extant second-order implementation~\cite{Pound2020,Miller2020}, which is based on a puncture scheme in the Lorenz gauge.
That scheme starts from the gauge-fixed version of the Einstein equations in Eqs.~\eqref{eq:Eh1 off gamma}--\eqref{eq:Eh2 off gamme}. In analogy with Eqs.~\eqref{eq:puncture_1st}--\eqref{eq:puncture_2nd}, the equations of the puncture scheme then become 
\begin{align}
    E^{\mu\nu}[h^{\calR 1*}] &= - (E^{\mu\nu}[{h}^{\calP 1*}])^\star, \label{eq:puncture_1st Lorenz} \\
    E^{\mu\nu}[h^{\calR 2*}] &= - (\delta^2 G^{\mu\nu}[h^{1*},h^{1*}] + E^{\mu\nu}[h^{\calP 2*}])^\star, \label{eq:puncture_2nd Lorenz} 
\end{align}
with the puncture moving on a trajectory governed by
\beq
\frac{{D}^{2}z^{\alpha}}{\dd{\tau^2}} = - \frac{1}{2}P^{\alpha\mu}(g_{\mu}{}^{\rho} - h^{\calR*}_{\mu}{}^{\rho})(2h^{\calR*}_{\rho\beta;\gamma} - h^{\calR*}_{\beta\gamma;\rho})u^{\beta}u^{\gamma}. \label{eq:2nd_eom_hres}
\eeq
The puncture and residual fields satisfy the gauge-fixed equation $E^{\mu\nu}[h^{\calR 1*}+h^{\calP1*}]=8\pi T^{\mu\nu}_1$ in the entire domain including $\gamma$, but $\e\deltaG^{\mu\nu}[h^{\calR 1*}+h^{\calP1*}]=\order{\e^2}\neq0$ at points away from $\gamma$.

For the purpose of modelling an inspiral into a black hole, these equations are solved with a two-timescale ansatz that splits the solution into slowly varying amplitudes and rapidly varying phases ~\cite{Pound2020,Miller2020,Pound2021,Flanagan2021}:
\begin{align}
h^{n*}_{\mu\nu} &= \sum_{n'\geq 0}\sum_{\bm{k}} \e^{n'}h^{nn'\bm{k}}_{\mu\nu}(\bm{p},M_A,\bm{x}_{\rm BL})e^{-i \bm{k}\cdot\bm{\varphi}},\label{eq:h multiscale}\\
z^i &= z^i_0(\bm{p},\bm{\varphi})+\e z^i_1(\bm{p},M_A,\bm{\varphi})+\order{\e^2}.\label{eq:z multiscale}
\end{align}
Here $\bm{x}_{\rm BL}=(r_{\rm BL},\theta_{\rm BL},\phi_{\rm BL})$ are Boyer-Lindquist spatial coordinates in the black hole spacetime; $z^i$ are the Boyer-Lindquist spatial coordinates of the puncture's trajectory; $\bm{k}=(k^r,k^\theta,k^\phi)$, and each $k^i$ runs over all integers; $\bm{k}\cdot\bm{\varphi}:=\sum_i k^i\varphi_i$; $\bm{p}$ is a set of three orbital parameters that slowly evolve due to dissipation (e.g., orbital energy, angular momentum, and Carter constant); $M_A=(M^1,J^1)$ are corrections to the central black hole's mass and spin that slowly evolve due to gravitational-wave absorption; and $\bm{\varphi}=(\varphi_r,\varphi_\theta,\varphi_\phi)$ are a set of three phase variables describing the  radial, polar, and azimuthal motion of the small object around the central black hole.

We will not need the technical details of this two-timescale puncture scheme for our discussion, but its general structure will help to highlight some of the subtleties that arise in converting our results into a usable puncture in a highly regular gauge.

The singular field we have obtained has the form
\beq
h^\S_{\mu\nu} = \e \mathring{h}^{\S1}_{\mu\nu}+ \e^2 \mathring{h}^{\S2}_{\mu\nu} +\order{\e^3},
\eeq
where $\mathring{h}^{\S n}_{\mu\nu}$ is given by Eqs.~\eqref{eq:hS1transform} and \eqref{eq:hS2 HR} with Eqs.~\eqref{eq:hS1Prime}, \eqref{eq:hrgaugeSR}, and \eqref{eq:hrgaugeSS}. This differs in two significant ways from the Lorenz-gauge case. First, as we reiterated at several points in our presentation, the coefficients here are not in precise correspondence with the coefficients $h^{\S n}_{\mu\nu}$ in a self-consistent expansion in an $\e$-independent coordinate system. Second, even if we had access to $h^{\S n}_{\mu\nu}$, it would satisfy a locally defined gauge condition, while we wish our residual field (and therefore $h^{\R n}_{\mu\nu}$) to satisfy some gauge that simplifies the linearized Einstein tensor in the external background. If $h^{\S n}_{\mu\nu}$ and $h^{\R n}_{\mu\nu}$ satisfy different gauge conditions, it is not obvious how one would write the field equations in a gauge-fixed form.

To see the impact of these differences, suppose we define punctures
\begin{align}
h^{\calP n}_{\mu\nu} \coloneqq \mathring{h}^{\mathrm{S}n}_{\mu\nu},
\end{align}
with some choice of windowing to set $h^{\calP n}_{\mu\nu}$ to zero outside some region around $\gamma$, and that we then solve the field equations~\eqref{eq:puncture_1st} and \eqref{eq:puncture_2nd} for residual fields $h^{\calR n}_{\mu\nu}$. The left-hand side of these field equations can be in any convenient gauge. For concreteness, take it to be the Lorenz gauge, such that
\begin{align}
    E^{\mu\nu}[h^{\calR 1}] &= - (\deltaG^{\mu\nu}[{h}^{\calP 1}])^\star,  \\
    E^{\mu\nu}[h^{\calR 2}] &= - (\delta^2 G^{\mu\nu}[h^{1},h^{1}] + \deltaG^{\mu\nu}[h^{\calP 2}])^\star.  
\end{align}
Counter to the fields in the scheme~\eqref{eq:puncture_1st Lorenz}--\eqref{eq:puncture_2nd Lorenz}, the fields here do not satisfy \beq
E^{\mu\nu}[h^{\calR 1}]+\delta G^{\mu\nu}[h^{\calP 1}]=8\pi T^{\mu\nu}_1.
\eeq
There is, effectively, an additional, order-$\e^2$ source that extends away from $\gamma$. We can understand the form of this source by noting that $\delta G^{\mu\nu}[h^{\calP1}]\sim \mathring{\delta G}{}^{\mu\nu}[h^{\calP1}]+a^i\partial h^{\calP1}_{\alpha\beta}$. Since $\mathring{\delta G}{}^{\mu\nu}[\mathring{h}^{\S1}]=0$, this  implies that $\deltaG^{\mu\nu}[{h}^{\calP 1}]$ contains singular terms $\propto a^i/r^2$. As a consequence, $h^{\calR 1}_{\mu\nu}$ will be discontinuous at $r=0$. In principle, $h^{\calR 2}_{\mu\nu}$ will precisely cancel this discontinuity, since $\e h^{\calR 1}_{\mu\nu}+\e^2 h^{\calR2}_{\mu\nu}$ will still sum to $h^\R_{\mu\nu} + \order{\e^3}$ on $\gamma$. But $h^{\calR 1}_{\mu\nu}$ cannot be used in the equation of motion for $\gamma$ or in the term $\mathring{h}^{\S\R}_{\mu\nu}$ in the second-order puncture. Additional work would be required to correctly formulate a self-consistent puncture scheme in a highly regular gauge.

Fortunately, our results do suffice for other practical formulations of the field equations. The expansions~\eqref{eq:h multiscale}--\eqref{eq:z multiscale} automatically include an expansion of the acceleration, meaning that our local results can be incorporated directly into a two-timescale implementation. After performing the expansion
\beq
\mathring{h}^{\S n}_{\mu\nu} = \sum_{n'\geq 0}\sum_{\bm{k}} \e^{n'}\mathring{h}^{\S nn'\bm{k}}_{\mu\nu}(\bm{p},M_A,\bm{x}_{\rm BL})e^{-i \bm{k}\cdot\bm{\varphi}},\label{eq:multiscale hSn}
\eeq
we can define punctures
\begin{align}
h^{\calP1}_{\mu\nu} &= \sum_{\bm{k}} \mathring{h}^{\S 1,0,\bm{k}}_{\mu\nu}e^{-i \bm{k}\cdot\bm{\varphi}}\label{eq:new hP1}\\
h^{\calP2}_{\mu\nu} &= \sum_{\bm{k}} \left(\mathring{h}^{\S 2,0,\bm{k}}_{\mu\nu} + \mathring{h}^{\S 1,1,\bm{k}}_{\mu\nu}\right)e^{-i \bm{k}\cdot\bm{\varphi}}.\label{eq:new hP2}
\end{align}
These punctures will fit directly into the two-timescale field equations, with residual fields that are regular on $\gamma$. In practical terms, the punctures would be constructed by substituting the expansion of the trajectory, Eq.~\eqref{eq:z multiscale}, into our formulas for the singular field and then performing a decomposition into the Fourier modes $e^{-i \bm{k}\cdot\bm{\varphi}}$.

The two-timescale expansion is specialized to the inspiral phase of bound binary systems, somewhat limiting the generality of our result. An alternative that could be used in generic spacetimes would be an ordinary Taylor series expansion in powers of $\e$. The field equations would then be Eqs.~\eqref{eq:puncture_1st}--\eqref{eq:puncture_2nd}, and the residual fields would again be regular on $\gamma$. One can obtain the punctures in this scheme from our singular field by substituting an ordinary Taylor series $z^\mu(\tau,\e) = z^\mu(\tau)+\e z^\mu_1(\tau)+\order{\e^2}$; such an expansion is detailed in Ref.~\cite{Pound2015}. Although this expansion breaks down on long time scales, it should suffice for many purposes.

Regardless of whether a two-timescale expansion or Taylor expansion is used, several other steps are required to construct a practical puncture. One must first convert our Fermi--Walker 
coordinate expressions for \(\mathring{h}^{\S n}_{\mu\nu}\) into a covariant form. This can be done using Synge's world function and near-coincidence expansions, as detailed in Ref.~\cite{Pound2014}. Following this, because the field equations are typically solved using a decomposition into a basis of angular harmonics, the singular field must be decomposed into that basis.

With the punctures in the highly regular gauge, we will have achieved our aim: the second-order source term $\delta^2 G^{\mu\nu}[h^1,h^1]$ will be far less singular than it is in a generic gauge, substantially reducing the numerical cost of second-order computations.

\subsection{Mode-sum regularisation}\label{sec:app_modesum}

The calculation of the second-order stress-energy tensor, \(T_{2}^{\mu\nu}\), in Sec.~\ref{sec:o2stress} opens up another avenue for second-order implementations: mode-sum regularisation.
Here, instead of directly solving for the regular field (by way of the residual field in the puncture scheme), we solve for the entirety of \(h^n_{\mu\nu}\) and then subtract \(h^{\S n}_{\mu\nu}\) from it to leave \(h^{\R n}_{\mu\nu}\).
The difficulty of subtracting one divergent quantity from another is avoided by decomposing each field into multipole modes (which are finite) and performing the subtraction at this level.
Schematically, following the notation in the introduction,
\begin{equation}
    \hrtext{R}_{\mu\nu}(z) = \sum_{ilm} \left[h_{ilm}(z) - h^{\S}_{ilm}(z)\right]Y^{ilm}_{\mu\nu}(z), \label{eq:hModeSum}
\end{equation}
where $z$ is a point on $\gamma$, and \(Y^{ilm}_{\mu\nu}\) denotes a basis of angular harmonics.
Analogous equations can be written for any quantity constructed from derivatives of \(\hrtext{R}_{\mu\nu}\), such as the self-force. This mode-sum method has been the  basis for most first-order implementations, and it is typically more efficient than a puncture scheme.

To date, using this method has not been possible at second order due to the strong divergence at the worldline. The second-order field generically behaves as $\sim 1/r^2$, leading to individual modes that diverge as $\sim \log|r_{\rm BL}-r_o|$, making the mode-sum formula~\eqref{eq:hModeSum} incoherent.
But it should now be possible to implement this method using the weaker divergence of the highly regular gauge and the knowledge of \(T_2^{\mu\nu}\).
As described in the introduction, in the highly regular gauge the most singular part of the source for the second-order field is \(\sim 1/r^2\) and has individual modes $\delta^2 G_{ilm}\sim \log|r_{\rm BL}-r_o|$; see the discussion surrounding Eq.~\eqref{eq:ddG_ilm HR}.
This suggests that, at worst, the most singular part of the solution has modes that behave as 
\beq
h^2_{ilm} \sim (r_{\rm BL}-r_o)^2\log|r_{\rm BL}-r_o|.\label{eq:h2ilm}
\eeq
This is \(C^1\) differentiable, which is sufficiently smooth to calculate one derivative of $h^{\R2}_{\mu\nu}$ (and hence the second-order self-force) using mode-sum regularization.
Therefore, this should now be a viable approach.

Following the discussion in the previous section, we assume the field equations are written in either a Taylor expansion or two-timescale expansion. We write
\begin{align}
\deltaG^{\mu\nu}[h^1] &= 8\pi T^{\mu\nu}_1,\label{eq:dGh1 HR}\\
\deltaG^{\mu\nu}[h^2] &= -\delta^2G^{\mu\nu}[h^1,h^1] +8\pi T^{\mu\nu}_2 \label{eq:dGh2 HR}
\end{align}
with the understanding that the metric perturbations and stress-energy have been re-expanded and re-combined [e.g., in analogy with Eqs.~\eqref{eq:multiscale hSn}--\eqref{eq:new hP2}]. The specifics of these re-expansions are not important for this discussion, but we refer interested readers to Sec. 7.1 of Ref.~\cite{Pound2021} for details. 

We must now formulate and solve Eqs.~\eqref{eq:dGh1 HR}--\eqref{eq:dGh2 HR} in such a way that (i) $h^1_{\mu\nu}$ on the right-hand side of Eq.~\eqref{eq:dGh2 HR} is in a highly regular gauge, and (ii) both equations can be solved mode by mode in a numerically convenient gauge. Combining these requirements is nontrivial because
we do not have a prescription for solving the field equations globally in a highly regular gauge; these gauges are inherently a local construction near the worldline. 

To sketch a suitable method, we start by assuming that the the first-order modes are computed in  some convenient gauge (e.g., the Lorenz, Regge--Wheeler--Zerilli, or radiation gauge). Such computations are now routine~\cite{Barack2018}. We label the computed modes $h^{1,\rm num}_{ilm}$.
From this starting point, we can perform a first-order gauge transformation to the highly regular gauge, mode by mode, so that
\begin{equation}
    h^{1}_{ilm} = h^{1,\mathrm{num}}_{ilm} + (\lie_{\xi_{1}} g)_{ilm}. \label{eq:worldtube_HR_1st}
\end{equation}
The vector $\xi^\mu_{1}$ can be found as a local expansion near the worldline, in four dimensions, to some finite order in $r$, with any convenient extension away from that local neighbourhood.
The gauge perturbation \(\lie_{\xi_1}g_{\mu\nu}\) can then be decomposed into the chosen basis of modes using the methods described in, e.g., Refs.~\cite{Heffernan2012,WardellWarburton2015,Miller2016}. An alternative method of computing suitable modes $h^{1}_{ilm}$ would be to use our puncture scheme in the highly regular gauge at first order; one could still use mode-sum regularization at second order.

From the first-order modes in the highly regular gauge, we can calculate the source modes in the field equation for the full second-order perturbation,
\begin{equation}
    \deltaG_{ilm}[h^{2}] = - \delta^2 G_{ilm}[h^{1},h^1] + 8\pi T^{2}_{ilm}. \label{eq:worldtube_HR_2nd}
\end{equation}
Because the modes of $h^1_{\mu\nu}$ are in the highly regular gauge, the right-hand side has the desired regularity. We once again suppose this field equation is solved in a convenient gauge to obtain $h^{2,\rm num}_{ilm}$. In a well-behaved gauge such as Lorenz, these modes will have the form~\eqref{eq:h2ilm}; note that in this scenario, $h^{2,\rm num}_{\mu\nu}$ satisfies the Lorenz gauge condition but $h^{1}_{\mu\nu}$ does not.

In order to subtract the singular field from the total field, we next must put $h^{2}_{ilm}$ and $h^{\S2}_{ilm}$ in the same gauge. There are numerous ways of achieving this. For simplicity, we assume that we do the same at second order as at first: transform $h^{2,\rm num}_{ilm}$ to the highly regular gauge, in the same manner we transformed $h^{1,\rm num}_{ilm}$, such that
\begin{equation}
    h^{2}_{ilm} = h^{2,\rm num}_{ilm} + (\lie_{\xi_{2}}g)_{ilm}. \label{eq:HRToLor}
\end{equation}
As at first order, the vector $\xi^\mu_{2}$ can be found as a local expansion in four dimensions, after which $\lie_{\xi_{2}}g_{\mu\nu}$ can be decomposed into modes. The modes $h^{\S2}_{ilm}$ can be calculated from the local expressions in this paper, and $h^{\R 2}_{\mu\nu}$ can then be calculated using the mode-sum formula~\eqref{eq:hModeSum}.

The crux of this scheme is finding the vectors $\xi^\mu_n$. To elucidate how they might be found, we assume that $h^{n,\rm num}_{ilm}$  are computed in the Lorenz gauge. If we trace reverse $h^{n}_{\mu\nu}=h^{n,\rm num}_{\mu\nu}+\lie_{\xi_n}g_{\mu\nu}$, take the divergence, and use $\nabla^\nu \bar{h}^{n,\rm num}_{\mu\nu}=0$, then we find that the gauge vector is determined by
\begin{equation}
    \Box \xi^{\mu}_{n} = \nabla_{\nu} \bar{h}_{n}^{\mu\nu}, \label{eq:GaugeVectorEq}
\end{equation}
where \(\Box \coloneqq g^{\mu\nu}\nabla_{\mu}\nabla_{\nu}\).
This equation can be solved in Fermi--Walker coordinates using the expressions for $h^n_{\mu\nu}$ in this paper, and $\xi^{\mu}_{n}$ can then be converted to covariant form following Ref.~\cite{Pound2015}.

\section{Conclusion}\label{sec:summary}

In summary, we have derived two main results: (i) the local metric perturbation in a class of highly regular gauges, to sufficient order in $r$ to implement a puncture scheme, and (ii) the validity of the Detweiler stress-energy in these gauges. We expect both of these to enable more efficient calculations of the second-order self-force and related quantities in binary systems. To that end, we have also outlined how they might be utilized in concrete numerical schemes. 

Our presentation stressed the utility of the highly regular gauges as a means of overcoming a specific computational challenge: the problem of infinite mode coupling. This might seem to suggest that the challenge  is purely a symptom of a mode decomposition. But an analogous problem would arise in a \(3+1\) calculation. The two source terms $\delta^2 G^{\mu\nu}[h^1,h^1]$ and $\delta G^{\mu\nu}[h^{\calP2}]$ in Eq.~\eqref{eq:puncture_2nd} would each diverge like $\sim 1/r^4$, and those two divergences would cancel each other to leave a regular remainder. To effect this cancellation, one would have to calculate each of the terms to extreme precision at small $r$, just as one would have to go to extreme mode numbers to calculate $\delta^2 G_{ilm}[h^1,h^1]$ as a sum over first-order modes. This high-precision problem would be alleviated in a highly regular gauge.

Besides these pragmatic aspects, our results have  clarified a sense in which point masses remain a well-defined consequence of matched asymptotic expansions beyond linear order. To further bolster this, we have shown that the Detweiler stress-energy is valid outside the highly regular gauges, at least in the Lorenz gauge but probably far more broadly, if one adopts a canonical distributional definition of the second-order Einstein tensor, \(\delta^2 G^{\mu\nu}\). On one hand, this result is not entirely as compelling as the result in highly regular gauges. The canonical definition requires one to know the local solution for $h^2_{\mu\nu}$ before one can use \(\delta^2 G^{\mu\nu}\) as a source for the global solution; in this sense, there is little distinction between a puncture scheme and solving the field equation~\eqref{eq:dGh2 HR} with the canonical \(\delta^2 G^{\mu\nu}\). Yet, on the other hand, the canonical definition does provide a compelling physical interpretation: a small object not only moves as a test body in the effective metric, it also has the stress-energy of such an object. There may seem to be a conflict between the object behaving as a test body and simultaneously having a gravitating stress-energy, but this seeming contradiction is alleviated by the fact that the field equation has the local form of a {\em linearized} Einstein equation in the effective metric, given by Eq.~\eqref{eq:Detweiler EFE} and previously written by Detweiler. Just as in the ordinary linearized Einstein equation with a point source, there is a one-order mismatch between the test-body orbit and the gravitational field it creates.

There are several ways one might extend our results. Our calculations are only applicable to the case of a nonspinning and spherically symmetric object; they should be generalized to the more astrophysically relevant case of a spinning, non-spherically symmetric body. It would also be conceptually interesting, at the least, to extend them to higher perturbative order in $\e$. At a more practical level, it may be possible to make \(\hrtext{SS}_{\mu\nu}\) even more regular by removing the order \(r^0\) piece of the perturbation so that \(\hrtext{SS}_{\mu\nu}\egamma = 0\).
We have so far been unsuccessful in our attempts to find a gauge transformation that achieves this, hinting that these $\order{r^0}$ terms may contain invariant information about the coupling between external tides and the object's local gravity. But we have also not ruled out the possibility of gauging the terms away. All of these extensions might draw on the work of Harte~\cite{Harte:2014ooa,Harte:2016vwo}, who has shown how nonlinearities can be reduced by adopting Kerr--Schild or generalized Kerr--Schild gauges.

\begin{acknowledgments}

This work was supported by a Royal Society University Research Fellowship and a Royal Society Research Grant for Research Fellows.

\end{acknowledgments}

\appendix

\section{Linear and quadratic Einstein tensors and their adjoints}\label{sec:Einstein tensors}

For a generic metric of the form $g_{\mu\nu}+h_{\mu\nu}$ satisfying $G_{\mu\nu}[g]=0$, the Einstein tensor $G^{\mu\nu}[g+h]$ can be expanded in powers of $h_{\mu\nu}$ and its derivatives as
\beq
G^{\mu\nu}[g+h] = \delta G^{\mu\nu}[h] + \delta^2 G^{\mu\nu}[h,h] + \order{|h|^3},
\eeq
where the linear term is
\begin{equation}
    \deltaG^{\mu\nu}[h] = h_{\alpha}{}^{(\mu;\nu)\alpha} + g^{\mu\nu}h^{\alpha}{}_{[\alpha;\beta]}{}^{\beta} - \frac{1}{2}(h^{\mu\nu;\alpha}{}_{\alpha} + h_{\alpha}{}^{\alpha;\mu\nu}), \label{eq:deltaG}
\end{equation}
and the quadratic term is
\begingroup
\allowdisplaybreaks
\begin{align}
    \delta^2 G^{\mu\nu}[h,h] ={}& \tfrac{1}{2}h^{\mu\nu}{}_{;\alpha}h^{\alpha\beta}{}_{;\beta} - \tfrac{1}{4}h_{\beta}{}^{\beta;\alpha}h^{\mu\nu}{}_{;\alpha} + h^{\mu\nu}h_{\alpha}{}^{[\alpha;\beta]}{}_{\beta} \nonumber \\
        & + h^{\mu\alpha;\beta}h^{\nu}{}_{[\alpha;\beta]} + \tfrac{1}{2}h^{\beta}{}_{\beta;\alpha}h^{\alpha(\mu;\nu)} \nonumber \\
        & - h_{\alpha\beta}{}^{;\beta}h^{\alpha(\mu;\nu)} + \tfrac{1}{4}h_{\alpha\beta}{}^{;\mu}h^{\alpha\beta;\nu} \nonumber \\
        & + h_{\alpha\beta}(h^{\nu[\mu;\alpha]\beta} - h^{\alpha[\mu;|\nu|\beta]}) \nonumber \\
        & + g^{\mu\nu}\bigl(h_{\alpha}{}^{[\beta;\alpha]}h_{\beta\rho}{}^{;\rho} + \tfrac{1}{8}h^{\rho}{}_{\rho;\beta}h_{\alpha}{}^{\alpha;\beta} \nonumber \\
        & + \tfrac{1}{4}h_{\alpha\rho;\beta}h^{\alpha\beta;\rho} - \tfrac{3}{8}h_{\alpha\beta;\rho}h^{\alpha\beta;\rho} \nonumber \\
        & - h^{\alpha\beta}[h^{\rho}{}_{[\rho;\alpha]\beta} + h_{\alpha[\beta;\rho]}{}^{\rho}]\bigr) \nonumber\\
        & - 2h^{(\mu}{}_\rho\delta G^{\nu)\rho}[h]. \label{eq:delta2G}
\end{align}%
\endgroup

The quadratic Einstein tensor $\delta^2 G^{\mu\nu}[h,h]$ is not uniquely defined with two distinct arguments. For convenience we adopt a symmetric bilinear definition of it,
\beq
\delta^2 G^{\mu\nu}[h^\flat,h^\sharp]:=\frac{1}{2} \left.\frac{d^2}{d\lambda_1\lambda_2}G_{\mu\nu}[g+\lambda_1 h^\flat + \lambda_2 h^\sharp]\right|_{\lambda_i=0},
\eeq
which reduces to Eq.~\eqref{eq:delta2G} when $h^\flat_{\mu\nu}=h^\sharp_{\mu\nu}=h_{\mu\nu}$. We also define a linear operator
\beq
Q^{\mu\nu}_{\flat}[h^\sharp]:=\delta^2 G^{\mu\nu}[h^\flat,h^\sharp],\label{eq:Qdef general}
\eeq
which is the term bilinear in $h^\flat_{\mu\nu}$ and $h^\sharp_{\mu\nu}$ if we expand $G^{\mu\nu}[g+h^\flat+h^\sharp]$ in powers of $h^\sharp_{\mu\nu}$ and its derivatives. That is,
\begin{multline}
G^{\mu\nu}[g+h^\flat+h^\sharp] = G^{\mu\nu}[g+h^\flat] + Q^{\mu\nu}_{\flat}[h^\sharp] \\+ \order{|h^\sharp|^2,|h^\flat|^2|h^\sharp|}.
\end{multline}

In the body of the paper we make extensive use of the adjoints of these quantities. The linearized Einstein tensor is self-adjoint, satisfying $\delta G^{\dagger\mu\nu}[h]=\delta G^{\mu\nu}[h]$. The adjoint of $Q^{\mu\nu}_\flat$ is

\begin{widetext}
\begin{align}
    Q^{\dagger\mu\nu}_\flat [\phi] ={}& \frac{1}{2}\Big[\phi^{\alpha\beta}\big(h^{\mu\nu}_\flat{}_{;\alpha\beta} - 2h^{(\mu}_\flat{}_{\alpha;}{}^{\nu)}{}_{\beta} + g^{\mu\nu}\big\{h^{\flat}_{\alpha\rho;\beta}{}^{\rho} - \tfrac{1}{2}h^{\flat}_{\alpha\beta;\rho}{}^{\rho}\big\} + h^{\flat}_{\alpha\beta;}{}^{(\mu\nu)}\big) - h^{\alpha\beta}_\flat\big(2\phi_{\alpha}{}^{(\mu;\nu)}{}_{\beta} - \phi^{\mu\nu}{}_{;\alpha\beta} \nonumber \\
        & - \phi_{\alpha\beta;}{}^{(\mu\nu)} + g^{\mu\nu}\big\{\phi^{\rho}{}_{\rho;\alpha\beta} + \phi_{\alpha\beta;\rho}{}^{\rho} - 2\phi_{\alpha}{}^{\rho}{}_{;\rho\beta}\big\}\big) + \phi^{\alpha(\mu}h^{|\beta|}_\flat{}_{\beta;}{}^{\nu)}{}_{\alpha} + 2h^{\alpha(\mu}_\flat\phi^{|\beta|}{}_{\beta;}{}^{\nu)}{}_{\alpha} \nonumber \\
        & + \phi^{\beta}{}_{\beta;\alpha}\big(h^{\alpha(\mu;\nu)}_\flat - \tfrac{1}{2}h^{\mu\nu;\alpha}_\flat\big) - \frac{1}{2}g^{\mu\nu}\big(\phi^{\rho}{}_{\rho;\beta}h^{\flat}_{\alpha}{}^{\alpha;\beta} + 2h^{\flat}_{\alpha}{}^{\beta;\alpha}\big\{\phi^{\rho}{}_{\rho;\beta} - 2\phi_{\beta\rho;}{}^{\rho}\big\} + \phi^{\alpha}{}_{\alpha}h^{\flat}_{\beta}{}^{\beta;\rho}{}_{\rho} \nonumber \\
        & - 2\phi^{\alpha\rho;\beta}h^{\flat}_{\alpha\beta;\rho} + 3\phi^{\alpha\beta;\rho}h^{\flat}_{\alpha\beta;\rho}\big) + h^{\flat}_{\alpha}{}^{\beta;\alpha}\big(\phi^{\mu\nu}{}_{;\beta} - 2\phi_{\beta}{}^{(\mu;\nu)}\big) + \phi^{\mu\nu}\big(h^{\flat}_{\alpha\beta;}{}^{\alpha\beta} - \tfrac{1}{2}h^{\flat}_{\alpha}{}^{\alpha;\beta}{}_{\beta}\big) \nonumber \\
        & - 2\phi^{\alpha(\mu}h^{\nu)\beta}_\flat{}_{;\alpha\beta} - 2h^{\mu\nu}_\flat\phi_{\alpha}{}^{[\alpha;\beta]}{}_{\beta} + h^{\mu\nu;\alpha}_\flat\phi_{\alpha\beta;}{}^{\beta} - 2h^{\alpha(\mu}_\flat\phi^{\nu)\beta}{}_{;\alpha\beta} + 2\phi^{\alpha(\mu}h^{\nu)}_\flat{}_{\alpha;\beta}{}^{\beta} \nonumber \\
        & - \phi^{\alpha}{}_{\alpha}\big(h^{\beta(\mu;\nu)}_\flat{}_{\beta} - \tfrac{1}{2}h^{\mu\nu;\beta}_\flat{}_{\beta}\big) - 2\phi^{\alpha(\mu}h^{\flat}_{\alpha}{}^{|\beta|;\nu)}{}_{\beta} + \frac{1}{2}\phi^{\mu\nu}{}_{;\beta}h^{\flat}_{\alpha}{}^{\alpha;\beta} + \phi_{\beta}{}^{\beta;(\mu}h^{\flat}_{\alpha}{}^{|\alpha|;\nu)} \nonumber \\
        & + \phi^{\alpha\beta;(\mu}h^{\flat}_{\alpha\beta;}{}^{\nu)} - \phi^{\beta(\mu}{}_{;\beta}h^{\flat}_{\alpha}{}^{|\alpha|;\nu)} - 2\phi_{\alpha\beta;}{}^{\beta}h^{\alpha(\mu;\nu)}_\flat + 4\sym_{\mu\nu}\big(h_\flat^{\mu\alpha}\phi_{\alpha}{}^{[\nu;\beta]}{}_{\beta} + \phi^{\mu}{}_{[\alpha;\beta]}h_\flat^{\nu\alpha;\beta}\big)\Big]. \label{eq:Qadjoint}
\end{align}%
\end{widetext}

The expansion of the Einstein tensor with mixed indices or both indices down can be expressed in terms of the expansion with indices up. Again with $G_{\mu\nu}[g]=0$, we have 
\begin{multline}
G_{\mu}{}^\nu[g+h] = \delta(g_{\mu\rho}G^{\rho\nu})[h] + \delta^2(g_{\mu\rho}G^{\rho\nu})[h,h]\\ + \order{|h|^3},\label{eq:G mixed}
\end{multline}
where 
\begin{align}
    \delta(g_{\mu\rho}G^{\rho\nu})[h] &= g_{\mu\rho}\delta G^{\rho\nu}[h],\label{eq:dG mixed}\\
    \delta^2(g_{\mu\rho}G^{\rho\nu})[h,h] &= g_{\mu\rho}\delta^2G^{\rho\nu}[h,h] + h_{\mu\rho}\delta G^{\rho\nu}[h],
\end{align}
and
\begin{multline}
G_{\mu\nu}[g+h] = \delta(g_{\mu\rho}g_{\nu\sigma}G^{\rho\sigma})[h] + \delta^2(g_{\mu\rho}g_{\nu\sigma}G^{\rho\sigma})[h,h]\\ + \order{|h|^3},
\end{multline}
where 
\begin{align}
    \delta(g_{\mu\rho}g_{\nu\sigma}G^{\rho\sigma})[h] &= g_{\mu\rho}g_{\nu\sigma}\delta G^{\rho\sigma}[h],\\
    \delta^2(g_{\mu\rho}g_{\nu\sigma}G_{\rho\sigma})[h,h] &= g_{\mu\rho}g_{\nu\sigma}\delta^2G^{\rho\sigma}[h,h]\nonumber\\
    &\quad+2h_{\rho(\mu}g_{\nu)\sigma}\delta G^{\rho\sigma}[h].\label{eq:ddG down}
\end{align}

\section{Correspondence between STF expansion of the regular field and derivatives of the regular field}\label{app:reg_field}

This section details how to relate the STF tensors featured in the decomposition of the regular field in Sec.~\ref{sec:STF_gauge_vector} to derivatives of the field evaluated on the worldline.
The first two orders match those presented in Appendix~B of Paper I but with some STF labels switched.\footnote{Eq.~(B5h) in Paper I has the prefactor \(1/6\) which has been corrected here in Eq.~\eqref{eq:K11} to \(1/3\).}

At order \(r^0\)
\begingroup
\allowdisplaybreaks
\begin{align}
    \Ahat{0,0} ={}& \hronehr_{tt}\big|_{\gamma}, \label{eq:A00} \\
    \Bhat{0,0}_{a} ={}& \hronehr_{ta}\big|_{\gamma}, \label{eq:B00} \\
    \Ehat{0,0}_{ab} ={}& \hronehr_{\langle ab\rangle}\big|_{\gamma}, \label{eq:E00} \\
    \Khat{0,0} ={}&  \frac{1}{3}\delta^{ab}\hronehr_{ab}\big|_{\gamma}. \label{eq:K00}
\end{align}%
\endgroup
At order \(r\)
\begingroup
\allowdisplaybreaks
\begin{align}
    \Ahat{1,1}_{a} ={}& \hronehr_{tt,a}\big|_\gamma, \label{eq:A11} \\
    \Bhat{1,1}_{ab} ={}& \hronehr_{t\langle a,b\rangle}\big|_\gamma, \label{eq:B11} \\
    \Chat{1,1}_{a} ={}& \frac{1}{2}\e_{a}{}^{bc}\hronehr_{tb,c}\big|_\gamma, \label{eq:C11} \\
    \Dhat{1,1} ={}& \frac{1}{3}\hronehr_{ta,}{}^{a}\big|_\gamma, \label{eq:D11} \\
    \Ehat{1,1}_{abc} ={}& \hronehr_{\langle ab,c\rangle}\big|_\gamma, \label{eq:E11} \\
    \Fhat{1,1}_{ab} ={}& \frac{2}{3}\e^{cd}{}_{(a}\hronehr_{b)c,d}\big|_\gamma, \label{eq:F11} \\
    \Ghat{1,1}_{a} ={}& \frac{3}{5}\hronehr_{\langle ab\rangle,}{}^{b}\big|_\gamma, \label{eq:G11} \\
    \Khat{1,1}_{a} ={}& \frac{1}{3}\delta^{bc}\hronehr_{bc,a}\big|_\gamma. \label{eq:K11}
\end{align}%
\endgroup
Finally at order \(r^2\)
\begingroup
\allowdisplaybreaks
\begin{align}
    \Ahat{2,0} ={}& \frac{1}{6}\hronehr_{tt,a}{}^{a}\big|_\gamma, \label{eq:A20} \\
    \Ahat{2,2}_{ab} ={}& \frac{1}{2}\hronehr_{tt,\langle ab\rangle}\big|_\gamma, \label{eq:A22} \\
    \Bhat{2,0}_{a} ={}& \frac{1}{6}\hronehr_{ta,b}{}^{b}\big|_\gamma, \label{eq:B20} \\
    \Bhat{2,2}_{abc} ={}& \frac{1}{2}\hronehr_{t\langle a,bc\rangle}\big|_\gamma, \label{eq:B22} \\
    \Chat{2,2}_{ab} ={}& \frac{1}{3}\e^{cd}{}_{(a}\hronehr_{|tc|,b)d}\big|_\gamma, \label{eq:C22} \\
    \Dhat{2,2}_{a} ={}& \frac{3}{10}\hronehr_{t}{}^{b}{}_{,\langle ab\rangle}\big|_\gamma, \label{eq:D22} \\
    \Ehat{2,0}_{ab} ={}& \frac{1}{6}\hronehr_{\langle ab\rangle,c}{}^{c}\big|_\gamma, \label{eq:E20} \\
    \Ehat{2,2}_{abcd} ={}& \frac{1}{2}\hronehr_{\langle ab,cd\rangle}\big|_\gamma, \label{eq:E22} \\
    \Fhat{2,2}_{abc} ={}& \frac{1}{2}\STF_{abc}\bigl(\e_{a}{}^{pq}\hronehr_{\langle pb\rangle,qc}\big|_\gamma\bigr), \label{eq:F22} \\
    \Ghat{2,2}_{ab} ={}& \frac{6}{7}\STF_{ab}\bigl(\hronehr_{\langle ja\rangle,}{}^{j}{}_{b}\big|_\gamma\bigr), \label{eq:G22} \\
    \Hhat{2,2}_{a} ={}& \frac{1}{5}\e_{a}{}^{cd}\hronehr_{bc,d}{}^{b}\big|_\gamma, \label{eq:H22} \\
    \Ihat{2,2} ={}& \frac{1}{10}\hronehr_{\langle ab\rangle,}{}^{ab}\big|_\gamma, \label{eq:I22}\\
    \Khat{2,0} ={}& \frac{1}{18}\hronehr_{a}{}^{a,b}{}_{b}\big|_\gamma, \label{eq:K20} \\
    \Khat{2,2}_{ab} ={}& \frac{1}{6}\hronehr_{c}{}^{c}{}_{,\langle ab\rangle}\big|_\gamma. \label{eq:K22}
\end{align}%
\endgroup

\section{Lie derivative of the first-order stress-energy tensor}\label{app:LieDT1}

In our derivation of the Detweiler stress-energy in Sec.~\ref{sec:T_HR}, the transformation from the rest gauge to the generic highly regular gauge is worldline-preserving, meaning its flow orthogonal to the worldline vanishes on the worldline. 
However, in Sec.~\ref{sec:T2gaugeInvSmooth}, we consider the change in the stress-energy under a generic smooth transformation.
As discussed in Ref.~\cite{Pound2015}, this necessitates the introduction of another Lie derivative, \(\lieZ\), which drags points of the worldline, \(z^\mu\), relative to points of the field, \(x^\mu\).
Instead of Eq.~\eqref{eq:T2_transform}, the second-order stress-energy tensor now transforms as
\begin{equation}
    T^{\mu\nu}_{2} = T^{\mu\nu}_{2'} + \left(\lie_{\xi_{1}} + \lieZ_{\xi_{1}}\right)T^{\mu\nu}_{1}. \label{eq:T2transformWorldlineMove}
\end{equation}

The Lie derivatives of $T^{\mu\nu}_1$ were previously presented by one of us in Ref.~\cite{Pound2015}. Here we reproduce (and correct a small error in) that result, and we  derive analogous results for the Lie derivatives of $ T^{1}_{\mu}{}^{\nu}$ and $T^1_{\mu\nu}$.

\subsection{Lie derivatives of $T^{\mu\nu}_1$}

Eq.~\eqref{eq:stress1} may be written so that it is invariant under reparametrisation as~\cite{Poisson2011}
\begin{align}
    T^{\mu\nu}_{1}(x;z) ={}& m \int_\gamma g^{\mu}_{\mu'}(x,z)g^{\nu}_{\nu'}(x,z)\dot{z}^{\mu'}\dot{z}^{\nu'} \nonumber \\
        & \times \frac{\delta^{4}(x,z)}{\sqrt{-g_{\rho'\sigma'}(z)\dot{z}^{\rho'}\dot{z}^{\sigma'}}} \dd{s}, \label{eq:stress1Inv}
\end{align}
where \(g^{\mu}_{\mu'}(x,z)\) is a parallel propagator from \(x^{\mu'} \coloneqq z^{\mu'}\) to \(x^{\mu}\), and $\dot z^{\mu'}:=\frac{dz^{\mu'}}{ds}$.
This form is particularly useful for our calculations of Lie derivatives of $ T^{\mu\nu}_{1}$.

The ordinary Lie derivative of Eq.~\eqref{eq:stress1Inv} is evaluated in the standard way, so
\begin{multline}
    \lie_{\xi_{1}}T^{\mu\nu}_{1} = m \int_\gamma \lie_{\xi_{1}} \biggl(g^{\mu}_{\mu'}(x,z)g^{\nu}_{\nu'}(x,z)\dot{z}^{\mu'}\dot{z}^{\nu'} \\
        \times \frac{\delta^{4}(x,z)}{\sqrt{-g_{\rho'\sigma'}(z)\dot{z}^{\rho'}\dot{z}^{\sigma'}}}\biggr) \dd{s}. \label{eq:LieXT1}
\end{multline}
The Lie derivative of the Dirac delta is found by integrating against a test function and is given by
\begin{equation}
    \lie_{\xi_{1}} \delta^{4}(x,z) = -\bigl(\xi^{\alpha'}_{1}{}_{;\alpha'} + \xi^{\alpha'}_{1} \nabla_{\alpha'}\bigr) \delta^{4}(x,z). \label{eq:LieDDirac}
\end{equation}
The other term in Eq.~\eqref{eq:LieXT1} is
\begin{align}
    \int_\gamma (\lie_{\xi_{1}}W^{\mu\nu})\delta^{4}(x,z) \dd{s} ={}& \int_\gamma \left(\xi_{1}^{\rho}W^{\mu\nu}{}_{;\rho} - 2\xi^{(\mu}_{1}{}_{;\rho}W^{\nu)\rho}\right) \nonumber \\
        & \times \delta^{4}(x,z) \dd{s} \nonumber \\
    ={}& - 2 \int_{\gamma} \xi^{(\mu}_{1}{}_{;\rho}W^{\nu)\rho} \delta^{4}(x,z) \dd{s}, \label{eq:LieXNonDelta}
\end{align}
where
\begin{equation}
    W^{\mu\nu} \coloneqq \frac{g^{\mu}_{\mu'}g^{\nu}_{\nu'}\dot{z}^{\mu'}\dot{z}^{\nu'}}{\sqrt{-g_{\rho'\sigma'}\dot{z}^{\rho'}\dot{z}^{\sigma'}}}. \label{eq:Wdef}
\end{equation}
In the second line of Eq.~\eqref{eq:LieXNonDelta}, we have used the identity \(g^{\alpha}_{\beta';\beta}\delta^{4}(x,z) = 0\)~\cite{Poisson2011} to eliminate $W^{\mu\nu}{}_{;\rho}$.
Taking our parameter $s$ to be proper time, we see that
\begin{align}
    \MoveEqLeft[4] \int_\gamma \left(\lie_{\xi_{1}}W^{\mu\nu}\right)\delta^{4}(x,z) \dd{s} \nonumber \\
    ={}& - 2 \int_\gamma g^{(\mu}_{\mu'}\xi^{\nu)}_{1}{}_{;\rho}g^{\rho}_{\nu'}u^{\mu'}u^{\nu'} \delta^{4}(x,z) \dd{\tau} \nonumber \\
    ={}& - 2 \int_\gamma g^{\mu}_{\mu'}g^{\nu}_{\nu'}u^{(\mu'}\frac{{D}\xi^{\nu')}_{1}}{\dd{\tau}}\delta^{4}(x,z) \dd{\tau}. \label{eq:LieXNonDeltaProperTime}
\end{align}
The final line is obtained by integrating the previous line against a test field $\phi_{\mu\nu}$:
\begin{align}
    \MoveEqLeft[4] \int \phi_{\mu\nu}\int_\gamma g^{(\mu}_{\mu'}\xi^{\nu)}_{1}{}_{;\rho}g^{\rho}_{\nu'}u^{\mu'}u^{\nu'} \delta^{4}(x,z) \dd{\tau} \dd{V} \nonumber \\
    ={}& \int_\gamma \phi_{\mu'\nu'}u^{(\mu'}\xi^{\nu')}_{1}{}_{;\rho'}u^{\rho'} \dd{\tau} \nonumber \\
    ={}& \int \phi_{\mu\nu} \int_\gamma g^{\mu}_{\mu'}g^{\nu}_{\nu'}u^{(\mu'}\frac{{D}\xi^{\nu')}_{1}}{\dd{\tau}}\delta^{4}(x,z) \dd{\tau}\dd{V}. \label{eq:TestFunctionLieXRearrange}
\end{align}

By combining Eqs.~\eqref{eq:LieDDirac} and~\eqref{eq:LieXNonDeltaProperTime}, we find
\begin{align}
    \MoveEqLeft[4] \int_\gamma \lie_{\xi_{1}} \biggl(\frac{g^{\mu}_{\mu'}g^{\nu}_{\nu'}\dot{z}^{\mu'}\dot{z}^{\nu'}}{\sqrt{-g_{\rho'\sigma'}\dot{z}^{\rho'}\dot{z}^{\sigma'}}} \delta^{4}(x,z)\biggr) \dd{s} \nonumber \\
    ={}& -\int_\gamma g^{\mu}_{\mu'}g^{\nu}_{\nu'}\biggl[\biggl(2u^{(\mu'}\frac{{D}\xi^{\nu')}_{1}}{\dd{\tau}} + u^{\mu'}u^{\nu'}\xi^{\rho'}_{1}{}_{;\rho'}\biggr) \nonumber \\
        & \times \delta^{4}(x,z) + u^{\mu'}u^{\nu'}\xi^{\rho'}_{1}\nabla_{\rho'}\delta^{4}(x,z)\biggr] \dd{\tau}. \label{eq:LieXT1Integrand}
\end{align}
This can be simplified by decomposing \(\xi^{\alpha'}_{1}\) into parallel and orthogonal parts, 
\beq
\xi^{\alpha'}_{1} = -u^{\alpha'}\xi^{1}_{\parallel} + \xi^{\alpha'}_{1 \perp}, 
\eeq
where $\xi^{\alpha'}_{1 \perp}:=P^{\alpha'}_{\ \,\beta'}\xi^{\beta'}_1$.
With this decomposition, we obtain
\begin{align}
    \lie_{\xi_{1}}T^{\mu\nu}_{1} ={}& - m \int_\gamma g^{\mu}_{\mu'}g^{\nu}_{\nu'}\biggl[2u^{(\mu'}\frac{{D}\xi^{\nu')}_{1\perp}}{\dd{\tau}} \delta^{4}(x,z) \nonumber \\
        & + u^{\mu'}u^{\nu'}\biggl(\xi^{\rho'}_{1}{}_{;\rho'} - \dv{\xi^{1}_{\parallel}}{\tau}\biggr)\delta^{4}(x,z) \nonumber \\
        & + u^{\mu'}u^{\nu'}\xi^{\rho'}_{1\perp}\nabla_{\rho'}\delta^{4}(x,z)\biggr] \dd{\tau}, \label{eq:LieXT1Final}
\end{align}
which agrees with Eq.~(D1) in Ref.~\cite{Pound2015} (with the correction of a  minus sign described in footnote~\ref{footnote:LieDT1}).

As discussed in Ref.~\cite{Pound2015}, because \(T_{1}^{\mu\nu}\) can be written in the form
\begin{equation}
    A^{\mu\nu}(x;z) = \int_\gamma B^{\mu\nu}(x,z(s))\sqrt{-g_{\mu'\nu'}\dot{z}^{\mu'}\dot{z}^{\nu'}} \dd{s}, \label{eq:AintB}
\end{equation}
its Lie derivative with respect to the dependence on \(z^{\mu}\) is given by
\begin{equation}
    \lieZ_{\xi_{1}}A^{\mu\nu}(x;z) = \int_\gamma \xi^{\rho'}_{1\perp} \nabla_{\rho'}B^{\mu\nu}(x,z) \dd{\tau}. \label{eq:AintBPert}
\end{equation}
For \(T^{\mu\nu}_{1}\), we see that
\begin{equation}
    B^{\mu\nu} = m \frac{g^{\mu}_{\mu'}g^{\nu}_{\nu'}\dot{z}^{\mu'}\dot{z}^{\nu'}}{- g_{\rho'\sigma'}\dot{z}^{\rho'}\dot{z}^{\sigma'}} \delta^{4}(x,z), \label{eq:BofTUU}
\end{equation}
which implies
\begin{align}
    \lieZ_{\xi_{1}}T^{\mu\nu}_{1} ={}& m\int_\gamma g^{\mu}_{\mu'}g^{\nu}_{\nu'}\biggl(2u^{(\mu'}\frac{d\xi^{\nu')}_{1\perp}}{d\tau}\delta^{4}(x,z) \nonumber \\
        & + u^{\mu'}u^{\nu'}\xi^{\rho'}_{1\perp}\nabla_{\rho'}\delta^{4}(x,z)\biggr) \dd{\tau}, \label{eq:LieZT1}
\end{align}
where we have used \(g^{\alpha}_{\beta';\gamma'}\delta^{4}(x,z) = 0\) and \(\xi_{1\perp}^{\nu}\nabla_\nu\dot z^{\mu} = \dot{z}^{\nu}\nabla_\nu\xi^{\mu}_{1\perp}\). The latter identity follows from Eq.~(B1) in Ref.~\cite{Pound2015}.

Eqs.~\eqref{eq:LieXT1Final} and~\eqref{eq:LieZT1}  sum to give
\begin{multline}
    (\lie_{\xi_1} + \lieZ_{\xi_{1}})T^{\mu\nu}_{1} = -m\int_{\gamma} g^{\mu}_{\mu'} g^{\nu}_{\nu'} u^{\mu'} u^{\nu'} \delta^4(x,z) \\
        \times \biggl(\xi^\rho_{1;\rho} - \dv{\xi^{1}_{\parallel}}{\tau}\biggr) \dd{\tau}, \label{eq:LieXZT1}
\end{multline}
which matches Eq.~(D2) from Ref.~\cite{Pound2015} (again with the missing minus sign added).
Note that this is also the same as Eq.~\eqref{eq:LieXi1T1} because \(\lieZ_{\xi_{1}}T^{\mu\nu}_{1} = 0\) for a worldline-preserving transformation.

\subsection{Lie derivatives of  $T^1_{\mu\nu}$ and $ T^{1}_{\mu}{}^{\nu}$}\label{app:LieDT1IndexPos}

The first-order stress-energy tensor with both indices down is given by
\begin{multline}
    T_{\mu\nu}^{1}(x;z) = m \int_\gamma g_{\mu\alpha}g_{\nu\beta} g^{\alpha}_{\mu'}(x,z)g^{\beta}_{\nu'}(x,z)\dot{z}^{\mu'}\dot{z}^{\nu'} \\
        \times \frac{\delta^{4}(x,z)}{\sqrt{-g_{\rho'\sigma'}(z)\dot{z}^{\rho'}\dot{z}^{\sigma'}}} \dd{s} \label{eq:stress1InvDown}
\end{multline}
and with mixed indices by
\begin{multline}
    T^{1}_{\mu}{}^{\nu}(x;z) = m \int_\gamma g_{\mu\alpha}g^{\alpha}_{\mu'}(x,z)g^{\nu}_{\nu'}(x,z)\dot{z}^{\mu'}\dot{z}^{\nu'} \\
        \times \frac{\delta^{4}(x,z)}{\sqrt{-g_{\rho'\sigma'}(z)\dot{z}^{\rho'}\dot{z}^{\sigma'}}} \dd{s}. \label{eq:stress1InvMixed}
\end{multline}
To calculate the ordinary Lie derivatives of these quantities, we follow the same methods described above. The results are
\begin{align}
     \lie_{\xi_{1}}T_{\mu\nu}^{1} ={}& m \int_\gamma g_{\mu\alpha}g_{\nu\beta}g^{\alpha}_{\alpha'}g^{\beta}_{\beta'}\bigl(2\xi^{1}_{\rho';}{}^{(\alpha'}u^{\beta')}u^{\rho'} - u^{\alpha'}u^{\beta'} \nonumber \\
        & \times \bigl[\xi^{\rho'}_{1}{}_{;\rho'} + \dot{\xi}^{1}_{\parallel} + \xi^{\rho'}_{1\perp}\nabla_{\rho'}\bigr]\bigr) \delta^{4}(x,z) \dd{\tau}, \label{eq:LieXT1DD} \displaybreak[0] \\
    \lie_{\xi_{1}}T^{1}_{\mu}{}^{\nu} ={}& m \int_\gamma g_{\mu\alpha}g^{\alpha}_{\alpha'}g^{\nu}_{\nu'}\bigl(\xi^{1}_{\rho';}{}^{\alpha'}u^{\nu'}u^{\rho'} - \xi^{\nu'}_{1}{}_{;\rho'}u^{\alpha'}u^{\rho'} \nonumber \\
        & - u^{\alpha'}u^{\nu'}\bigl[\xi^{\rho'}_{1}{}_{;\rho'} + \dot{\xi}^{1}_{\parallel} + \xi^{\rho'}_{1\perp}\nabla_{\rho'}\bigr]\bigr) \delta^{4}(x,z) \dd{\tau}. \label{eq:LieXT1DU}
\end{align}
Here and below, an overdot denotes a derivative with respect to $\tau$.

The Lie derivatives at $z^\mu$ follow trivially from Eq.~\eqref{eq:LieZT1}. Since we can pass the contraction through the derivative, as in $g_{\mu\rho}\lieZ_{\xi_{1}}T^{\rho\nu}_1 = \lieZ_{\xi_{1}}(g_{\mu\rho}T^{\rho\nu}_1)$, we get
\begin{align}
    \lieZ_{\xi_{1}}T_{\mu\nu}^{1} ={}& \int_\gamma g_{\mu\alpha}g_{\nu\beta}g^{\alpha}_{\mu'}g^{\beta}_{\nu'}\biggl(2u^{(\mu'}\dot{\xi}^{\nu')}_{1\perp}\delta^{4}(x,z) \nonumber \\
        & + u^{\mu'}u^{\nu'}\xi^{\rho'}_{1\perp}\nabla_{\rho'}\delta^{4}(x,z)\biggr) \dd{\tau}, \label{eq:LieZT1DD} \displaybreak[0] \\
    \lieZ_{\xi_{1}}T^{1}_{\mu}{}^{\nu} ={}& \int_\gamma g_{\mu\alpha}g^{\alpha}_{\mu'}g^{\beta}_{\nu'}\biggl(2u^{(\mu'}\dot{\xi}^{\nu')}_{1\perp}\delta^{4}(x,z) \nonumber \\
        & + u^{\mu'}u^{\nu'}\xi^{\rho'}_{1\perp}\nabla_{\rho'}\delta^{4}(x,z)\biggr) \dd{\tau}. \label{eq:LieZT1DU}
\end{align}

Combining these results, we find
\begin{align}
    \left(\lie_{\xi_{1}} + \lieZ_{\xi_{1}}\right)T^{1}_{\mu\nu} ={}& m\int_\gamma g_{\mu\alpha}g_{\nu\beta}g^{\alpha}_{\alpha'}g^{\beta}_{\beta'}\bigl(2\xi^{1}_{\rho';}{}^{(\alpha'}u^{\beta')}u^{\rho'} \nonumber \\
        & + 2u^{(\alpha'}\dot{\xi}_{1}^{\beta')} + u^{\alpha'}u^{\beta'}\dot{\xi}^{1}_{\parallel} \nonumber \\
        & - u^{\alpha'}u^{\beta'}\xi^{\rho'}_{1;\rho'}\bigr)\delta^4(x,z) \dd{\tau}, \label{eq:LieXi1T1DD} \displaybreak[0] \\
    \left(\lie_{\xi_{1}} + \lieZ_{\xi_{1}}\right)T^{1}_{\mu}{}^{\nu} ={}& m\int_\gamma g_{\mu\alpha}g^{\alpha}_{\alpha'}g^{\nu}_{\nu'}u^{\nu'}\bigl(\xi^{1}_{\parallel;}{}^{\alpha'} - \xi^{\rho'}_{1;\rho'}u^{\alpha'} \nonumber \\
        & + \dot{\xi}^{\alpha'}_{1} + u^{\alpha'}\dot{\xi}^1_{\parallel}\bigr)\delta^{4}(x,z) \dd{\tau}. \label{eq:LieXi1T1DU}
\end{align}

\bibliography{bibfile}
\end{document}